\documentclass[times]{aastex701}

\newcommand\benspaper{J26}
\newcommand\staceyspaper{A26}
\usepackage{amsmath}
\usepackage{relsize}
\begin{document}

\title{JWST Advanced Deep Extragalactic Survey (JADES) Data Release 5: Photometric Catalog}

\author[orcid=0000-0002-4271-0364]{Brant E. Robertson}
\affiliation{Department of Astronomy and Astrophysics, University of California, Santa Cruz, 1156 High Street, Santa Cruz, CA 95064, USA}
\email[show]{brant@ucsc.edu}  

\author[orcid=0000-0002-9280-7594]{Benjamin D.\ Johnson}
\affiliation{Center for Astrophysics $|$ Harvard \& Smithsonian, 60 Garden St., Cambridge MA 02138 USA}
\email{benjamin.johnson@cfa.harvard.edu}  

\author[orcid=0000-0002-8224-4505]{Sandro Tacchella}
\affiliation{Kavli Institute for Cosmology, University of Cambridge, Madingley Road, Cambridge, CB3 0HA, UK}
\affiliation{Cavendish Laboratory, University of Cambridge, 19 JJ Thomson Avenue, Cambridge, CB3 0HE, UK}
\email{st578@cam.ac.uk}  

\author[orcid=0000-0002-2929-3121]{Daniel J.\ Eisenstein}
\affiliation{Center for Astrophysics $|$ Harvard \& Smithsonian, 60 Garden St., Cambridge MA 02138 USA}
\email{deisenstein@cfa.harvard.edu}  

\author[orcid=0000-0003-4565-8239]{Kevin Hainline}
\affiliation{Steward Observatory, University of Arizona, 933 N. Cherry Avenue, Tucson, AZ 85721, USA}
\email{kevinhainline@arizona.edu} 

\author[orcid=0000-0002-8909-8782]{Stacey Alberts}
\affiliation{Steward Observatory, University of Arizona, 933 N. Cherry Avenue, Tucson, AZ 85721, USA}
\affiliation{AURA for the European Space Agency (ESA), Space Telescope Science Institute, 3700 San Martin Dr., Baltimore, MD 21218, USA}
\email{salberts@stsci.edu} 

\author[orcid=0000-0001-7997-1640]{Santiago Arribas}
\affiliation{Centro de Astrobiolog\'ia (CAB), CSIC–INTA, Cra. de Ajalvir Km.~4, 28850- Torrej\'on de Ardoz, Madrid, Spain}
\email{arribas@cab.inta-csic.es} 

\author[orcid=0000-0003-0215-1104]{William M. Baker}
\affiliation{DARK, Niels Bohr Institute, University of Copenhagen, Jagtvej 155A, DK-2200 Copenhagen, Denmark}
\email{william.baker@nbi.ku.dk} 

\author[orcid=0000-0002-8651-9879]{Andrew J.\ Bunker}
\affiliation{Department of Physics, University of Oxford, Denys Wilkinson Building, Keble Road, Oxford OX1 3RH, UK}
\email{andy.bunker@physics.ox.ac.uk} 

\author[orcid=0000-0002-0450-7306]{Alex J. Cameron}
\affiliation{Department of Physics, University of Oxford, Denys Wilkinson Building, Keble Road, Oxford OX1 3RH, UK}
\email{alex.cameron@physics.ox.ac.uk} 

\author[orcid=0000-0002-6719-380X]{Stefano Carniani}
\affiliation{Scuola Normale Superiore, Piazza dei Cavalieri 7, I-56126 Pisa, Italy}
\email{stefano.carniani@sns.it} 

\author[orcid=0000-0001-6301-3667]{Courtney Carreira}
\affiliation{Department of Astronomy and Astrophysics, University of California, Santa Cruz, 1156 High Street, Santa Cruz, CA 95064, USA}
\email{ccarreir@ucsc.edu} 

\author[orcid=0000-0002-7636-0534]{Jacopo Chevallard}
\affiliation{Department of Physics, University of Oxford, Denys Wilkinson Building, Keble Road, Oxford OX1 3RH, UK}
\email{chevalla@iap.fr} 

\author[orcid=0000-0001-8522-9434]{Chiara Circosta}
\affiliation{Institut de Radioastronomie Millim\'etrique (IRAM), 300 Rue de la Piscine, 38400 Saint-Martin-d'H\'eres, France}
\email{circosta@iram.fr}

\author[orcid=0000-0002-9551-0534]{Emma Curtis-Lake}
\affiliation{Centre for Astrophysics Research, Department of Physics, Astronomy and Mathematics, University of Hertfordshire, Hatfield AL10 9AB, UK}
\email{e.curtis-lake@herts.ac.uk} 

\author[orcid=0000-0002-9708-9958]{A. Lola Danhaive}
\affiliation{Kavli Institute for Cosmology, University of Cambridge, Madingley Road, Cambridge, CB3 0HA, UK}
\affiliation{Cavendish Laboratory, University of Cambridge, 19 JJ Thomson Avenue, Cambridge, CB3 0HE, UK}
\email{ald66@cam.ac.uk} 

\author[orcid=0009-0009-8105-4564]{Qiao Duan}
\affiliation{Kavli Institute for Cosmology, University of Cambridge, Madingley Road, Cambridge, CB3 0HA, UK}
\affiliation{Cavendish Laboratory, University of Cambridge, 19 JJ Thomson Avenue, Cambridge, CB3 0HE, UK}
\email{qd231@cam.ac.uk} 

\author[orcid=0000-0003-1344-9475]{Eiichi Egami}
\affiliation{Steward Observatory, University of Arizona, 933 N. Cherry Avenue, Tucson, AZ 85721, USA}
\email{egami@arizona.edu} 

\author[orcid=0000-0002-8543-761X]{Ryan Hausen}
\affiliation{Department of Physics and Astronomy, The Johns Hopkins University, 3400 N. Charles St., Baltimore, MD 21218}
\email{rhausen@ucsc.edu} 

\author[orcid=0000-0003-4337-6211]{Jakob M. Helton}
\affiliation{Department of Astronomy \& Astrophysics, The Pennsylvania State University, University Park, PA 16802, USA}
\email{jakobhelton@psu.edu}

\author[orcid=0000-0001-7673-2257]{Zhiyuan Ji}
\affiliation{Steward Observatory, University of Arizona, 933 N. Cherry Avenue, Tucson, AZ 85721, USA}
\email{zhiyuanji@arizona.edu} 

\author[orcid=0000-0002-4985-3819]{Roberto Maiolino}
\affiliation{Kavli Institute for Cosmology, University of Cambridge, Madingley Road, Cambridge, CB3 0HA, UK}
\affiliation{Cavendish Laboratory, University of Cambridge, 19 JJ Thomson Avenue, Cambridge, CB3 0HE, UK}
\affiliation{Department of Physics and Astronomy, University College London, Gower Street, London WC1E 6BT, UK}
\email{rm665@cam.ac.uk}

\author[orcid=0000-0003-4528-5639]{Pablo G. P\'erez-Gonz\'alez}
\affiliation{Centro de Astrobiolog\'ia (CAB), CSIC–INTA, Cra. de Ajalvir Km.~4, 28850- Torrej\'on de Ardoz, Madrid, Spain}
\email{pgperez@cab.inta-csic.es} 

\author[orcid=0000-0001-8630-2031]{D\'avid Pusk\'as}
\affiliation{Kavli Institute for Cosmology, University of Cambridge, Madingley Road, Cambridge, CB3 0HA, UK}
\affiliation{Cavendish Laboratory, University of Cambridge, 19 JJ Thomson Avenue, Cambridge, CB3 0HE, UK}
\email{dp670@cam.ac.uk}

\author[orcid=0000-0002-7893-6170]{Marcia Rieke}
\affiliation{Steward Observatory, University of Arizona, 933 N. Cherry Avenue, Tucson, AZ 85721, USA}
\email{mrieke@gmail.com} 

\author[orcid=0000-0002-5104-8245]{Pierluigi Rinaldi}
\affiliation{Space Telescope Science Institute, 3700 San Martin Drive, Baltimore, Maryland 21218, USA}
\email{prinaldi@stsci.edu} 

\author[orcid=0000-0002-4622-6617]{Fengwu Sun}
\affiliation{Center for Astrophysics $|$ Harvard \& Smithsonian, 60 Garden St., Cambridge MA 02138 USA}
\email{fengwu.sun@cfa.harvard.edu}

\author[orcid=0000-0001-6561-9443]{Yang Sun}
\affiliation{Steward Observatory, University of Arizona, 933 N. Cherry Avenue, Tucson, AZ 85721, USA}
\email{sunyang@arizona.edu} 

\author[orcid=0000-0003-4891-0794]{Hannah \"Ubler}
\affiliation{Max-Planck-Institut f\"ur extraterrestrische Physik (MPE), Gie{\ss}enbachstra{\ss}e 1, 85748 Garching, Germany}
\email{hannah@mpe.mpg.de}

\author[orcid=0000-0002-9081-2111]{James A. A. Trussler}
\affiliation{Center for Astrophysics $|$ Harvard \& Smithsonian, 60 Garden St., Cambridge MA 02138 USA}
\email{james.trussler@cfa.harvard.edu}

\author[orcid=0000-0001-6917-4656]{Natalia C. Villanueva}
\affiliation{Department of Astronomy, The University of Texas at Austin, Austin, TX, USA}
\email{nataliavillanueva@utexas.edu}

\author[orcid=0000-0003-1432-7744]{Lily Whitler}
\affiliation{Kavli Institute for Cosmology, University of Cambridge, Madingley Road, Cambridge, CB3 0HA, UK}
\affiliation{Cavendish Laboratory, University of Cambridge, 19 JJ Thomson Avenue, Cambridge, CB3 0HE, UK}
\email{lw851@cam.ac.uk}  

\author[orcid=0000-0003-2919-7495]{Christina C. Williams}
\affiliation{NSF National Optical-Infrared Astronomy Research Laboratory, 950 North Cherry Avenue, Tucson, AZ 85719, USA}
\email{christina.williams@noirlab.edu} 

\author[orcid=0000-0001-9262-9997]{Christopher N. A. Willmer}
\affiliation{Steward Observatory, University of Arizona, 933 N. Cherry Avenue, Tucson, AZ 85721, USA}
\email{cnaw@as.arizona.edu} 

\author[orcid=0000-0002-4201-7367]{Chris Willott}
\affiliation{NRC Herzberg, 5071 West Saanich Rd, Victoria, BC V9E 2E7, Canada}
\email{chris.willott@nrc.ca}

\author[orcid=0000-0002-8876-5248]{Zihao Wu}
\affiliation{Center for Astrophysics $|$ Harvard \& Smithsonian, 60 Garden St., Cambridge MA 02138 USA}
\email{zihao.wu@cfa.harvard.edu} 

\author[orcid=0000-0003-3307-7525]{Yongda Zhu}
\affiliation{Steward Observatory, University of Arizona, 933 N. Cherry Avenue, Tucson, AZ 85721, USA}
\email{yongdaz@arizona.edu} 

\collaboration{all}{The JADES Collaboration}

\begin{abstract}

The JWST Advanced Deep Extragalactic Survey (JADES) provides deep, multi-wavelength imaging and spectroscopy of the GOODS-North and GOODS-South fields, enabling studies of galaxy formation and evolution from the local universe to the first few hundred million years after the Big Bang. We present the public release of the JADES Data Release 5 (DR5) photometric catalogs and describes the methodologies used for source detection, deblending, photometry, uncertainty estimation, and catalog curation. The catalogs are constructed from 35 space-based imaging mosaics obtained with JWST/NIRCam, JWST/MIRI, HST/ACS, and HST/WFC3, combining approximately 1250 hours of JADES imaging with extensive additional public JWST and HST observations in the GOODS fields. Sources are identified using custom signal-to-noise–based detection and deblending algorithms optimized for the depth, resolution, and complex point-spread-function structure of JWST imaging. Source centroids, shapes, and photometric apertures are determined using a new fast two-dimensional Gaussian regression method applied to detection-image profiles. We provide forced circular-aperture photometry, ellipsoidal Kron photometry, and curve-of-growth measurements for every source in every band. We introduce a new pixel-level regression framework to model photometric uncertainties as a function of aperture size and local mosaic properties, accounting for correlated noise in heterogeneous JWST mosaics. Photometric redshifts are computed using template-based fitting applied to both small-aperture photometry on unconvolved images and Kron photometry on common-PSF mosaics. The JADES DR5 catalogs supersede previous JADES photometric releases, and are publicly released through the Mikulski Archive for Space Telescopes and an interactive web interface.
\end{abstract}


\section{Introduction}
\label{sec:intro} 

In the span of a few short years,
James Webb Space Telescope (JWST) has completely
rewritten our view of the distant cosmos.
By discovering a host of galaxies at
distances beyond what was widely thought probable
\citep[e.g.,][]{castellano2022a,naidu2022a,finkelstein2022a,robertson2023a,curtis-lake2023a,hainline2024a,robertson2024a,carniani2024a,naidu2025a},
JWST has expanded the horizon of the universe
of tangible things past redshift $z\sim14$
to the first 300 million years of cosmic time.
Such discoveries have already addressed many
of the open questions about the epoch
of cosmic reionization left unanswered
before JWST \citep[for a review, see][]{robertson2022a}
and posed even more about the efficiency 
of early galaxy formation, the origins of the
first stars, and the beginnings of cosmic reionization.

The data used to search for the most distant
galaxies serves many scientific purposes, and the
uniform reduction and analysis of extragalactic
survey fields  provide a foundation for studies of galaxy formation and evolution across cosmic time \citep[e.g.,][]{treu2022a,windhorst2023a,oesch2023a,casey2023a,finkelstein2025a}, including the assembly of stellar mass \citep{weibel2024a,shuntonv2025b}, the growth of structure \citep[e.g.,][]{weibel2025a}, and the emergence of diverse galaxy populations \citep{donnan2024a}, while simultaneously enabling the discovery of faint sources in the nearby Universe, such as brown dwarfs in the Milky Way
\citep[e.g.][]{langeroodi2023a,burgasser2024a,hainline2025a}.
In this context, several major JWST/NIRCam imaging programs have released deep extragalactic datasets accompanied by public photometric catalogs, including CEERS \citep{cox2025a} and COSMOS-Web \citep{shuntov2025a}. Independent analyses have generated source catalogs from
public JWST imaging, such as the Astrodeep project \citep{merlin2024a}.
Together, these efforts span a wide range of survey strategies in depth, area, and filter coverage, and have rapidly transformed studies of galaxy evolution, from the statistical properties of galaxy populations to the identification of rare objects at both high and low redshift.

As part of the JWST Advanced Deep Extragalactic Survey \citep[JADES;][]{rieke2020a,bunker2020a,eisenstein2023a}
Data Release 5 (DR5), this paper presents the
detection, deblending, and source characterization methods,
as well as catalogs containing forced aperture photometry,
photometric redshifts, and curves-of-growth for 
35 space-based mosaic images acquired with JWST NIRCam,
JWST MIRI, HST ACS, and HST WFC3 over the Great Observatories
Origins Deep Survey (GOODS) North and South Fields.
The catalogs are constructed from sources detected in the
JADES DR5 mosaics detailed in a companion paper 
\citep[][hereafer \benspaper{}]{johnson2026a}.

We provide an overview of imaging data from \benspaper{}
used for cataloging in Section \ref{sec:imaging}. Since these
images leverage public JWST community surveys in GOODS-N and GOODS-S
in addition to the $\sim1250$ JWST observation hours of JADES
imaging in these fields, we provide detailed references to all the
JWST programs used in constructing our catalogs in Section \ref{sec:imaging}.
As discussed in \benspaper{} and Section \ref{sec:bithash}, 
the JADES DR5 release provides
a new methodology for hash encoding the list of
JWST programs contributing to the measurements for every
source and reports that information in the catalogs.
The methods for constructing detection and deblending images,
identifying and separating sources, and curating the resulting 
source catalogs are detailed in Section \ref{sec:detection}.
We present the techniques we employ to model the individual source properties, including
a new method for fast two-dimensional Gaussian regression for
modeling the source flux profiles, and determine the \citet{kron1980a} photometry
ellipsoidal apertures in Section \ref{sec:source-catalog}.
Section \ref{sec:psf} details the model JWST NIRCam and MIRI and 
empirical HST point spread function construction, followed by
a presentation of the methodology for creating common-PSF images 
in Section \ref{sec:common-psf-mosaics}. The schema for data
quality and context flagging for the catalogs is discussed
in Section \ref{sec:flagging}. 
In Section \ref{sec:uncertainties},
we detail a new regression model for generating mosaics of the photometric
uncertainty scaling, using the pipeline variance image as a
template for the single-pixel limit of the sky background.
Forced circular aperture and Kron photometry on all sources are
detailed in Section \ref{sec:forced-photometry}, and the
curve-of-growth measurements for all sources discussed in 
Section \ref{sec:curve-of-growth}. The photometric
redshift catalog generation is presented in Section \ref{sec:photometric-redshifts}.
We summarize the main text and provide data access instructions in 
Section \ref{sec:summary}. The Appendix \ref{sec:catalog-format} provides detail on
the format for all components of the catalog.

\section{Overview of Imaging Data}
\label{sec:imaging} 

Table \ref{tab:imaging} summarizes the filter set, areal coverage, and median depths of the HST/ACS, HST/WFC3, JWST/NIRCam, and JWST/MIRI imaging data. The JADES NIRCam Data Release 5 imaging data are presented by
\benspaper{}, including all the NIRCam filter mosaics
that form the core dataset for the detection, deblending,
and photometric analysis presented in this work. 
The
NIRCam data are drawn primarily from the JADES
JWST Programs 1180, 1181, 1210, 1286, and 1287 \citep{eisenstein2023a}, and the 
JADES Origins Field Program 3215 \citep{eisenstein2025a}.
We also include deep NIRCam imaging in the same fields from the 
Next Generation Deep Extragalactic Exploratory Public
(NGDEEP) Survey \citep{bagley2024a}, \citep{ostlin2025a}, the
MIRI Deep Imaging Survey \citep[MIDIS;][]{perez-gonzalez2024a,ostlin2025a},
the JWST Extragalactic Medium-band Survey \citep[JEMS;][]{williams2023a}, Prime Extragalactic Areas for Reionization and Lensing Science (PEARLS) Survey
\citep{windhorst2023a}, the  Bias-free Extragalactic Analysis for Cosmic Origins with NIRCam survey \citep{morishita2025a}, and JWST Program 6511 \citep{perez-gonzalez2025a}.
The mosaics incorporate wider area data from the First Reionization Epoch Spectroscopically Complete
Observations (FRESCO) Survey \citep{oesch2023a}, the 
Parallel
wide-Area Nircam Observations to Reveal And Measure the
Invisible Cosmos (PANORAMIC) Survey \citep{williams2025a},
 the Public Observation Pure Parallel Infrared Emission-Line Survey
\citep[POPPIES;][]{kartaltepe2024a},
and the Slitless Areal Pure-Parallel HIgh-Redshift
Emission Survey \citep[SAPPHIRES;][]{sun2025a}.

For the MIRI data, the JADES MIRI parallel imaging described
by 
\citet[][hereafter \staceyspaper{}]{alberts2026a}
is combined
with the Systematic Mid-infrared Instrument Legacy Extragalactic Survey
\citep[SMILES;][]{rieke2024a} public release mosaics \citep{alberts2024a},
following the mosaic stacking methods 
described in \benspaper{}. JADES MIRI parallel imaging includes
deep F770W imaging in GOODS-S, medium F770W, F1280W, and F1500W
parallels in GOODS-S, and F770W and F1280W parallels in GOODS-N.
The F560W, F1800W, F2100W, and F2550W imaging used for our catalogs
are taken directly from the SMILES release and reprojected
to our mosaic pixel footprint before processing.

For the HST data, we use the Hubble Legacy Fields (HLF) mosaics
\citep{illingworth2016a,whitaker2019a} and include HST/ACS F435W,
F606W, F775W, F814W, F850LP and HST/WFC3 F105W, F125W,
F140W, and F160W imaging in our analysis. 
We use the HLF GOODS-S v2.0 and GOODS-N v2.5 mosaics,
re-registered to the JWST NIRCam mosaic astrometric frame,
reprojected to the JWST mosaic pixel scale,
and culled to the same mosaic footprint.

\begin{deluxetable}{lccccc}[!ht]
\tablecaption{Summary of Imaging Data from \benspaper{} \label{tab:imaging}}
\tablehead{
\colhead{Filter} & \colhead{Wavelength} & \colhead{GOODS-S Area} & \colhead{GOODS-S Depth} & \colhead{GOODS-N Area} & \colhead{GOODS-N Depth} \vspace{-0.2cm}\\
\colhead{} & \colhead{[$\mu$m]} & \colhead{[arcmin$^{2}$]} & \colhead{[AB]} & \colhead{[arcmin$^{2}$]} & {[AB]}
}
\startdata
HST/ACS     F435W   & 0.433 &272.19&28.55&222.64&28.65\\
HST/ACS     F606W   & 0.592 &481.81&28.24&233.37&28.67\\
HST/ACS     F775W   & 0.769 &284.31&28.12&241.40&28.35\\
HST/ACS     F814W   & 0.806 &390.25&27.99&295.73&28.52\\
HST/ACS     F850LP  & 0.904 &479.14&27.08&240.16&27.77\\
HST/WFC3    F105W   & 1.055 &118.05&27.89&153.72&27.61\\
HST/WFC3    F125W   & 1.249 &181.41&27.77&170.90&27.41\\
HST/WFC3    F140W   & 1.392 &151.56&26.40&117.46&26.66\\
HST/WFC3    F160W   & 1.537 &221.71&27.21&170.78&27.22\\
JWST/NIRCam F070W   & 0.705 &72.18&29.18&30.40&28.87\\
JWST/NIRCam F090W   & 0.902 &159.42&29.66&125.42&29.43\\
JWST/NIRCam F115W   & 1.154 &210.47&29.78&180.37&28.91\\
JWST/NIRCam F150W   & 1.501 &213.93&29.83&139.16&29.04\\
JWST/NIRCam F162M   & 1.627 &33.31&29.65&14.77&28.87\\
JWST/NIRCam F182M   & 1.845 &87.81&29.08&109.50&28.73\\
JWST/NIRCam F200W   & 1.988 &233.71&29.73&174.31&29.40\\
JWST/NIRCam F210M   & 2.096 &76.63&28.67&78.98&28.41\\
JWST/NIRCam F250M   & 2.503 &36.17&29.29&---&---\\
JWST/NIRCam F277W   & 2.777 &210.44&30.07&115.01&29.61\\
JWST/NIRCam F300M   & 2.996 &34.50&29.85&9.07&29.07\\
JWST/NIRCam F335M   & 3.362 &119.45&29.71&78.62&29.12\\
JWST/NIRCam F356W   & 3.565 &214.16&30.07&155.14&29.38\\
JWST/NIRCam F410M   & 4.084 &157.52&29.42&110.79&28.85\\
JWST/NIRCam F430M   & 4.281 &10.07&28.71&8.74&28.20\\
JWST/NIRCam F444W   & 4.402 &231.18&29.60&207.53&28.84\\
JWST/NIRCam F460M   & 4.630 &10.08&28.28&8.77&27.78\\
JWST/NIRCam F480M   & 4.817 &18.07&28.65&---&---\\
JWST/MIRI   F560W   & 5.645 &35.30&25.65&---&---\\
JWST/MIRI   F770W   & 7.639 &58.34&25.77&9.46&26.50\\
JWST/MIRI   F1000W  & 9.953 &34.57&24.51&---&---\\
JWST/MIRI   F1280W  & 12.810 &52.48&23.81&9.28&24.71\\
JWST/MIRI   F1500W  & 15.064 &58.13&23.53&---&---\\
JWST/MIRI   F1800W  & 17.983 &36.06&22.31&---&---\\
JWST/MIRI   F2100W  & 20.795 &36.02&21.76&---&---\\
JWST/MIRI   F2550W  & 25.364 &36.23&19.57&---&---\\
\enddata
\tablecomments{We report aperture-corrected 5-$\sigma$ point source depths, computed as the median value of the model uncertainty image for $r=0.1"$ apertures calculated following the method described in Section \ref{sec:uncertainty-mosaics}. As discussed in \citet{eisenstein2025a}, for NIRCam these depths agree well with expectations from the JWST Exposure Time Calculator \citep{pontoppidan2016a}.}
\end{deluxetable}

\section{Detection}
\label{sec:detection}

The depth and resolution of the JWST imaging
provides an opportunity to construct
sensitive multi-band stacks to detect
faint sources and low-surface brightness
exteriors of extended objects, separate
complex interacting and overlapping sources,
and accurately measure the shapes and
sizes of galaxies across cosmic time. 
At a high-level our approach for object
detection and segmentation follows the
previous approaches employed in prior
JADES data releases and analyses
\citep[][]{rieke2023a,eisenstein2025a,deugenio2025a,ji2024a,robertson2024a},
but  some of the details are presented
thoroughly here for the first time.

The key features of the detection process
include the construction of the detection and
deblending images, which are composed
of signal-to-noise ratio (SNR)  stacks of the
mosaics presented by \benspaper{}, and the
detection and deblending algorithms. Separate
signal (i.e., flux) and noise stacked
images are generated, and then the ratio
of the images used to construct an SNR
mosaic when appropriate. We present
the details of how
these signal and noise mosaics are constructed
and combined into detection and deblending images
below in Sections \ref{sec:noise-images}-\ref{sec:deblending-image}.
The custom detection and deblending methods
that identify and separate distinct sources
are described in Sections \ref{sec:detection-algorithm},
\ref{sec:blended-segmentation}, and
\ref{sec:deblending}. These methods have been
demonstrated previously to provide high-quality
detection completeness for faint sources 
\citep{robertson2024a} and have found success in 
identifying very distant high-redshift sources that 
are both compact and isolated \citep{robertson2023a}
or extended and confused with foreground interlopers
\citep[e.g.,][]{robertson2024a,hainline2024a,carniani2024a}.
Once the sources have been detected and deblended, the
catalogs are curated as described in Section \ref{sec:catalog-curation},
source structural properties measured (Section \ref{sec:source-properties}),
and then photometry performed (Section \ref{sec:forced-photometry}).

\begin{figure*}[ht!]
\includegraphics[width=1.0\textwidth]{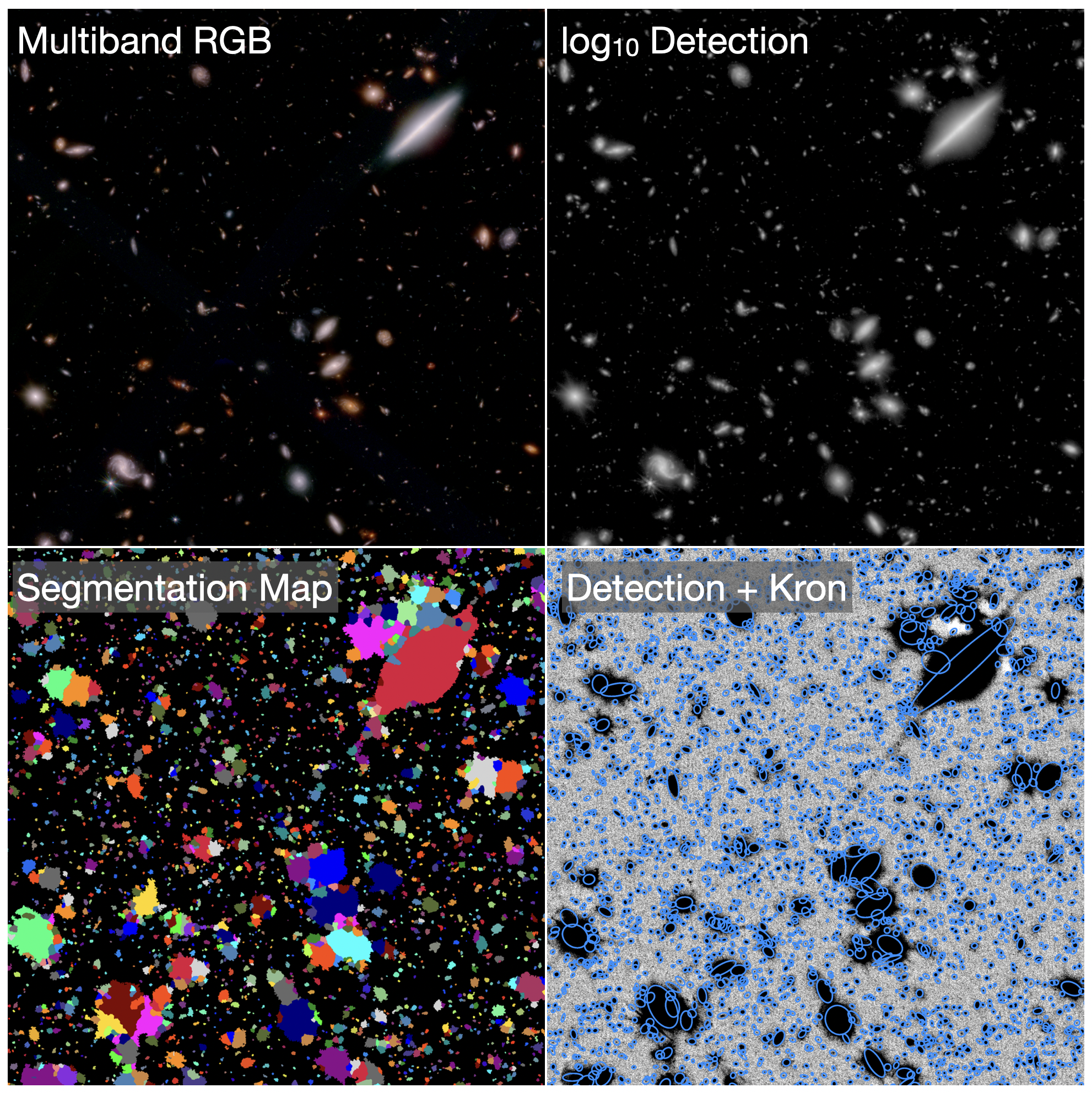}
\caption{Overview of the detection and deblending process. The multi-band NIRCam imaging from \benspaper{} forms the core data used to
detect and isolate sources, shown as an RGB composite of the JADES Origins Field region \citep{eisenstein2025a} of GOODS-S (upper left). The long-wavelength NIRCam images are stacked to form a deep multi-band detection image that provides the pixel-level signal-to-noise ratio of sources (upper right; logarithmic projection), while a multi-band stack
of short-wavelength NIRCam filters near $\lambda\sim2\mu$m are used to deblend sources. Using the detection and deblending methods described in Section \ref{sec:detection}, a segmentation map of pixel assignments to individual sources is created (lower left). The properties of these pixels are used in a new Gaussian regression model to determine source sizes and \citet{kron1980a} photometry apertures for each source, shown as ellipsoidal overlays on a harsh stretch of the detection image in the lower right panel and described in Section \ref{sec:source-catalog}.
\label{fig:detection}}
\end{figure*}

Figure \ref{fig:detection} illustrates the process of detection
and source characterization detailed in this Section. The upper
left panel shows an RGB composite of the multi-band imaging for
a portion of the JADES Origins Field \citep{eisenstein2025a} in GOODS-S from \benspaper{}, including the source
JADES-GS-z14-0 \citep{carniani2024a}. The long-wavelength NIRCam
filters are stacked to create the detection image (a logarithmic
projection is shown in the upper right panel of Figure \ref{fig:detection}) and the short-wavelength NIRCam filters near $\lambda\approx2\mu$m are used
during deblending. The result of the detection and deblending 
algorithm produces a segmentation map, as shown in the lower left
panel. The Gaussian regression modeling of the source properties
then (see Section \ref{sec:source-catalog} below) allow for the estimation of Kron photometry apertures, as shown as
overlays on a harsh stretch of the detection image in the lower right panel of Figure \ref{fig:detection}. We now turn to describing the
detection image construction, and the detection and deblending methods.

\subsection{Noise Images}
\label{sec:noise-images}

To construct a weighted SNR image
for the JWST NIRCam images, the
signal and noise image for every filter
must be defined. The noise image can
be directly estimated from the \texttt{ERR}
HDU image extension supplied by the \texttt{jwst}
pipeline mosaics constructed as 
described by \benspaper. However, since
the noise image appears in the denominator,
unexpected pixel-level variations in the noise
can lead to artificial spikes in the 
significance level of detected objects.
For various versions of the \texttt{jwst}
pipeline this indeed occurred, with 
occasional small but non-zero pixels present
in the \texttt{ERR} images leading to
spikes in the resulting SNR images
used for detection. To avoid such possible
issues, we construct a separate ``noise
mosaic'' from the \texttt{ERR} mosaics.
For each filter, the \texttt{ERR} mosaic is squared to form a variance mosaic that is median filtered with a $5\times5$ boxcar window. The square root of the filtered variance mosaic is then computed to obtain a filtered noise mosaic. Finally, the ratio of the original \texttt{ERR} mosaic and the filtered noise mosaic is calculated, and any non-zero pixel in the original \texttt{ERR} mosaic with a ratio $<0.9$ is replaced by the corresponding pixel in the filtered noise mosaic. This
procedure successfully eliminated the \texttt{ERR} 
mosaic divots occasionally present in the 
images processed using the \texttt{jwst} pipeline
and prevented the corresponding SNR spikes that
could lead to spurious detections.

\subsection{Detection Image}
\label{sec:detection-image}

For our detection images, we compute
inverse-variance-weighted signal-to-noise
ratio mosaics by combining the individual
mosaics from JWST long-wavelength filters.
The variance images used in weighting are determined
by squaring the noise mosaics constructed as
described in Section \ref{sec:noise-images}.
The combined signal mosaic is constructed by summing
the \texttt{SCI} mosaic for each filter,
inversely weighted by its variance image, and
then dividing by the sum of the inverse variance
images. The combined noise mosaic is computed
by taking the inverse square root of the sum
of the inverse variance images. The SNR image
used for detection is the ratio of these combined
signal and noise mosaics. We note that we use
the mosaics without PSF-matching to construct the
SNR images. Using common-PSF images could reduce
chromaticity in the detection image at the expense
of reduced signal-to-noise in low surface brightness
regions around galaxies.

The GOODS-N and GOODS-S fields have differing
filter coverage. In GOODS-N, we combine
F277W, F335M, F356W, F410M, F430M, F444W, and
F460M
where available in constructing the
detection image. In GOODS-S, we additionally
include F480M where available.

In GOODS-N, in the region of the field
covered with LW data only from the POPPIES program (PID 5398),
the F444W imaging is shallow. For improved
signal-to-noise, we therefore add the F200W
subregion mosaics from the \texttt{jw053980.gn\_pa220},
\texttt{jw053980.gn\_pa227}, and \texttt{jw053980.gn\_pa230}
regions to the detection image. See \benspaper{}
for a description of the subregion mosaics.

The JADES imaging includes MIRI parallel observations
in GOODS-N and MIRI parallels in GOODS-S that are stacked
with the SMILES MIRI mosaics, leading to 
regions of the fields where there is
MIRI imaging without NIRCam coverage. For these regions,
we include MIRI-generated SNR image subregions
constructed from the \texttt{SCI} and \texttt{ERR}
HDUs of the MIRI mosaics processed with the \texttt{jwst}
pipeline. 

Lastly, the ordering and placement
of the JADES NIRSpec observations required some targeting
to be performed based on HST-only regions of the field
that do not have JWST coverage. To support these NIRSpec
observations, we generate SNR detection images based on the
HLF HST mosaics. For the signal mosaic, we combine the flux HDUs
from the HST ACS F775W, F814W, and F850LP filter mosaics using
inverse-variance weighting. The noise images used in the SNR
image and for weighting are computed following the uncertainty
mosaic method described in Section \ref{sec:uncertainty-mosaics}
below, computing the random aperture 
uncertainties in circular apertures of radius $r=0.25"$.
We limit the regions where HST-based SNR detection image
contributions are added to areas where JADES NIRSpec
spectroscopic targeting from HST-detected sources was used.

The final SNR detection image used in the detection algorithm
is then the composite of these separate SNR images, using only
the JWST NIRCam SNR image in any area where the NIRCam, MIRI, or
HST SNR detection submosaics overlap.

\subsection{Deblending Image}
\label{sec:deblending-image}

While the detection image is constructed from the
long wavelength mosaic data, there are higher
resolution NIRCam data available at shorter wavelengths over most of the
GOODS-S and GOODS-N fields with sufficient depth
to support a deblending image to match the
detection image. Given the expected redshift
distribution and reddening of known sources,
we expect that the majority of sources detected
in the bluest long wavelength NIRCam filters
will also be detected in the reddest short wavelength
NIRCam filters at wavelengths $\lambda\approx2\mu$m.
We therefore construct a deblending image as an
inverse-variance-weighted SNR image combining the
F182M, F200W, and F210M in each field. Where
these short wavelength NIRCam filters are unavailable,
we substitute the NIRCam detection SNR image, the
HST SNR detection image, and then MIRI detection image,
in that order of preference, ensuring that all
regions of the detection image are covered by valid
pixels in the composite deblending image. In what
follows, we
will use the symbol $SNR_D$ when
referring to the SNR of a pixel or source in the
deblending image.

\subsection{Detection and Deblending Algorithms}
\label{sec:detection-algorithm}

The source catalog generated from the JADES DR5
imaging uses a segmentation map-based approach
where a series of detection criteria are
applied to pixels in a detection image to define
regions containing sources, deblending criteria
are applied to detected regions to define separate
sources within contiguous detected regions, and
then an integer segmentation map labeling pixels
with unique source IDs defines the pixels
belonging to each object. The segmentation maps
are used with customized versions of 
the \texttt{photutils} routine
\texttt{SourceCatalog} to generate object
catalogs, as described in Section \ref{sec:source-catalog}.
The rest of this Section describes how the
segmentation maps that feed into the source catalog
procedure are generated, including the detection
and deblending methods.

\subsubsection{Initial Blended Segmentation}
\label{sec:blended-segmentation}

A first blended segmentation is created 
using
the \texttt{photutils} routine \texttt{detect\_sources}
applied to the maximum of the detection and deblending
SNR images smoothed with a gaussian with a half-pixel
standard deviation, using a threshold of $SNR>1.5$ 
over a minimum of 1 pixel regions.
The resulting segments are refined then using a series 
of morphological processing algorithms.
We apply the \texttt{scipy.ndimage} routine
\texttt{binary\_fill\_holes} to fill any gaps
in the segments and relabel the segments using
\texttt{scipy.ndimage.label}. The segment exteriors
are reshaped by the following sequence: 1) dilation
using \texttt{scipy.ndimage.expand\_labels},
2) eroded twice using a four-connected kernel 
with \texttt{scipy.ndimage.binary\_erosion},
3) dilated once, 4) opened twice using an
eight-connected kernel with 
\texttt{scipy.ndimage.binary\_opening},
4) dilated once, 5) holes filled, 
and then 6) relabeled. This series of 
operations helps to refine the boundaries of the
source segments, but small sources can be eliminated
by the erosion and opening operations. Using the
resulting segmentation map as a mask, a 
\texttt{detect\_sources} pass with $SNR>3.5$ over
a minimum of 4 0.03" pixel regions is performed, any
additional recovered sources are dilated once using
\texttt{expand\_labels}, and the segmentation maps 
merged.

Some additional processing of the initial segmentation
map is performed before detailed deblending. An
initial source catalog from the segmentation map
and detection image is generated. Looping over the
objects, thin tenuous connections between distinct
objects are broken by eroding each object's 
segmentation with four iterations. If this erosion
splits segments into multiple objects, the
objects are relabeled and \texttt{expand\_labels}
is used to regrow their segments to fill the original
extent of their parent's footprint.

Once the tenuous
connections are broken, the source catalog is again
rebuilt. The sources are each investigated to remove
any segmentation map fragments that would have failed
the original detection conditions. Approximately, this
means fragments that do not have a median $SNR>3$
in either the detection or deblending image. Such
sources are removed, as are objects with areas
below six pixels. The sources in the
resulting segmentation map are then considered for
possible deblending.

\subsubsection{Deblending}
\label{sec:deblending}

Once the initial blended segmentation map has
been generated, a catalog is created and each
source inspected. The region of
the deblending image
delineated by the bounding box of each
object's segment is extracted, a floor
of $SNR_D=1.5$ applied, and then \texttt{detect\_sources} 
used to find any sources in the deblending
image within the segmentation with $SNR_D>1.5$ and
larger than four pixels. Holes in the
segmentation map of these deblended objects are
filled and a single \texttt{binary\_opening}
pass with an eight-connected kernel applied.
Any sources with $SNR_D>3.5$ and areas larger
than four pixels removed by the opening
algorithm are added back to the segmentation map,
and two dilation iterations using
\texttt{expand\_labels} are applied. If no additional sources are found by the deblending algorithm within the segmentation derived from the detection image, then the object is not deblended
and its original segment is retained.  Otherwise, the
object is marked for deblending.

If the object is marked for deblending, the
deblending algorithm identifies locally dominant
peaks in the deblending images and iteratively
assigns pixels within the object segment to
the deblended peaks following a set of conditions.
Local maxima in the deblending image are
determined using the \texttt{scipy.ndimage.maximum\_filter}
routine with a disk filter of radius $r=3$ pixels.
These SNR peaks in the deblending image are then
candidate locations for newly deblended objects,
but the algorithm only retains the peaks that
dominate their locale in the deblending image. For
the purposes of the algorithm, we define local
dominance for a peak as having the largest
maximum among peaks in a region defined by a minimum
SNR that is some multiplicative fraction $\xi$ of
the peak maximum $p_{SNR}$. In other words, a peak
with maximum SNR $p_{SNR}$ is locally dominant if 
it is the highest peak within an iso-significance
contour $\xi p_{SNR}$ that surrounds it. For this
catalog, we take $\xi=0.5$.

Procedurally, the algorithm considers the list of peaks
sorted by descending SNR. Starting with the highest peak
with SNR $p_{SNR,0}$,
any candidate peaks within an iso-significance contour
$\xi p_{SNR,0}$ are discarded from further consideration as they are 
locally subdominant.
If there are no remaining candidate
peaks, there is only one dominant peak, all the remaining pixels are assigned to
the singular peak such that the original
and deblended segmentations for the region match, and the
algorithm moves on to deblending the next object from the
initial blended segmentation map. 
Otherwise, the
next highest peak with SNR $p_{SNR,1}$
in the list of candidate subregions
is considered. First, all pixels within a
contour $p_{SNR,1}$ about the first peak
can be safely assigned to it. Then, 
all lower-SNR 
peaks within an
iso-significance contour $\xi p_{SNR,1}$ about
the second peak are removed from consideration.

Once more than one peak has been deblended,
decisions about how to assign pixels to distinct
peaks must be made. Choices in the literature
range from the flux contrast peak assignment
method from \texttt{Source Extractor} \citep{bertin1996a},
shared assignment between peaks using
forward modeling of source surface brightness profiles
as in \texttt{SCARLET} \citep{melchior2018a}, and
AI/machine learning-based partial attribution
models \citep{hausen2022a}. Here, we use a
deterministic approach where we iteratively
consider the pixels sorted in descending SNR. 
Starting with pixels with an SNR equal to the
most recently considered peak SNR $p_{SNR,i}$,
which are the lowest SNR pixels yet unassigned,
they inherit the peak assignment of their highest SNR
neighbor, or become the start of the current peak 
if they are at its local maximum. These pixel
assignments continue until $SNR=p_{SNR,i+1}$ of the
next lowest peak under consideration is reached.
Once the deblending pixel value $SNR=p_{SNR,i+1}$
of the next lowest peak is reached, its local $\xi p_{SNR,i+1}$
iso-significance contour is cleared of lower-significance
peaks, and the process of pixel assignments continues
until all peaks have been considered. 

At this stage in the algorithm, all the locally dominant
peaks in the filtered deblending image within the parent
segmentations have been identified and pixels
greater than a fraction of $\xi$ of their peak SNR
assigned to them, accounting for the local proximity of
pixels to multiple peaks. Before the remaining pixels
are assigned to peaks, we take two additional steps
to recover possible objects of interest that may
otherwise be lost. First, faint satellites in the
exteriors of bright galaxies may fail the locally
dominant peak criterion but may be visually distinguishable.
To recover these objects, we apply a Scharr filter
to the deblending image to attempt a rotationally-invariant
edge detection. Within the parent segmentation, 
we apply \texttt{detect\_sources} with a threshold of
10 to the Scharr-filtered deblending SNR image over
a minimum area of four pixels. The segments from
any resulting detections are dilated using 
\texttt{expand\_labels} and added to the list of
peaks. Second, since the deblending image is
bluer than the detection image there are rare
sources where the contrast in significance
between the $\lambda\approx2\mu$m deblending
image and the $\lambda\gtrsim2.7\mu$m detection
image leads to missed or poorly deblended extremely red 
sources in crowded regions. To recover these regions,
we look for peaks in the detection image that have
an $SNR>30$ and exceed the deblending SNR by a factor
of $10\times$ over an area of at least 10 pixels.
Of these regions, we remove those with high eccentricity
which are diffraction spikes with chromatic shifts in
local SNR maxima. Any remaining red regions are morphologically
opened, dilated, filled, and expanded by three pixels, 
and then merged to augment the existing deblended segmentations.

Once the final collection of deblended peaks within
the initial blended detection segmentations have been
determined, grown to occupy the local region of the
deblending image they dominate, and augmented with
edge-detected sources and highly-reddened sources, the
deblended segmentations have to be grown to fill the 
detection segmentations. A continuation of the algorithm
used to assign pixels to peaks during the initial
deblending would not fill the segmentation, since
pixels in locally subdominant peaks would be left unassigned.
Instead, we use a modified version of the \texttt{expand\_labels}
routine that only expands segments into regions that
satisfy a user-specified condition. We set this condition to
be a SNR threshold, and then iteratively lower the threshold
while expanding the deblended peak segments to fill their
parent blended segmentations while
following iso-significance contours of the deblending image.

One additional heuristic choice is made regarding the growth
of deblended segmentations, which was primarily necessitated
by the impact of long, high SNR diffraction spikes from stars
on the growth of deblended segmentations. With narrow, large-angle
extensions of sources intersecting other bright sources in the
detection image, we should expect that deblended segmentation
maps should be discontinuous and that locally dominant sources
should not necessary end their spatial extent when they encounter
another peak that spans narrow regions of the parent segmentation.
Accordingly, when a secondary deblended peak encounters the
edge of a parent segmentation during the growth phase, it is allowed
to grow by another 20 pixels into the parent segmentation. At that
point its growth is halted and the dominant peak segmentation is allowed
to grow around or past it. 
The deblended segmentations are cleaned to remove any possible low
SNR fragments or segments with areas less than five pixels, and then
the deblended segments are grown within the initial blended parent
segments to fill any remaining unassigned pixels. This final fill
completes the deblending process.

\subsubsection{Second-Pass for Isolated Sources}
\label{sec:second-pass-detection}

Once the initial segmentation map has been
generated, a second simplified detection is performed
to improve recovery of faint, compact objects in isolated
regions of the imaging.
The segmentation map and the mosaic of the subregion
mask are dilated by 10 pixels and used to mask the SNR
detection image. A second pass of detection in unmasked regions
is performed using
the \texttt{photutils} routine \texttt{detect\_sources}
with a $SNR>3$ threshold over a minimum of 5 pixel regions.
The resulting segmentation map is dilated using two
iterations and a disk footprint of radius 2 pixels,
holes are filled, and two iterations of binary opening
performed. The resulting second-pass segmentation map is 
then added to the original deblended segmentation map.
This combined
segmentation map forms the segmentation map that
defines the detection source catalog used as
input to the catalog curation stage that defines
our final source catalog, described in Section
\ref{sec:catalog-curation} below.

\subsection{Catalog Curation}
\label{sec:catalog-curation}

\begin{figure*}[ht!]
\includegraphics[width=1.0\textwidth]{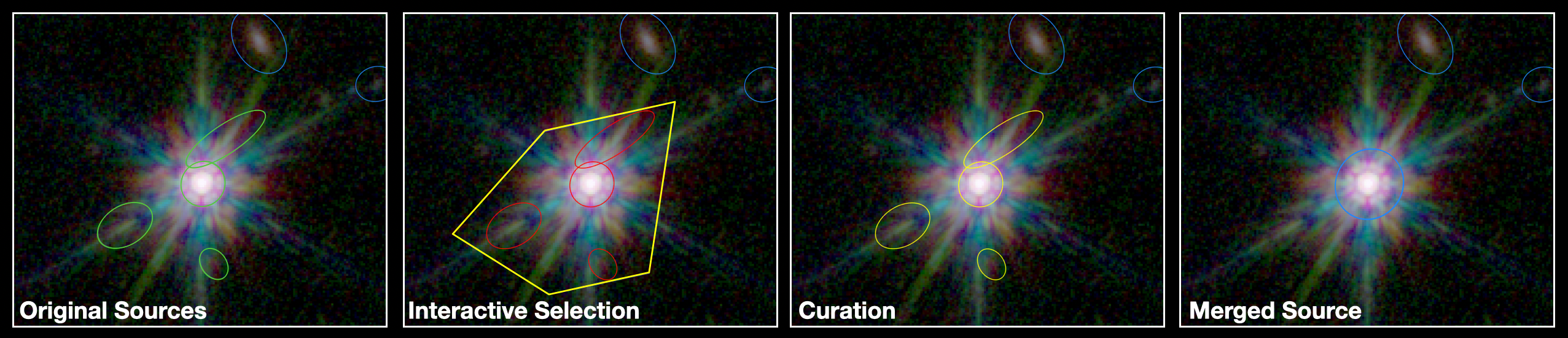}
\caption{Illustration of the interactive
catalog curation tool in FitsMap \citep{hausen2022b} applied
to amend the JADES DR5 photometric catalog. Using the source catalog overlay (green ellipses)
on the JWST mosaic (RGB), clearly spurious sources in need of curation, such as shredded diffraction spikes, can be
identified visually (far left panel). Using the interactive selection
tool in FitsMap (left center panel), sources within a polygonal bounding box (yellow line)
can be selected (red ellipses) for further processing.
Once selected, the object properties can be recorded and the sources
 marked (yellow ellipses in center right panel)) for curation (e.g., merging or deletion. In this case, the diffraction spike shreds are re-merged into the central star (blue ellipse in far right panel).
\label{fig:catalog-curation}}
\end{figure*}

While the implemented algorithms for
detection and deblending find considerable
success in identifying faint objects and
disentangling complex, crowded regions 
populated with extended sources, the
methods do experience failures. Notably,
the deblending algorithm can shred diffraction
spikes and the exteriors of extended but
high SNR objects. These failure modes
can be addressed directly by re-merging
the fragmented segments of shredded
objects back into their parent segmentation.
However, the scale of the images and
catalog volume required a method 
for performing this catalog curation
quickly and efficiently. Figure \ref{fig:catalog-curation}
illustrates our solution to this issue.
For convenient viewing of the 
JADES DR5 mosaic data, overlays of our catalog
data onto images, and inspection of NIRSpec
slit placements, we use an interactive
FitsMap \citep{hausen2022b} website.
We have extended the FitsMap interface
to create a selection tool that highlights
all objects interior to a user-defined polygon
and record their source IDs. These IDs
could then be submitted via an online
form to a cloud-hosted spreadsheet and
updated by volunteers from the collaboration
who marked collections of objects
for re-merging or, occasionally, deletion.
The spreadsheet could be automatically
downloaded and parsed such that the
segmentation map defining sources in the 
catalogs could be updated to reflect the
input user curation. The catalog
curation process was performed for 
both GOODS-S and GOODS-N, and involved
less than a percent of all originally
detected objects by number. 

Figure \ref{fig:catalog-curation} illustrates
the catalog curation process. Using an overlay
of detected sources on the FitsMap, objects
that are clearly spurious fragmentations,
such as shredded diffraction spikes extending
from the JWST PSF of bright stars can be
identified (far left panel). The interactive
selection tool in FitsMap allows for these
objects to be bounded by a polygon and
isolated from other sources (center left panel).
Once selected, the properties of these
sources can then be recorded and their
properties curated (e.g., merged or
deleted) appropriately (center right panel).
In the example shown, the shredded diffraction spikes
are re-merged into the central star (far right panel).

\section{Source Catalog}
\label{sec:source-catalog}

Once the deblended
segmentation maps have been
computed following the methods described
in Section \ref{sec:detection-algorithm},
the source properties required for
forced photometry are computed and cataloged.
Iterating through each source segmentation,
the cataloging method computes a source
centroid (Section \ref{sec:centroiding})
about a local maximum in the detection image.

\subsection{Centroiding}
\label{sec:centroiding}

The centroids for objects are determined
from the regions near their peaks, corresponding
approximately to the areas in the detection image where
the peaks are locally dominant. Within each deblended
subregion, the dominant peak in the detection image is located.
We restrict the dominant peak not to be adjacent to the edge of 
the deblended segment to avoid fitting centroids of satellites to
peaks in the surrounding background from their host central
galaxies. In the vicinity of a peak with detection SNR $p_{SNR}$,
the pixels within iso-significance  contour higher than a fraction 
$\xi=0.5$ of $p_{SNR}$ are used to compute a SNR-weighted
barycenter, which is adopted as the centroid.  For objects smaller
than 25 pixels, the entire object is instead used to 
compute a barycenter for the centroid.  For reference,
we also compute and record barycenters for the entire deblended
pixel segmentation from the detection image SNR distribution.

\subsection{Source Properties}
\label{sec:source-properties}

With the centroids computed, the source properties can then be determined
from pixels around the centroid locations. We adopt similar methods to
those used by the \texttt{SourceCatalog} routines implemented
by \texttt{photutils}. Since we have computed customized
centroids, for convenience we re-implement the \texttt{SourceCatalog} 
routines for computing source properties relative to these
chosen locations. The central moments are calculated from the object
centroid, the object segmentation,
 and the corresponding detection signal image data. The
covariance of the detection signal pixels is then computed from
their central moments, following the \texttt{SourceCatalog} routine
\texttt{\_covariance}.

Following the \texttt{SourceCatalog} methods, we
continue by computing and storing the approximate
second moments of the detection signal pixel
distribution. The semimajor axis size $A_p$ and semiminor axis size
$B_p$ are computed from the eigenvalues of the covariance matrix.
The eccentricity is computed as $\epsilon_p = \sqrt{1 - B_p^2/A_p^2}$.
Here, we denote quantities computed approximating the \texttt{photutils}
methods with a $p$ subscript. The circularized pixel radius about the
centroids are computed from $A_p$, $B_p$, and the on-sky
orientation $\theta_p$ angle computed from the covariance matrix.
A \citet{kron1980a} aperture
radius is computed from the first moment of the detection
signal pixels within an aperture with a circularized radius six times
the second moment, and following \texttt{SourceCatalog}
the unscaled Kron radius \texttt{kron\_radius\_p} calculated as the
ratio of the detection signal-weighted normalized first moment and the 
sum of the detection signal within the aperture. A minimum
unscaled Kron radius is enforced as $\min($\texttt{kron\_radius\_p}$)=1.4$, adopting the \texttt{photutils} default
value.

\subsubsection{Gaussian Regression Model for Source Property Estimation}
\label{sec:sb-model}

In computing the approximate second moments of the detection signal
distribution for each object, the extent of the segmentation used
in selecting which pixels contribute to the integrals can influence
the resulting object sizes. While we retain and record the object
sizes and unscaled Kron radii computed approximating the 
\texttt{photutils} methods, we perform two-dimensional Gaussian
regression fits to the detection image profiles of each object
to determine directly the Gaussian second moments, orientation
on the sky, and the corresponding aperture for determining the 
Kron radius.

Once the object centroid has been determined, a two-dimensional
Gaussian model for an object can be defined as a function 
of the $x$ and $y$ pixel coordinates relative to the centroid as
\begin{equation}
\label{eqn:simple-gaussian}
z(x,y) = K \exp[-(ax^2 + 2bxy + cy^2)].
\end{equation}
The amplitude $K$ sets the normalization of the model, and the
parameters $a$, $b$, and $c$ set the shape of the multivariate
elliptical isophotes. The parameters $a$, $b$, and $c$
can be related to the Gaussian variances $A^2$ and
$B^2$ and position angle $\theta$
through the formulae
\begin{equation}
\label{eqn:semimajor}
A^2 = \frac{1}{2(a \cos^2\theta + 2 b \cos\theta \sin\theta  + c \sin^2\theta)},
\end{equation}
\begin{equation}
\label{eqn:semiminor}
B^2 = \frac{1}{2(a \sin^2\theta - 2 b \cos\theta \sin\theta  + c \cos^2\theta)},
\end{equation}
\noindent
and
\begin{equation}
\label{eqn:pa}
\theta = \frac{1}{2}\arctan \left(\frac{2b}{a-c}\right).
\end{equation}
\noindent
For convenience, we can re-write Equation \ref{eqn:simple-gaussian} as
\begin{equation}
\label{eqn:gaussian-model}
z(x,y) = \exp[\alpha x^2 + 2\beta xy + \gamma y^2 + \delta]
\end{equation}
\noindent
with the substitutions $\alpha = -a$, $\beta = -b$, $\gamma = -c$ and $\delta = \log K$.
The model represented by Equation \ref{eqn:gaussian-model} can then be fit using an
iterative regression scheme to the detection image profile of each object.
Here, we extend the one-dimensional
weighted Gaussian regression method presented by \citet{guo2011a}
to multidimensional Gaussian distributions. We can define a metric we wish
to optimize when fitting the Gaussian to the detection image profile as 
\begin{equation}
\chi_W^2 = \sum_i [z_i\log z_i - z_i(\alpha x_i^2 + 2\beta x_i y_i + \gamma y_i^2 + \delta)]^2.
\end{equation}
\noindent
Here, the sum over index $i$ runs over the pixels in the object segmentation, $x_i$
and $y_i$ are pixel coordinates relative to the object centroid, and $z_i$ is the
value of the $i$-th pixel.
Taking the partial derivatives of this metric with respect to the parameters
and setting them equal to zero results in a system of linear equations 
\begin{equation}
\left[\begin{matrix}
\sum z^2 x^4 & 2 \sum z^2 x^3 y & \sum z^2 x^2 y^2 &  \sum z^2 x^2 \\
\sum z^2 x^3 y & 2 \sum z^2 x^2 y^2 & \sum z^2 x y^3 &  \sum z^2 xy \\
\sum z^2 x^2 y^2 & 2 \sum z^2 x y^3 & \sum z^2 y^4 &  \sum z^2 y^2 \\
\sum z^2 x^2  & 2 \sum z^2 x y & \sum z^2 y^2 &  \sum z^2
\end{matrix}
\right]
\left[
\begin{matrix}
\alpha\\
\beta\\
\gamma\\
\delta\\
\end{matrix}
\right]
=
\left[
\begin{matrix}
\sum z^2 x^2 \log z \\
\sum z^2 xy \log z \\
\sum z^2 y^2 \log z \\
\sum z^2 \log z \\
\end{matrix}
\right]
\end{equation}
\noindent
where we have suppressed the index $i$, but the sums run over the pixels in the object segmentations. 
In the presence of noise in the $z_i$, \citet{guo2011a} introduced an iterative scheme 
for performing a weighted one-dimensional Gaussian regression. We can write
an iterative scheme for performing the equivalent two-dimensional Gaussian regression as
\begin{equation}
\label{eqn:regression}
\left[\begin{matrix}
\sum z_{(k-1)}^2 x^4 & 2 \sum z_{(k-1)}^2 x^3 y & \sum z_{(k-1)}^2 x^2 y^2 &  \sum z_{(k-1)}^2 x^2 \\
\sum z_{(k-1)}^2 x^3 y & 2 \sum z_{(k-1)}^2 x^2 y^2 & \sum z_{(k-1)}^2 x y^3 &  \sum z_{(k-1)}^2 xy \\
\sum z_{(k-1)}^2 x^2 y^2 & 2 \sum z_{(k-1)}^2 x y^3 & \sum z_{(k-1)}^2 y^4 &  \sum z_{(k-1)}^2 y^2 \\
\sum z_{(k-1)}^2 x^2  & 2 \sum z_{(k-1)}^2 x y & \sum z_{(k-1)}^2 y^2 &  \sum z_{(k-1)}^2
\end{matrix}
\right]
\left[
\begin{matrix}
\alpha_k\\
\beta_k\\
\gamma_k\\
\delta_k\\
\end{matrix}
\right]
=
\left[
\begin{matrix}
\sum z_{(k-1)}^2 x^2 \log z \\
\sum z_{(k-1)}^2 xy \log z \\
\sum z_{(k-1)}^2 y^2 \log z \\
\sum z_{(k-1)}^2 \log z \\
\end{matrix}
\right].
\end{equation}
As before, the summations run over the pixels in the segmentation. The quantity $\log z$
is computed from the detection image pixel values $z_i$. The iterated quantities $z_{(k)}$
are computed as
\begin{equation}
z_{(k)} = \left\{
\begin{matrix}
z & \mathrm{for}\, k =0 \\
e^{\alpha_k x^2 + 2\beta_k x y + \gamma_k y^2 + \delta_k} & \mathrm{for}\, k>0
\end{matrix}
\right.
\end{equation}
\noindent
In each iteration, the vector on the right hand side of Equation \ref{eqn:regression}
is multiplied by inverse of the matrix on the left hand side to compute the vector
of parameter values. The values $z_{(k)}$ are updated at each iteration with improved solutions for the parameters $\alpha_k$, $\beta_k$, $\gamma_k$, and $\delta_k$.
The parameter values converge quickly, to within floating-point precision
before the maximum number of ten iterations that we enforce on the algorithm.

\begin{figure*}[ht!]
\includegraphics[width=1.0\textwidth]{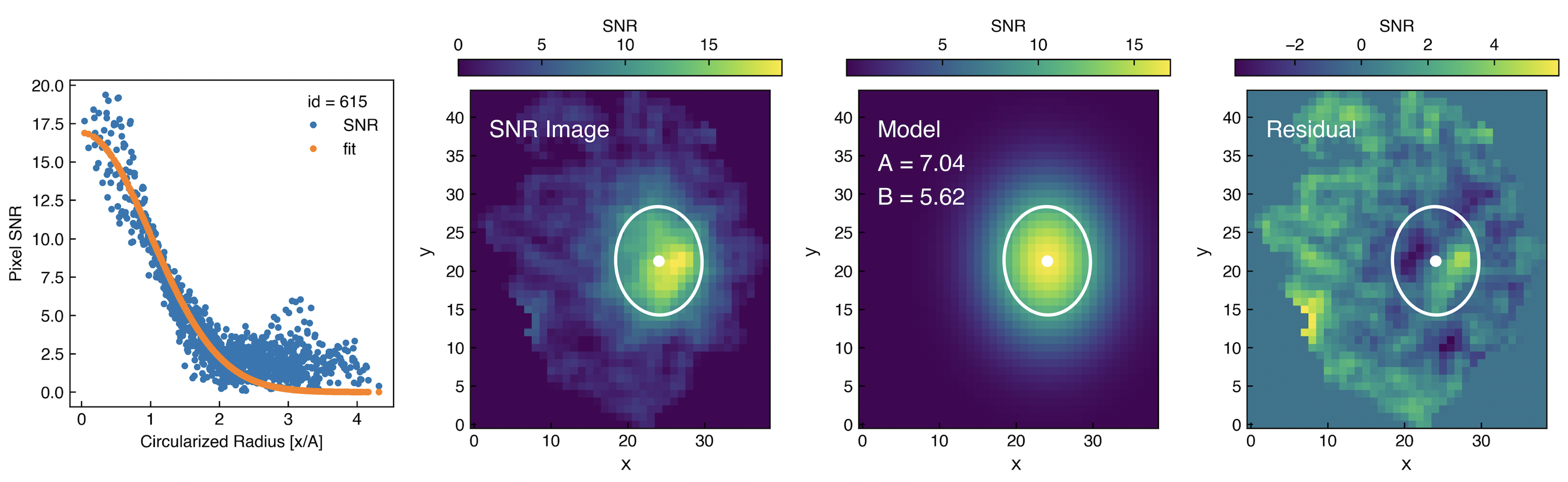}
\caption{Example of Gaussian regression modeling, shown for
object \texttt{ID=615} from the JADES DR5 GOODS-S region. The pixel SNR
data from the detection image (center left panel), masked by the object
segmentation. The regression method described in Section 
\ref{sec:sb-model} is used to fit a two-dimensional Gaussian profile to
the source, as shown in the center right panel. The semimajor and semiminor axes of the best-fit are indicated in pixels, and the 1$\sigma$ isodensity contour shown as an overlaid white ellipse in the center left, center right, and far right panels. The goodness of fit can be judged from the residual of the SNR image less the model, as shown in the far right panel, and by over plotting the circularized  SNR pixel data (blue points) and model (orange curve) as shown in the far left panel.
\label{fig:gaussian-regression}}
\end{figure*}

Figure \ref{fig:gaussian-regression} shows an example of the Gaussian regression 
method applied to an object in GOODS-S (\texttt{ID}=615). The leftmost panel of the
figure shows the distribution of pixel SNR values as a function of the circularized
pixel radius from the object centroid, along with the circularized model of the
two-dimensional Gaussian over plotted. The middle left panel shows the SNR detection
image of pixels in the segmentation assigned to the object, with the $1-\sigma$
ellipse defined by the semimajor axis $A$, semiminor axis $B$, and position angle $\theta$
for the object. The middle right panel shows the Gaussian model determined via regression,
with the same ellipse overlaid along with annotations indicating the best-fit values of
$A$ and $B$, and the rightmost panel shows the residual between the
SNR image and model. As Figure \ref{fig:gaussian-regression} illustrates, the Gaussian
regression method we employ both provides an estimate of the source sizes that can
be used for further constraints on photometric apertures but also provides an actual
model for the source SNR profile that can be compared with the detection image and
assessed directly for accuracy via its residuals.

Once the semimajor axis $A$ (Equation \ref{eqn:semimajor}), semiminor axis
$B$ (Equation \ref{eqn:semiminor}), and position angle $\theta$ (Equation \ref{eqn:pa})
are determined, the \texttt{photutils} procedures for computing the Kron apertures
are replicated. Pixels within a circularized radius of six times the Gaussian second
moments are used compute the unscaled Kron radius, with a minimum value of 1.4, and
the resulting aperture is used in Kron photometry (see Section \ref{sec:kron-aperture-photometry} below).

\section{Point Spread Functions}
\label{sec:psf}

The process of computing aperture 
corrections for photometric measurements
and generating common-PSF images requires
point spread functions for each image.
Here we describe our methods for
constructing the PSFs we use for our
HST, JWST NIRCam, and JWST MIRI photometric
analysis.

\subsection{Apodization}
\label{sec:apodization}

The two-dimensional PSFs are used for
both constructing common-PSF images and
for photometric aperture corrections.
The methods for constructing empirical
and model PSFs that we use generate
rectangular image footprints for the
PSFs.
We note that for each PSF, we apply a 
circular apodization to ensure that the
PSF normalization is computed within a 
projected radius, rather than within a
integrated rectangular area, and to
ameliorate possible side lobe artifacts
during convolutions
caused by effective square wave truncations
of the PSF. The multiplicative apodization we apply
is a circular Tukey filter that can
be written as
\begin{equation}
T(r,\alpha,\lambda) = \left\{
\begin{matrix}
1 & r < (1-\alpha)\lambda \\
\frac{1}{2} \left[1 - \cos\left(\frac{\pi r}{\alpha \lambda} - \frac{\pi}{\alpha}\right)\right] & (1-\alpha)\lambda \leq r < \lambda \\
0 & r \geq \lambda
\end{matrix}
\right.
\end{equation}
\noindent
where $r$ is the fractional radius of a pixel from the
center of the PSF in terms of the
half-width of the square PSF image, and we choose
$\alpha=0.1$ and $\lambda=0.995$. The PSFs are renormalized
after apodization. Other choices could
be made to further reduce the power in side lobes, but
we chose the Tukey filter as a compromise to preserve
the strength of the prominent JWST diffraction spikes.

\subsection{HST Effective Point Spread Functions}
\label{sec:hst-epsf}

The HST effective point spread functions
(ePSF) used to compute aperture corrections
and common-PSF image convolutional kernels
are determined from the HLF ACS and WFC3
mosaics using
methods from the \texttt{photutils} package.
Lists of known stars in GOODS-N and GOODS-S
are used as input to create separate ePSFs in each
HST filter for each field, limiting to stars
$18<m_{AB}<21$ for ACS filters and
$18<m_{AB}<22$ for WFC3 filters. Briefly the
stars are extracted from the HLF mosaics
using the \texttt{extract\_stars} routine.
An \texttt{EPSFBuilder} instance is initialized
with no oversampling and a maximum of ten iterations,
and then applied to the star locations to generate
an ePSF, which is then circularly apodized as
described in Section \ref{sec:apodization} above.
The resulting ePSFs for each filter
are visually inspected for artifacts.

\subsection{NIRCam Model Point Spread Functions}
\label{sec:nircam-mpsf}

As described in \benspaper{} and following methods developed from
those presented in \citet{ji2024a},
we generate model point spread functions (mPSFs)
for each subregion filter mosaic, acquired at a specific
position angle, in the GOODS-N
and GOODS-S fields. The \texttt{STPSF} code \citep{perrin2014a}
is used to simulate PSFs for each filter and at a given
position angle. These simulated PSFs are tiled on a pattern across the sky, embedded appropriately into
exposures arrayed at the locations at the actual exposures
that comprise each subregion mosaic, and then combined using the \texttt{jwst} pipeline as described in \benspaper.
The resulting simulated, noiseless subregion mosaics consist of a grid of carefully placed stars with the same dither pattern as the observed program subregion mosaics. The mPSF
for each subregion mosaic is then measured from these simulated
stars using the \texttt{EPSFBuilder} methods from 
the \texttt{photutils} package in a manner similar to that
used for the HST ePSFs, but with exact known locations and
without noise. The resulting mPSFs are then circularly apodized as described in Section \ref{sec:apodization}. For GOODS-S,
there are 384 distinct subregion NIRCam filter mosaic mPSFs.
For GOODS-N, there are 169 distinct subregion NIRCam filter
mosaic mPSFs. These distinct mPSFs are all used for common-PSF
image generation (see Section \ref{sec:common-psf-mosaics}
below), but for aperture corrections for convenience we
select a single mPSF in each filter to use. For GOODS-S,
we select mPSFs from the \texttt{jw012100},  \texttt{jw012860},
\texttt{jw019630},
and \texttt{jw032150} subregion mosaic mPSFs, and in GOODS-N 
we select from the \texttt{jw011810.hst}, \texttt{jw025140.gn\_pa133}, 
\texttt{jw018950}
subregion mosaic mPSFs, depending on the filter.

As discussed by both \benspaper{} and
\citet{ji2024a}, this method produces an
mPSF that reproduces extremely well the encircled energy
curves measured directly from stars in mosaic images while
maintaining the diffraction spike structure that can be
difficult to measure at low signal-to-noise for a large
number of stars in an observed mosaic.

\subsection{MIRI Point Spread Functions}
\label{sec:miri-mpsf}

To compute aperture corrections for the JWST/MIRI photometry,
we adopt the MIRI PSFs from \citet{alberts2024a}.
The F560W and F770W PSFs (A. Gaspar, private communication)
are empirical and use high-SNR JWST commissioning observations
that highlight the ``cruciform" feature induced by internal reflection
in the detector
\citep{gaspar2021a,libralato2024a,dicken2024a}.
The longer wavelength MIRI PSFs (F1000W-F2550W) are
modeled using STPSF \citep{perrin2014a}. Each MIRI PSFs
is resampled at the JADES DR5 mosaic resolution (0.03" pixels)
and apodized as described in Section \ref{sec:apodization}.

\section{Common PSF Mosaics}
\label{sec:common-psf-mosaics}

With the definitions of the effective
point spread functions for each filter, as
described in Section \ref{sec:psf},
the process for creating common-PSF images can
then proceed. Our approach for generating
common-PSF images is informed by the 
structure of the JWST PSF, which famously
features prominent diffraction spikes.
Given that the mosaics involve stacks
of subregion mosaics acquired at
a variety of position angles, applying
a common convolutional kernel to the 
final mosaic stack in generating a common
PSF image could lead to additional
artifacts related to the diffraction
spikes. Correspondingly, we build
common-PSF images for subregion mosaics
in each filter at separate position
angles by constructing a convolutional kernel
using the apodized, two-dimensional
PSF image appropriate for that region,
and then stack the common-PSF
subregion mosaics into a global
common-PSF mosaic for each filter.
The exceptions to this approach are
the HST mosaics, for which we use
realigned versions of the HLF mosaics
that are constructed from HST subimages
with a variety of effective PSFs. In
generating our common-PSF images, we
bring each filter to the F444W PSF,
leaving all redder JWST NIRCam and MIRI
images with their native PSFs. 
For the long-wavelength NIRCam filters,
the subregion mosaics are processed
for each NIRCam detector module as
separate subregion mosaics, since each
detector module has a slightly different
PSF in each filter.
In
the JADES DR5 catalogs, properties measured
on the common-PSF images are designated
with a \texttt{\_CONV} suffix.

Procedurally, the common-PSF images
are constructed using a series of
operations on the subregion filter
mosaics. These subregion mosaics
are prepared by splitting them into
portions with appropriate zero padding
to allow for rapid GPU-accelerated
convolutions. This optimization was
primarily necessitated by bringing 
the HLF mosaics to a common PSF with the
JWST mosaics, as those images have the
full mosaic pixel footprint in memory.
The circularly-apodized versions of the
native PSF of the input image and the
target F444W PSF are prepared for convolutions
with the subregion mosaic, which may involve
padding to an optimal size for rapid convolution.

The common-PSF images are then generated
by computing a convolutional kernel that
deconvolves the original subregion mosaic
using the input PSF, and then convolves
the result with the target F444W PSF after
regularizing the image with a small level
of power from white noise (i.e., the variance of a white-noise random process).
If the input data is $y$ of size $K \times M$, we can define
\begin{equation}
S_y = \frac{|\tilde{y}|^2}{KM},
\end{equation}
which provides the power spectrum of the image.
In the presence of Poisson noise, some fraction
of this power will be contributed by a constant
flat power spectrum $N_y$, which will correspond
to the variance of the typical Poisson background.
Denoting the input filter PSF as $H_\mathrm{in}$,
the regularized deconvolution kernel can be approximated
as
\begin{equation}
\tilde{G} = \frac{\tilde{H}_\mathrm{in}^\star (S_y-N_y)}{|\tilde{H}_\mathrm{in}|^2(S_y-N_y) + N_y}
\end{equation}
\noindent
To convolve the image to the target output PSF $H_{\mathrm{out}}$, we compute
\begin{equation}
\tilde{y}' = \tilde{H}_{\mathrm{out}}\tilde{G}\tilde{y}
\end{equation}
\noindent
and then take the inverse Fourier transform to find the
common-PSF image $y'$.

We further filter each common-PSF image with a 2D
cosine bell window function $B(\alpha)$ multiplied in frequency
space, using the \texttt{photutils}
implementation to create the filter and GPU
acceleration to perform the convolution. Smaller
values of the argument $\alpha$ correspond to
low-pass filtering of the image, such that frequencies
greater than $\alpha$ times the Nyquist frequency
are suppressed. For HST ACS
images, we use $\alpha=0.3$. For HST WFC3
images, we set $\alpha=0.4$. For the JWST
NIRCam common-PSF mosaics, the value $\alpha=0.9$
has a limited effect on the output mosaics and
is intended to help reduce numerical noise in
regions where the PSF matching is imperfect.

\section{Data Quality and Context Flagging}
\label{sec:flagging}

The increasing large volume of data imaging
and catalog products available from JADES
and similar surveys can provide a challenge
for the rapid analysis and understanding
of galaxy populations. By providing additional
catalog-level information on the local image
data quality near objects, endeavoring to
propagate consistent source information 
across versions of our reductions, and
informing the user about the visual
scene around sources of interest through
context flags, we have attempted to
make the catalog data more readily useful
without having to analyze the much larger
volume of image mosaics. This section
describes some of these methods and 
data quality flagging provided in the catalogs.

\subsection{Source ID Restoration}
\label{sec:source-id-restoration}

The complex evolution of the upstream
data processing of the JWST imagery, the
constant addition of data from new programs,
and the desire to maintain consistent
designations for hundreds of thousands of
objects across multiple versions of input
datasets has required the development of 
techniques to propagate object source 
identifications using pixel-level information
rather than catalog matching. Since the
JWST images are crowded in deep imagery
and the deblending algorithms can successfully
disentangle complicated, multi-component
projected systems, relying on catalog-level
source matching based on coordinate proximity
can become unreliable even when the same
algorithm is applied to 
different reductions of the same dataset.
For instance, with new data added between
different reductions of mosaics in the same field
either the detection or deblending images, or both,
may change in a manner that alters the
character of the segmentation map. A choice
then has to be made as to how to propagate a
source identification from one version of the
reduction to the next. 

Here, we choose directly
to propagate source identification based on the
largest number of shared pixels between segments
across different versions of the reduction.
Beginning with a reference segmentation map
from a prior version of the reduction and the
segmentation map from the current version of our
processing, we create
a pixel-aligned reference segmentation map on the same
\texttt{WCS} header as the current version. Then,
for every object, the
detection catalog generated from the current version's
segmentation map is used to extract the bounding box
of the object's segment. The pixels corresponding to this
segment are inspected in the reference segmentation map
from the prior version, and any source ID values are recorded.
The segments for each of these source IDs in the reference
segmentation map are then examined in the current version's
segmentation map, and the current version 
segments that inherit the most
pixels from a given reference segment also inherit its
source identification. Any segment in the current
reduction's segmentation map whose corresponding reference
segmentation source ID has been assigned to another object
is instead assigned a new source ID in the new version,
and usually such objects correspond to a newly deblended object.

While this pixel-level source ID restoration usually
works well, there are limitations. When applied across
multiple successive versions of reductions, if objects
are deblended, reblended, and deblended again, their
source IDs are not guaranteed to propagate unless the
source ID restoration method is applied iteratively.
As with proximity-based catalog-matching, the method
can encounter difficulty in regions with large
numbers of spurious objects from systematic noise sources
(e.g., regions heavily affected by the ``weave'', see \benspaper{}).
We plan to explore ameliorations for these issues in
future work.

\subsection{Bad Pixel Flagging}
\label{sec:bad-pixels}

For each object, local pixel-level
information about the
single-band mosaics recorded in the catalogs can
help provide simple quality assurance and
sample filtering. Indicating whether a 
source could be affected by bad or missing
data directly in its segmentation would
be useful, but additional context for being
{\it near} missing or bad data can also 
assist in identifying potential systematic
uncertainties such as enhanced noise near
the edge of a detector boundary or a source
truncated by a chip gap. In these latter
cases, the source itself may not show any
bad pixels or missing data within its
segmentations if simply because its segmentation
may not extend into a region without data.
To enable some additional context flagging,
for every filter (e.g., F090W) we define a flag 
(e.g., \texttt{FLAG\_F090W}) that records
the number of bad or missing pixels in a region
consisting of the bounding box for each object's
segmentation extended by 10 pixels in each direction.
Defining the bad pixel flag in this manner for
each object for each filter enables the end
user to employ quantitative cuts based on
a tolerance for potential bad or missing pixels.
By comparing the value of the flag with the 
size of the object's segmentation bounding box
plus frame, the severity of the bad or missing
data in each filter can be assessed directly from
the catalog information.

\subsection{Bright Neighbors}
\label{sec:bright-neighbors}

At the catalog level, information about
relative brightness of proximate objects
can provide useful context about a source's
photometric properties. For instance,
a compact, faint source near a substantially
brighter neighbor may have a less reliable
background subtraction or photometric redshift
than an isolated object. Such systematic
uncertainties can be difficult to quantify
directly in the photometric uncertainties,
but can be captured qualitatively with
context flagging. Correspondingly, we
implement a bright neighbor flag (\texttt{FLAG\_BN})
for each object that indicates whether any proximate source
has a Kron flux at least
twice larger than the primary source (\texttt{FLAG\_BN}$=1$) or
ten times larger (\texttt{FLAG\_BN}$=2$).
To determine whether to set \texttt{FLAG\_BN},
the bounding box of each object's segmentation
is extended by 10 pixels (0.3") in each direction and
inspected for the presence other proximate sources.
If the Kron flux of the proximate sources exceeds
a factor of two or ten, \texttt{FLAG\_BN} is
set accordingly.

\subsection{Parent Segmentation Identification}
\label{sec:parent-segmentation-identification}

For each deblended object in the catalog,
the relation between the object and
its fully blended parent detection segment 
is recorded and retained as the
object's \texttt{PARENT\_ID}. 
The \texttt{PARENT\_ID} can be used to
provide spatial context for objects in the
catalog in tasks such as associating
satellite and parent galaxies, identifying objects
that may be merging, or analyzing gravitational
lens systems.
First, note
that the parent detection segmentation map
and the deblended segmentation map have
the same shape and are pixel-aligned.
Then, to
determine an object's \texttt{PARENT\_ID}
from its blended parent segmentation, the
pixels covering the blended parent segment
are extracted from the deblended segmentation
map, and their unique values determined and counted.
The deblended object \texttt{ID} that comprises 
the most pixels of the blended parent segment 
becomes the \texttt{PARENT\_ID} for all objects
deblended from that parent segment.
In effect, the \texttt{ID} of largest 
object in pixel area among deblended subregions from a given
blended segment is assigned as the \texttt{PARENT\_ID}
for all its deblended siblings.

\subsection{JWST Program Bithash}
\label{sec:bithash}
The JADES DR5 products include an integer image that
hash encodes which JWST programs contribute data to every
pixel in the imaging data. The method for
generating the program bithash is detailed in \benspaper{},
and involves assigning each subregion mosaic a 
bit in a binary representation of the integer that indicates
whether the program contributes data to a given pixel.
The value of the integer hash at the location of every
source in the JADES DR5 catalogs is measured and
recorded in the source's \texttt{PID\_HASH} field. By decoding
the \texttt{PID\_HASH} to generate the list of contributing
programs at the object's location, the origin of the data can be
cited appropriately. When using the JADES DR5 catalogs, we
encourage use of the \texttt{PID\_HASH} to find the original
references for contributing data and more readily enable their
citations. A full list of contributing programs, the
bits assigned to each program, and the original references
to the data are provided in \benspaper{}.

\section{Photometric Uncertainty Framework}
\label{sec:uncertainties}

The uncertainties associated with the source
photometry should be comprised of the
sky background and Poisson uncertainty from
the electronic signal of the source flux.
While the latter contribution can be
computed directly from the integrated
source flux and the calibrated gain of the
detector, the sky background has to be modeled
owing to pixel-level correlations induced 
during the mosaicing process. Dithering of
exposures and the combination of subregion
mosaics at differing position angles causes
resampling of pixels during the construction
of the mosaics, and the individual 
exposure pixel values become
coupled across multiple pixels in the final mosaics.
These correlations are complicated by the
character of the JWST PSF, with large-angle
diffraction spikes, and even by the JWST
detector interpixel capacitance \citep{rauscher2014a}.

As a result of the pixel-level correlations, 
the scaling
of photometric uncertainty from the sky
background with the size of
an aperture used to integrate the source flux
should scale superlinearly with the
size of the aperture
\citep[e.g.,][]{labbe2005a,quadri2007a,whitaker2011a,skelton2014a,whitaker2019a}.
The scaling of the uncertainty with the linear
aperture size can be approximately modeled
with a power-law as
\begin{equation}
\label{eqn:error-scaling}
\sigma = \sigma_{1} N^{\beta}
\end{equation}
\noindent
where $N$ is the linear aperture size in pixels,
$\sigma_{1}$ is the single-pixel uncertainty, and
$1\leq\beta\leq2$ characterizes the pixel covariance
through the increase of uncertainty with aperture
size and is bounded between uncorrelated ($\beta=1$)
and perfectly correlated ($\beta=2$) limits.
Since the correlations arise from details of the
image combination process, the local
number, relative orientation on the sky,
and the composite dither pattern of the exposures
will affect the value of $\beta$ spatially. 
We therefore expect the relation expressed
in Equation \ref{eqn:error-scaling} to vary
spatially throughout the image. The degree
of inhomogeneity of the mosaic pattern
and depth will in turn affect the level of
spatial variation in the correlations. Assuming
a uniform sky, if
the mosaic is fairly homogeneous, the distribution
of random aperture fluxes measured on source-free
regions of the sky will also be fairly uniform
across the mosaic. The scaling in Equation
\ref{eqn:error-scaling} can be computed by measuring
an outlier-insensitive estimate of the
spread of random aperture fluxes, such as the 
mean absolute deviation estimator for the normal
distribution standard deviation, as a function of
aperture size and then fitting the resulting
trend. The fit can be used to assign model 
photometric uncertainties to flux measurements
made using different aperture sizes.
This same method can be used to compute the
photometric uncertainties for sources in common-PSF mosaics,
which will have a different pixel covariances and
associated $\beta$ values compared with
their corresponding native PSF mosaics.

For a more complicated mosaic that is inhomogeneous in depth and
composition, instead of
a dominant trend of uncertainty with aperture
size there will be a variety of subtrends. The degree of
pixel covariance will be a local quantity. In \citet{rieke2023a},
the uncertainty scaling Equation \ref{eqn:error-scaling} was
measured for the \texttt{jw01180.deep} and \texttt{jw019630}
subregion mosaics
by measuring the sky flux in random apertures across the image
and grouping the apertures by the local exposure time. Many
uncertainty scalings were estimated, and the photometric uncertainties
for a given source were assigned from these fitted scalings
based on the exposure time 
at the source location and the aperture used for photometry (see
their Figure 3). The same technique was used in previous
JADES imaging data releases and analyses, where the combined
mosaics where not highly inhomogeneous in depth or composition
\citep[e.g.,][]{eisenstein2025a,deugenio2025a,robertson2024a}.

For the current analysis, as described by \benspaper{}, the JADES DR5 data release
mosaics display dramatic variations in depth and composition
in their exposures. As the bithash images presented in \benspaper{}
indicate, the transition between regions with substantial depth or 
composition variation can be quite sharp and the character
of the pixel covariance could be expected to change discontinuously.
These features suggest that the approach of fitting a 
trend of uncertainty with aperture size, or even fitting multiple
trends of uncertainty with aperture size conditioned on the local
mosaic properties, should be adapted for a more complex
mosaic pattern. Here we introduce such an approach. First,
mosaics of the uncertainty measured from the distribution
of local sky flux measured in random apertures are constructed
for each filter and for a range of aperture sizes
(Section \ref{sec:uncertainty-mosaics}). We then present
a regression method for creating a model of the
photometric uncertainty at each pixel location in our mosaics
as a function of aperture size, which uses the background
variance mosaic propagated from the \texttt{jwst} pipeline
as a spatial template for the single-pixel uncertainty 
and jointly fits to the random aperture uncertainty mosaics
as a power-law with increasing aperture size
(Section \ref{sec:uncertainty-model-regression}).

\subsection{Uncertainty Mosaics}
\label{sec:uncertainty-mosaics}

The uncertainty mosaics provide a map of the local
spread in sky flux values measured in apertures
randomly placed across the image. To create the uncertainty
mosaics, we begin by selecting a low resolution grid
of pixel locations where the local uncertainty will be estimated
at a 50:1 ratio relative to the full mosaic resolution. We then distribute across each filter
mosaic 4,000,000 random apertures at valid locations masked
by the detection segmentation map and where the
filter WHT$>1$. A KDTree is built using the
\texttt{scipy} library and used to search for a target of
1000 nearest neighbor aperture locations to each 
pixel location where the local sky background distribution
will be estimated. In the tree search, we use a bithash mosaic
constructed for each filter individually to inform which random
apertures have been drawn from regions that have the same
combination of subregion mosaics and are therefore likely to
show the same depth and pixel covariance properties. We prioritize
random aperture locations from the same bithash as the pixel
location of interest in targeting a 1000 neighbors, but if at least 500
neighbor apertures cannot be selected from the same bithash value 
(e.g., the pixel is near a border between subregion mosaics) then
we augment those apertures with the nearest random sky apertures of
any bithash value. The RMS of these random sky fluxes are
measured and recorded, and then the low resolution uncertainty
mosaic is reinterpolated to the full resolution of the DR5
mosaic using the GPU-accelerated \texttt{cupy} implementation
of the \texttt{scipy.ndimage.zoom} routine. Any pixel-level
interpolation errors (zeros or negative errors) are replaced with the
median valid uncertainty mosaic value from the surrounding $21\times21$
pixel footprint.
Uncertainty mosaics are generated for circular apertures
with radii $r=[0.1",0.15",0.25",0.3",0.35",0.5",0.9",1.5"]$
for every native resolution and common-PSF filter mosaic.

\subsection{Regression Model for Uncertainty Mosaics}
\label{sec:uncertainty-model-regression}

With the uncertainty mosaics computed, an extension of the power-law
scaling model Equation \ref{eqn:error-scaling} can be constructed
for every pixel. Indeed, for the HST and MIRI mosaics we fit
simple power-laws at the pixel level throughout the mosaic to compute
model uncertainty mosaics at aperture sizes of
$r=[0.1",0.15",0.25",0.3",0.35",0.5"]$. For smaller apertures, we 
extrapolate the power-law toward the single-pixel limit (0.03"). 
Mosaics of the power-law amplitude and index $\beta$ are retained.
For larger 
apertures we scale linearly from the $r=0.5"$ value as the pixel
covariances begin to decorrelate on larger scales. 

For the NIRCam mosaics, we can additionally leverage the computation
of the pixel-level
source-free background uncertainty image, which is stored in the
WHT HDU as described in \benspaper{}. This mosaic quantity replaces
the WHT value from the standard \texttt{jwst} pipeline data model
and represents the expected pixel-level
uncertainty in the final associated with the quadrature sum of 
read and sky background noise. While we expect this image to not
perfectly reflect the true amplitude of the pixel-level background, 
it does correctly encode the relative expected depths of the pixel-level
background associated with the various subregions of the mosaic and the
discontinuity of their spatial topology. We therefore wish to include
this pixel-level uncertainty as a template for the single-pixel uncertainty
mosaic, allowing for rescaling by a multiplicative coefficient $\exp(\alpha)$, 
and then perform a joint fit of the coefficient and all the power-law
indices $\beta_j$ for the pixel-by-pixel uncertainty scalings. Given a 
number of $m$ pixels in the mosaic, this template-based
fitting of the uncertainty model for NIRCam ($m+1$ parameters)
is much more constrained than for the HST and MIRI mosaics ($2m$ parameters).
The fit of a single value for $\alpha$
does couple the uncertainty scaling of all pixels together, although somewhat weakly.
Since we newly introduce this method here, below we explicitly work through the
analytical solution of the system of equations for $\alpha$ and $\beta_j$.
Readers interested in the utilization of the uncertainties in the
forced photometry of sources can find the related discussion in
Section \ref{sec:forced-photometry}.

Consider the mosaic with $m$ pixels labeled with pixel index $j$,
each with $n$ aperture 
RMS measurements labeled with aperture index $i$. A model for the uncertainty scaling with
the linear aperture sizes $N_{ij}$ for every pixel location
based on a single-pixel aperture template can be written as
\begin{equation}
\sigma_{ij} = A P_{j} N_{ij}^{\beta_{j}}
\end{equation}
where $A P_{j}$ provides the single-pixel RMS, $A$ is a multiplicative coefficient
common to every pixel, $P_{j}$ is the template for the RMS uncertainty for a single-pixel
aperture, and the $1\leq \beta_j \leq 2$ are power-law indices. Taking the logarithm of
both sides, the model can be written as
\begin{equation}
y_{ij} = p_{j} + \alpha + \beta_j x_{ij} 
\end{equation}
\noindent
where $\alpha = \log A$, $p_j = \log P_j$, $x_{ij} = \log N_{ij}$ nd $y_{ij} = \log \sigma_{ij}$. The goal then is to compute best fit values of $\alpha$ and $\beta_j$ that best
reproduce $y_{ij}$ for each $x_{ij}$.

The summed squared difference between the model and the measured RMS uncertainties
can then be written
\begin{equation}
\chi^2 = \Sigma_{j}^{m} \Sigma_{i}^{n} [y_{ij} - (p_{j} + \alpha + \beta_{j} x_{ij})]^2
\end{equation}
\noindent
where $i$ runs over the number of $n$ apertures, and 
$j$ runs over the number of pixels $p_j$ in the template. 
The metric $\chi^2$ can be minimized with respect to the parameters $\alpha$
and $\beta_j$ by taking partial derivatives
\begin{equation}
\frac{\partial\chi^2}{\partial \alpha} = 2\Sigma_{j}\Sigma_{i} [y_{ij} - (p_j + \alpha + \beta_j x_{ij})](-1) = 0
\end{equation}
\noindent
and
\begin{equation}
\frac{\partial\chi^2}{\partial \beta_j} = 2\Sigma_{i} [y_{ij} - ( p_j + \alpha + \beta_j x_{ij})](-x_{ij}) = 0
\end{equation}
\noindent
After noting that $\Sigma_i \Sigma_j \alpha = nm\alpha$, the system of equations can be written as a matrix equation
\begin{equation}
\mathbf{M} \vec{\phi} \equiv
\begin{bmatrix}
\Sigma_{i} x_{i1}^2   & 0                   & \cdots & 0                   & \Sigma_{i} x_{i1} \\ 
0                     & \Sigma_{i} x_{i2}^2 & \ddots & \vdots              & \vdots      \\
\vdots                & \ddots              & \ddots & \vdots              & \vdots \\
0                     & \cdots              & \cdots & \Sigma_{i} x_{ij}^2 & \Sigma_{i} x_{ij} \\
    \Sigma_{i} x_{i1} & \cdots              & \cdots & \Sigma_{i} x_{ij}   & n m      
\end{bmatrix}
\begin{bmatrix}
\beta_1 \\
\vdots  \\
\vdots  \\
\beta_j \\
\alpha
\end{bmatrix}
=
\begin{bmatrix}
\Sigma_{i} x_{i1} (y_{i1} - p_1)\\
\vdots \\ 
\vdots \\
\Sigma_{i} x_{ij} (y_{ij} - p_j)\\
\Sigma_j \Sigma_{i} (y_{ij}-p_j)\\
\end{bmatrix}
\equiv
\vec{z}
\end{equation}
\noindent
The system can be solved for the parameters $\vec{\phi} = [\beta_j,\alpha]$ by
inverting the left-most matrix $\mathbf{M}$, whose block-like structure makes an analytical
solution feasible.  The solution for the parameters is then just
$\vec{\phi} = \mathbf{M}^{-1} \vec{z}$.

The matrix $\mathbf{M}$ can be split into a block-wise form as
\begin{equation}
\mathbf{M} = 
\begin{bmatrix}
\mathbf{A} & \mathbf{B} \\
\mathbf{C} & \mathbf{D}
\end{bmatrix}
\end{equation}
\noindent
where
\begin{equation}
\mathbf{A} = 
\begin{bmatrix}
\Sigma_{i} x_{i1}^2   & 0      & \cdots & 0 \\ 
0                     & \Sigma_{i} x_{i2}^2 & \ddots & \vdots \\
\vdots                & \ddots & \ddots & 0 \\
0                    & \cdots & \cdots & \Sigma_{i} x_{ij}^2  
\end{bmatrix},\,\,
\mathbf{B} = 
\begin{bmatrix}
 \Sigma_{i} x_{i1} \\ 
\vdots                \\
\Sigma_{i} x_{ij}
\end{bmatrix},\,\,
\mathbf{C} = 
\begin{bmatrix}
\Sigma_{i} x_{i1} & \cdots & \Sigma_{i} x_{ij}
\end{bmatrix},\,\,
\mathbf{D} = 
\begin{bmatrix}
n m
\end{bmatrix}.
\end{equation}
\noindent
The inverse of this matrix is computable because $\mathbf{A}$ is 
diagonal and $\mathbf{D}$ is a scalar constant. Note that 
$\mathbf{B} = \mathbf{C}^\mathrm{T}$, but each submatrix is treated
distinctly for clarity. The blockwise inverse is
complicated but analytical, and can be written as
\begin{equation}
\label{eqn:M-inverse}
\mathbf{M}^{-1} = 
\begin{bmatrix}
\mathbf{A}^{-1} + \mathbf{A}^{-1} \mathbf{B}(\mathbf{M}/\mathbf{A})^{-1} \mathbf{C} \mathbf{A}^{-1} & - \mathbf{A}^{-1} \mathbf{B} (\mathbf{M}/\mathbf{A})^{-1} \\
- (\mathbf{M}/\mathbf{A})^{-1} \mathbf{C} \mathbf{A}^{-1}  & (\mathbf{M}/\mathbf{A})^{-1}
\end{bmatrix}
\end{equation}
\noindent
where
\begin{equation}
\label{eqn:A-inverse}
\mathbf{A}^{-1} =
\begin{bmatrix} 
(\Sigma_{i} x_{i1}^{2})^{-1}   & 0      & \cdots & 0 \\ 
0                     & (\Sigma_{i} x_{i2}^{2})^{-1} & \ddots & \vdots \\
\vdots                & \ddots & \ddots & 0 \\
0                    & \cdots & \cdots & (\Sigma_{i} x_{ij}^{2})^{-1}  
\end{bmatrix}
\end{equation}
\noindent
is simple because $\mathbf{A}$ is 
diagonal, and
\begin{equation}
\label{eqn:M-A}
\mathbf{M}/\mathbf{A} = \mathbf{D} - \mathbf{C} \mathbf{A}^{-1} \mathbf{B}.
\end{equation}
\noindent
Note that $\mathbf{M}/\mathbf{A}$ is actually a scalar constant, so $(\mathbf{M}/\mathbf{A})^{-1}$ is also a scalar constant.
The quantity 
\begin{equation}
\mathbf{C} \mathbf{A}^{-1} \mathbf{B} =
{\mathlarger{\sum\limits_{j}^{} }} \frac{ \left(\Sigma_i x_{ij}\right)^2}{\Sigma_i x_{ij}^2}
\end{equation}
\noindent
expresses ratio of square of the summed logarithmic aperture sizes to the sum of their squares, totaled over the mosaic. This scalar is then subtracted from $\mathbf{D}$,
and the inverse taken to provide the lower-right submatrix inverse
\begin{equation}
\label{eqn:M-A-inverse}
(\mathbf{M}/\mathbf{A})^{-1} = [\mathbf{D}-\mathbf{C} \mathbf{A}^{-1} \mathbf{B}]^{-1} =
\left[ nm - {\mathlarger{\sum\limits_{j}^{} }} \frac{ \left(\Sigma_i x_{ij}\right)^2}{\Sigma_i x_{ij}^2} \right]^{-1}.
\end{equation}
\noindent
The submatrix $(\mathbf{M}/\mathbf{A})^{-1}$ is used to compute the
upper right and lower left submatrices as
\begin{equation}
\label{eqn:A-B-M-A-inv}
-\mathbf{A}^{-1} \mathbf{B} (\mathbf{M}/\mathbf{A})^{-1}
= -\left[ nm - {\mathlarger{\sum\limits_{j}^{} }} \frac{ \left(\Sigma_i x_{ij}\right)^2}{\Sigma_i x_{ij}^2} \right]^{-1}
\begin{bmatrix}
 \dfrac{\Sigma_{i} x_{i1}}{\Sigma_i x_{i1}^2}  \\ 
\vdots                \\
 \dfrac{\Sigma_{i} x_{ij}}{\Sigma_i x_{ij}^2} 
\end{bmatrix}
\end{equation}
\noindent
and
\begin{equation}
\label{eqn:M-A-C-A-inv}
-(\mathbf{M}/\mathbf{A})^{-1} \mathbf{C} \mathbf{A}^{-1}
= -\left[ nm  - {\mathlarger{\sum\limits_{j}^{} }} \frac{\left(\Sigma_i x_{ij}\right)^2}{\Sigma_i x_{ij}^2} \right]^{-1}
\begin{bmatrix}
\dfrac{\Sigma_{i} x_{i1}}{\Sigma_{i} x_{i1}^2}  & \cdots & \dfrac{\Sigma_{i} x_{ij}}{\Sigma_{i} x_{ij}^2}.
\end{bmatrix}
\end{equation}
\noindent
The remaining upper left submatrix in Equation \ref{eqn:M-inverse}
for $\mathbf{M}^{-1}$ is the sum of $\mathbf{A}^{-1}$ (Equation \ref{eqn:A-inverse})
and the matrix product of Equations \ref{eqn:M-A}, \ref{eqn:A-B-M-A-inv}, and 
\ref{eqn:M-A-C-A-inv}.

Once the best-fit parameters $\alpha$ and $\beta_j$ are determined,
uncertainty model mosaics are computed for every filter and circular aperture 
size $r=[0.1",0.15",0.25",0.3",0.35",0.5"]$. The sky background contribution
to the photometric uncertainty of each object is then log-linearly interpolated
based on the aperture size of the photometric measure and the object's position
on the uncertainty model mosaic. These uncertainties are added in quadrature
with the Poisson uncertainty associated with the electronic signal of the source,
and reported as the uncertainties with the `\texttt{\_e}' suffixes in the
released photometric catalogs. We also measure the uncertainty estimates from the
\texttt{ERR} HDU from the \texttt{jwst} pipeline as the uncertainties with the `\texttt{\_ei}' suffixes in the catalogs, noting that these `\texttt{\_ei}'
uncertainties do not account for 
possible correlated pixel affecting the uncertainty scaling with aperture size and
therefore underestimate the photometric uncertainties.

\section{Forced Photometry}
\label{sec:forced-photometry}

The source population in the JADES images display a wide range of
spectral energy distributions. The approach to measuring
photometry for sources involves the forced sum of
pixel values within apertures centered at the
locations of detections, where the detected
sources are discovered in deep long-wavelength
stacks where the sensitivity to many sources
is expected to be high. By performing forced
photometry for detections from the long-wavelength
stack, high-redshift sources can be
discovered even as their fluxes in dropout
filters approach the sky background.

In this Section, we describe fixed aperture
forced photometry including circular aperture 
photometry (Section \ref{sec:circular-aperture-photometry})
and ellipsoidal Kron aperture photometry
(Section \ref{sec:kron-aperture-photometry}),
where the aperture size is selected or determined
from the source properties in the detection image.
Section \ref{sec:curve-of-growth} below describes
curve-of-growth photometric measures for every object.
We note here that
when using the forced photometry on unconvolved mosaics
for scientific applications, 
the use of a fixed aperture with point source aperture corrections will induce anomalous colors for extended objects.
Analyses that want unbiased colors (for objects with no intrinsic color gradients) should use the common-PSF mosaics.  
Analyses that want maximum SNR for compact objects and in particular maximum SNR on the measurement of differences in flux between nearby bands should use photometry performed on the
unconvolved mosaics.

\subsection{Circular Aperture Photometry}
\label{sec:circular-aperture-photometry}

For every source, forced 
circular aperture photometry is reported
for apertures of radii $r=[0.1",0.15",0.25",0.3",0.35",0.5"]$.
These photometric measures are labeled as
\texttt{CIRC1}-\texttt{CIRC6} in the catalogs,
with the filter of the measurement included
as a prefix. For instance, the $r=0.3$"
circular aperture measurements for
the F090W filter are labeled as \texttt{F090W\_CIRC4}.
The photometry is aperture corrected by measuring the
fractional encircled energy of the corresponding filter's
PSF within the aperture (see Section \ref{sec:psf}).
For each filter, an additional circular aperture photometric
measure \texttt{CIRC0} is computed within a radius that encircles 
80\% of the PSF energy for that band, such that the aperture correction
for a point source should be constant with wavelength.
The photometric uncertainties computed from the 
model regression on the random aperture uncertainty
mosaics, as described in Section \ref{sec:uncertainties},
are reported with the \texttt{\_e} suffix (e.g., these F090W
uncertainties are \texttt{F090W\_CIRC4\_e}) and the
uncertainties estimated from the \texttt{ERR} HDU are
labeled with the \texttt{\_ei} suffix (e.g., \texttt{F090W\_CIRC4\_ei}).
The forced circular aperture photometry is reported
for the filter and common-PSF mosaics in the \texttt{CIRC}
and \texttt{CIRC\_CONV} HDUs of the catalogs, respectively.

We include photometry on the direct images and also
include corrections for a local background. For circular
aperture photometry, the local background is measured
in a circular annulus over the radii $1.5"<r<1.55"$.
As with the source flux measurement, pixels from
segmentations belonging to other sources are censored.
The median pixel value within the annulus is computed,
and then the corresponding interior flux within the
circular aperture area is subtracted from the
integrated flux measurement. The background flux removed
is recorded in the field labeled with the \texttt{\_bkg} suffix
(e.g., \texttt{F090W\_CIRC1\_bkg}). The circular aperture measurements
that include background subtraction are provided in the 
\texttt{CIRC\_BSUB}
and \texttt{CIRC\_BSUB\_CONV} HDUs of the catalogs.

\begin{deluxetable}{lccccccccc}[!ht]
\tabletypesize{\scriptsize}
\tablecaption{Circular Aperture Size and Correction Information \label{tab:aperture-corrections}}
\tablehead{
\colhead{Filter} & \multicolumn{2}{c}{\texttt{CIRC0}} & \texttt{CIRC1}, $r=0.1$" & \texttt{CIRC2}, $r=0.15$"& \texttt{CIRC3}, $r=0.25$"& \texttt{CIRC4}, $r=0.3$"& \texttt{CIRC5}, $r=0.35$"& \texttt{CIRC6}, $r=0.5$" \vspace{-0.2cm}\\
\colhead{}  & \colhead{$r$ [arcsec]} & \colhead{Aper. Corr.} & \colhead{Aper. Corr.}&  \colhead{Aper. Corr.} & \colhead{Aper. Corr.}& \colhead{Aper. Corr.}& \colhead{Aper. Corr.} & \colhead{Aper. Corr.}
}
\startdata
HST/ACS     F435W   & 0.433 &1.25& 1.532& 1.277& 1.161&1.141&1.128&1.100\\
HST/ACS     F606W   & 0.592 &1.25& 1.628& 1.304& 1.181&1.151&1.134&1.104\\
HST/ACS     F775W   & 0.769 &1.25& 1.620& 1.285& 1.166&1.140&1.116&1.088\\
HST/ACS     F814W   & 0.806 &1.25& 1.667& 1.312& 1.165&1.139&1.115&1.082\\
HST/ACS     F850LP  & 0.904 &1.25& 2.067& 1.522& 1.259&1.218&1.185&1.123\\
HST/WFC3    F105W   & 1.055 &1.25& 2.772& 1.738& 1.266&1.209&1.181&1.120\\
HST/WFC3    F125W   & 1.249 &1.25& 2.955& 1.865& 1.305&1.222&1.185&1.132\\
HST/WFC3    F140W   & 1.392 &1.25& 3.100& 1.954& 1.357&1.251&1.203&1.152\\
HST/WFC3    F160W   & 1.537 &1.25& 3.309& 2.064& 1.413&1.277&1.210&1.153\\
JWST/NIRCam F070W   & 0.161 &1.25& 1.355& 1.267& 1.144&1.114&1.095&1.060\\
JWST/NIRCam F090W   & 0.141 &1.25& 1.319& 1.240& 1.156&1.122&1.101&1.066\\
JWST/NIRCam F115W   & 0.129 &1.25& 1.296& 1.226& 1.155&1.130&1.118&1.071\\
JWST/NIRCam F150W   & 0.125 &1.25& 1.324& 1.223& 1.152&1.135&1.128&1.079\\
JWST/NIRCam F162M   & 0.128 &1.25& 1.346& 1.223& 1.154&1.137&1.124&1.084\\
JWST/NIRCam F182M   & 0.135 &1.25& 1.386& 1.225& 1.160&1.137&1.124&1.091\\
JWST/NIRCam F200W   & 0.141 &1.25& 1.410& 1.234& 1.166&1.138&1.124&1.092\\
JWST/NIRCam F210M   & 0.147 &1.25& 1.430& 1.242& 1.173&1.141&1.124&1.096\\
JWST/NIRCam F250M   & 0.176 &1.25& 1.573& 1.318& 1.179&1.160&1.136&1.103\\
JWST/NIRCam F277W   & 0.189 &1.25& 1.623& 1.351& 1.184&1.166&1.146&1.105\\
JWST/NIRCam F300M   & 0.201 &1.25& 1.675& 1.384& 1.189&1.170&1.156&1.109\\
JWST/NIRCam F335M   & 0.221 &1.25& 1.762& 1.428& 1.206&1.174&1.162&1.115\\
JWST/NIRCam F356W   & 0.231 &1.25& 1.811& 1.444& 1.220&1.178&1.164&1.120\\
JWST/NIRCam F410M   & 0.259 &1.25& 1.965& 1.488& 1.270&1.196&1.170&1.139\\
JWST/NIRCam F430M   & 0.270 &1.25& 2.043& 1.504& 1.291&1.207&1.173&1.145\\
JWST/NIRCam F444W   & 0.275 &1.25& 2.062& 1.514& 1.299&1.215&1.176&1.145\\
JWST/NIRCam F460M   & 0.290 &1.25& 2.176& 1.542& 1.330&1.235&1.182&1.151\\
JWST/NIRCam F480M   & 0.302 &1.25& 2.262& 1.567& 1.351&1.254&1.191&1.152\\
JWST/MIRI   F560W   & 1.334 &1.25& 5.945& 3.246& 1.989&1.781&1.638&1.434\\
JWST/MIRI   F770W   & 0.792 &1.25& 5.443& 2.901& 1.739&1.609&1.541&1.341\\
JWST/MIRI   F1000W  & 0.632 &1.25& 6.954& 3.500& 1.826&1.595&1.489&1.369\\
JWST/MIRI   F1280W  & 0.756 &1.25& 9.964& 4.803& 2.218&1.812&1.596&1.401\\
JWST/MIRI   F1500W  & 0.864 &1.25& 12.871& 6.082& 2.644&2.082&1.765&1.423\\
JWST/MIRI   F1800W  & 1.020 &1.25& 17.626& 8.180& 3.371&2.564&2.092&1.509\\
JWST/MIRI   F2100W  & 1.157 &1.25& 22.524& 10.352& 4.137&3.085&2.462&1.645\\
JWST/MIRI   F2550W  & 1.159 &1.25& 32.483& 15.761& 5.697&4.151&3.227&1.967\\
\enddata
\tablecomments{The JWST NIRCam circular aperture corrections are computed from the model PSFs presented by \benspaper{}, while the JWST MIRI circular aperture corrections are calculated from the model PSFs presented by \staceyspaper{} (see Section \ref{sec:psf}. The HST aperture corrections are computed from the empirical PSFs computed in Section \ref{sec:hst-epsf}.}
\end{deluxetable}

\begin{figure*}[ht!]
\begin{centering}
\includegraphics[width=0.8\textwidth]{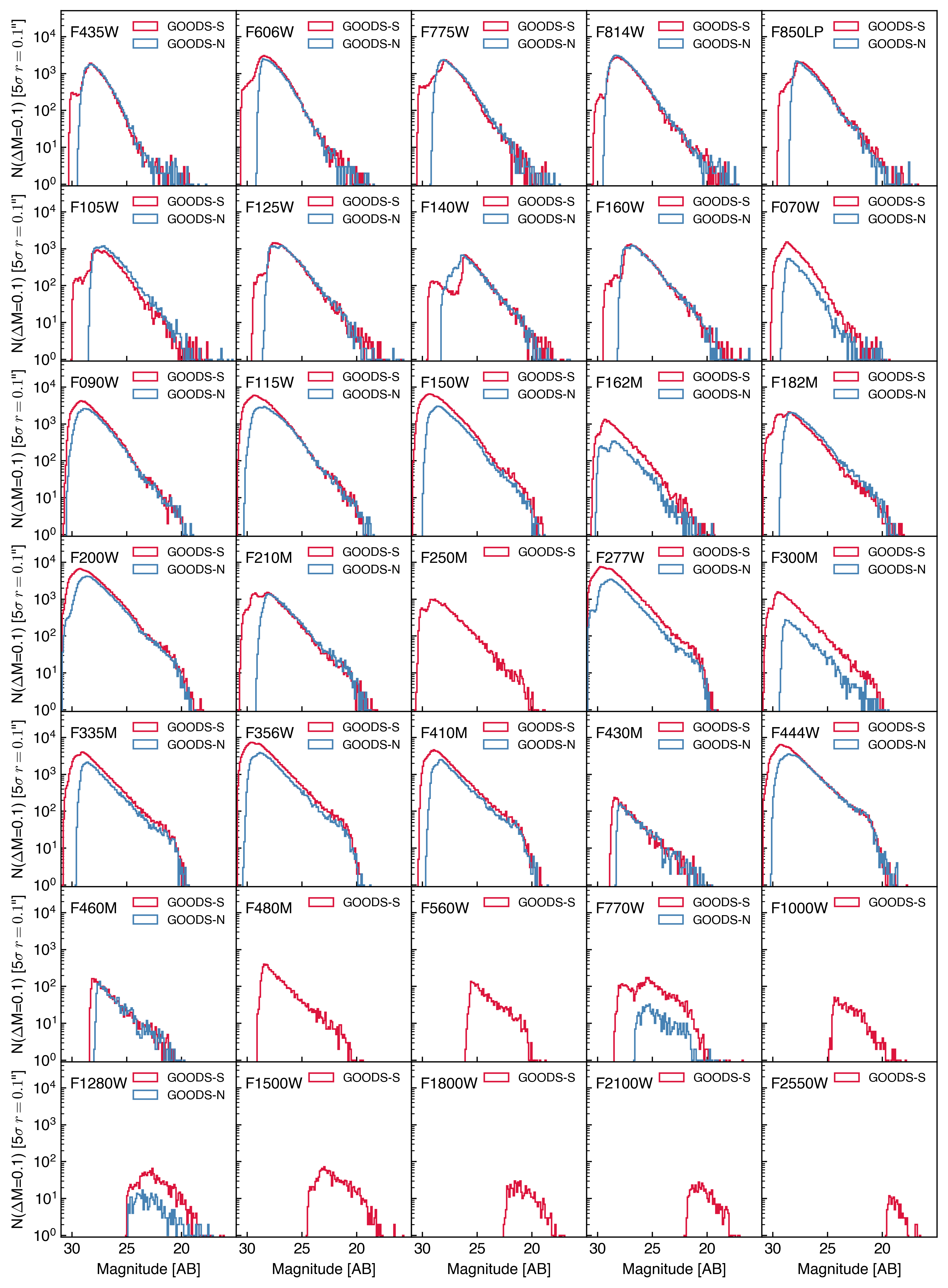}
\caption{Histograms of the source counts for 5$\sigma$-significant
objects measured in $r=0.1$" circular apertures, for the 35
HST/ACS, HST/WFC3, JWST/NIRCam, and JWST/MIRI filters used in the
JADES DR5 photometric catalogs. Shown in each panel are the
source counts in the GOODS-N (blue) and GOODS-S (red) fields in bins of width $\Delta M=0.1$ magnitudes,
and the filter mosaic measured in each panel is indicated in the upper left corner. All detected sources are considered for inclusion in each
panel, but only sources with SNR$>5$ \texttt{CIRC1} photometry in a
given filter are included in the histograms for that band. The source
counts between GOODS-S and GOODS-N are consistent once differences
in areal coverage are considered.
\label{fig:number-counts}}
\end{centering}
\end{figure*}

Figure \ref{fig:number-counts} presents histograms
of the JADES DR5 catalog sources with SNR$>5$ $r=0.1$" circular
aperture fluxes
in each
of the 35 HST/ACS, HST/WFC3, JWST/NIRCam, and JWST/MIRI
filter mosaics used in this work, spanning a factor
of $\sim60\times$ in wavelength. Each panel shows the
source count histograms for the
GOODS-N (blue) and GOODS-S (red) fields, plotted over the
range of $31>m_{AB}>15$. 
The number of SNR$>5$
sources range from $N>300,000$ in F200W, F277W, and F356W across
GOODS-S and GOODS-N to only $N=127$ in F2550W (from SMILES).
The GOODS-S and GOODS-N source counts
are consistent once the areal filter coverage is considered.

\begin{figure*}[ht!]
\begin{centering}
\includegraphics[width=0.8\textwidth]{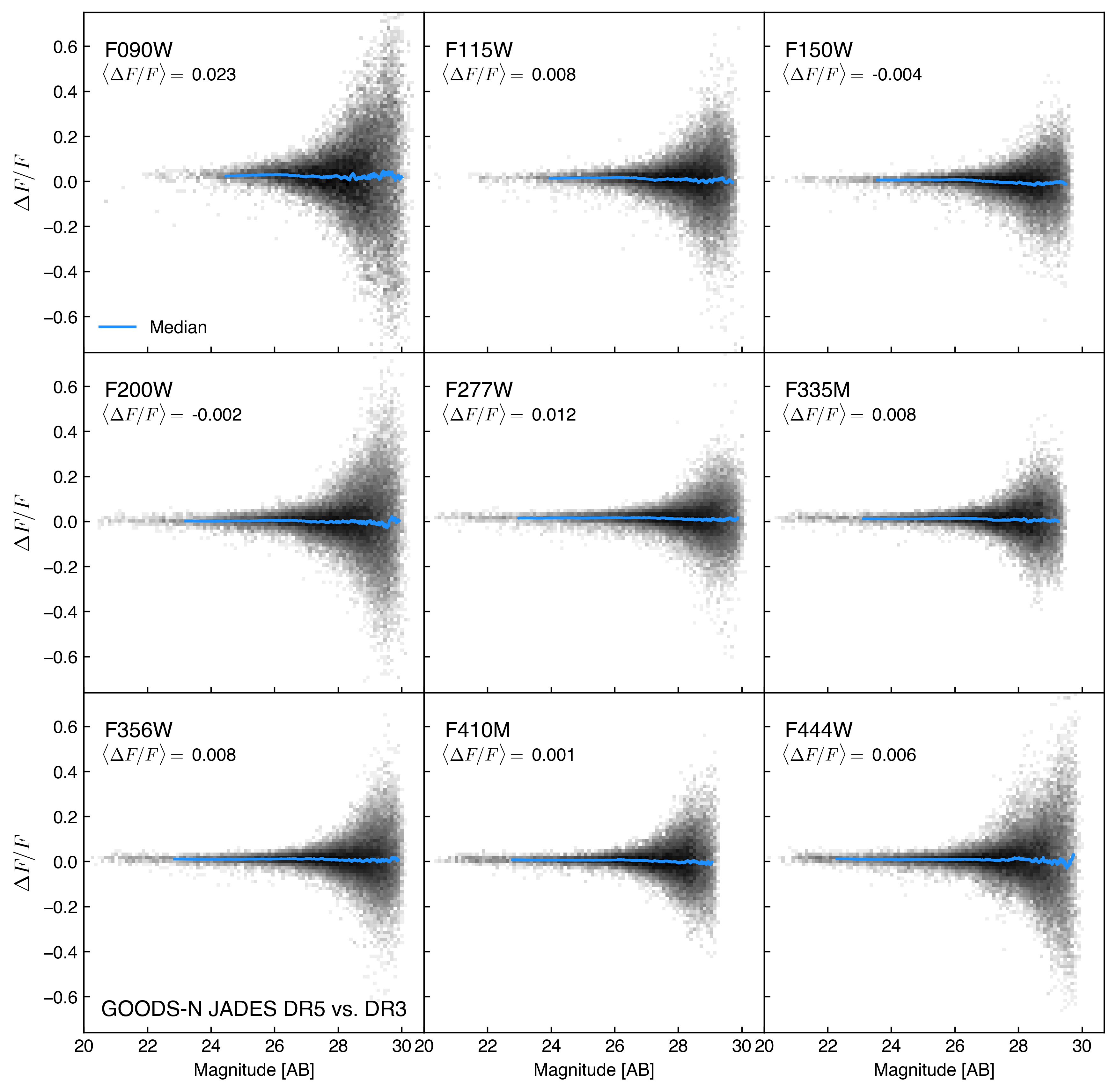}
\caption{Comparison between the JADES DR3 vs. DR5 
photometry. Shown are the fractional flux differences in
F090W, F115W, F150W, F200W, F277W, F335M, F356W, F410M, and
F444W between the DR3 and DR5 $r=0.1$" circular aperture photometry (\texttt{CIRC1}), measured relative to the DR5 flux, 
for sources with DR5 SNR$>5$ in each filter, as a function 
of the \texttt{CIRC1} AB magnitude in the filter. The distribution of
flux differences are shown as normalized two-dimensional histograms
 (shaded regions). Running windowed medians of the histogram values are computed and reported as a blue line, and the median of the
 running medians is reported as an annotation in each panel. The
 typical flux differences between DR3 and DR5 for well-matched sources
 is better than 1\%.
\label{fig:phot-comp-goods-n}}
\end{centering}
\end{figure*}

Given the release of prior photometric catalogs by the
JADES collaboration, a comparison between source photometric
measures between catalog versions is warranted. Figure \ref{fig:phot-comp-goods-n} presents a direct comparison between
the JADES DR3 photometric catalog for GOODS-N
released by \citet{deugenio2025a}
and the JADES DR5 photometric catalog described in this work.
Shown are the differences between the DR3 \texttt{CIRC1}
photometry and the DR5 \texttt{CIRC1}
photometry, plotted as a fraction change in flux
relative to the DR5 \texttt{CIRC1}
photometry, as a function of the DR5 \texttt{CIRC1} AB magnitude
in the F090W, F115W, F150W, F200W, F277W, F335M, F356W, F410M,
and F444W JWST NIRCam filters. Differences are measured for sources
whose centroids match between catalogs better than the \texttt{CIRC1}
aperture size of $r=0.1$" and have SNR$>5$ in DR5.
Normalized
two-dimensional histograms of these histograms 
are presented (shaded regions) along with running windowed
medians (blue lines).  The median value of the running median
differences for each filter are reported in the image panels.
Typically, measured flux
differences are better than 1\% with the largest
differences in F090W.

\subsection{Kron Aperture Photometry}
\label{sec:kron-aperture-photometry}

Forced ellipsoidal aperture photometry
is performed about the centroid locations
of every object in every filter. For
the size of the apertures, to approximate
a total flux we adopt the Kron aperture
sizes determined from the Gaussian profile
regression fits to the detection image
data described in Section \ref{sec:source-properties}.
The semimajor axis \texttt{A\_KRON}
and semiminor axis \texttt{B\_KRON} of these
ellipsoidal apertures are set to be proportional
to the 2D Gaussian profile semimajor and
semiminor sizes, times the dimensionless unscaled
Kron radius \texttt{R\_KRON\_U} and the
Kron parameter $k=2.5$. The position angle
of the ellipsoidal aperture on the
sky \texttt{THETA} is computed during the
Gaussian model regression and kept
fixed when forcing photometry. We additionally
compute ``short'' Kron photometry using
a smaller parameter $k=1.4$ to set the corresponding
semimajor and
semiminor aperture axes \texttt{A\_KRON\_S}
and \texttt{B\_KRON\_S}.
When measuring the photometry, a background
subtraction is performed where the local
background per pixel is measured from an ellipsoidal
annular aperture of width 4 pixels, inner
semimajor axis 2\texttt{A\_KRON}, and the
same axis ratio as the Kron aperture.
When measuring the photometry or background,
pixels belonging to the segmentations of other
sources are masked.  The uncertainties
are interpolated from uncertainty model
mosaics (see Section \ref{sec:uncertainty-mosaics}) appropriately for the Kron aperture size
of each source (\texttt{\{BAND\}\_KRON\_e} uncertainties)
and also estimated from the \texttt{ERR} HDU if available.
The fluxes and uncertainties are aperture corrected
by integrating the enclosed energy of the filter PSF
over the ellipsoidal aperture of the source.
The forced Kron aperture photometry is measured
for every source in every filter, using both the
native (\texttt{\{BAND\}\_KRON}) and common-PSF mosaics
(\texttt{\{BAND\}\_KRON\_CONV}).

\section{Curve of Growth}
\label{sec:curve-of-growth}

The adoption of Kron apertures generating
forced total flux estimates is heuristic
but follows the previous JADES photometric
releases \citep[e.g.,][]{rieke2023a,eisenstein2025a,deugenio2025a}
and benefits from convenient implementations in the
\texttt{photutils} library. Here we introduce
a new photometric measurement for the JADES
photometric catalog releases, where for every
source in every filter mosaic and 
common-PSF image the curve-of-growth of 
the flux distribution is computed. The 
curves-of-growth are computed in ellipsoidal apertures
with maximum semimajor and semiminor extent of
2\texttt{A\_KRON} and  2\texttt{B\_KRON}, respectively.
This aperture defines the region within which 
the \texttt{FLUX\_TOT} for the source is computed.
Just outside these apertures, the background flux
is computed and subtracted from the total interior
flux within the aperture. A series of fixed
separation apertures with the same axis ratio and
position angle are placed within the maximum aperture
size, and the aperture fluxes are computed. An interpolation
table is generated, and then the aperture sizes that enclose
fractional increases of 5\% of the \texttt{FLUX\_TOT}
are determined. The semimajor axes of these apertures
are recorded as \texttt{A\_5}, \texttt{A\_10}, \texttt{A\_15}...
all the way through 
\texttt{A\_100}, where \texttt{A\_100} is the inner most 
aperture size that first contains a flux \texttt{FLUX\_TOT},
which can be smaller than  2\texttt{A\_KRON} if the curve-of-growth
flattens at large distance.
These aperture sizes
then determine the curve-of-growth as the interior flux
within them grows to reach \texttt{FLUX\_TOT}.
The uncertainties on the \texttt{FLUX\_TOT} are computed 
from the uncertainty mosaics for apertures the size of \texttt{A\_100},
and the sky background contribution separately recorded in addition
to the total uncertainty.  The uncertainty estimate from the 
\texttt{ERR} HDU image is also measured.

\begin{figure*}[ht!]
\begin{centering}
\includegraphics[width=0.8\textwidth]{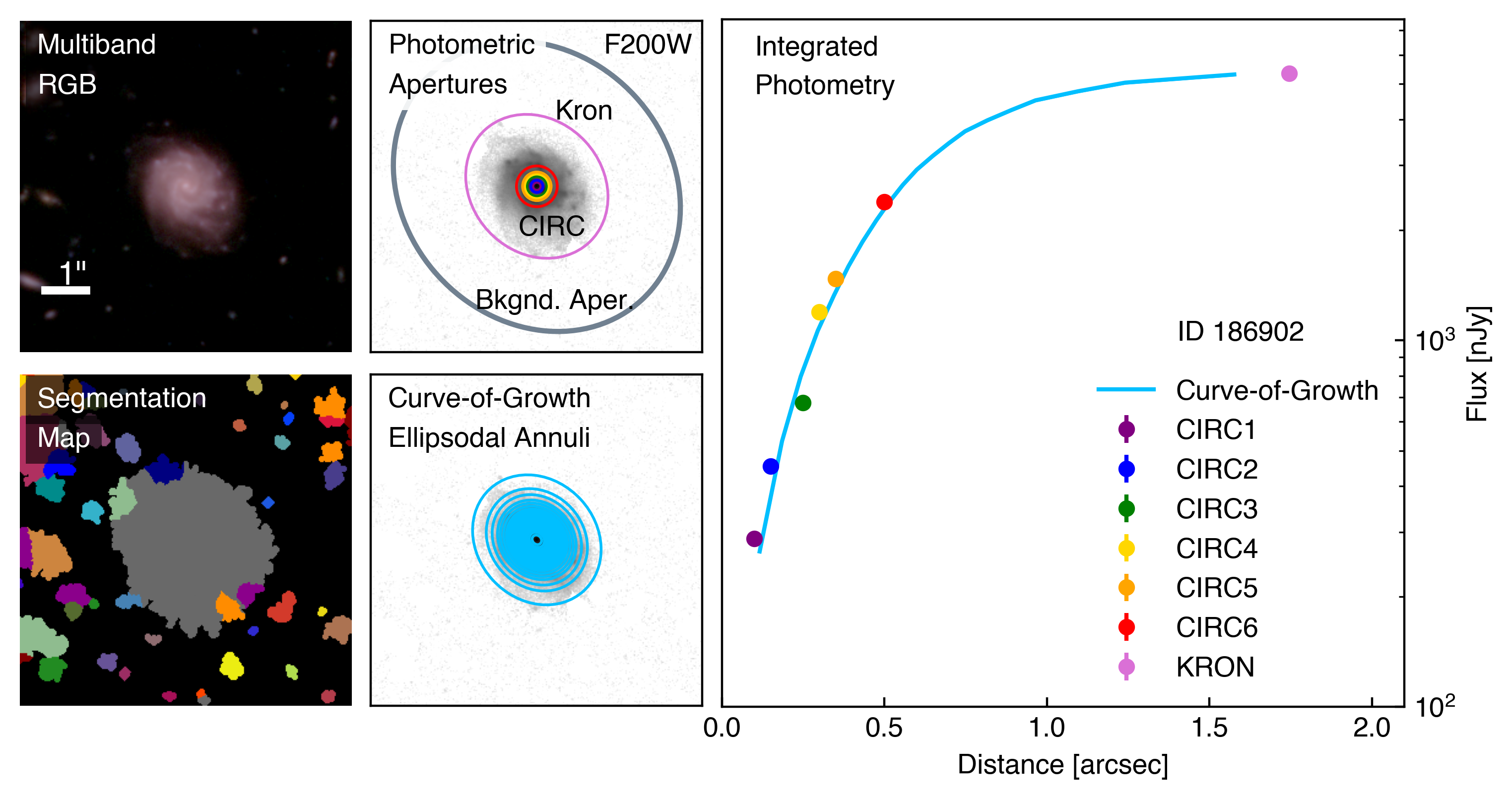}
\caption{Overview of the photometric measurements provided in the JADES
DR5 catalogs. The multi-band DR5 mosaics (upper left panel, shown as RGB) provide the pixel data for measuring the source photometry (see \benspaper{}). After the detection and deblending methods in Section 
\ref{sec:detection} are applied, a segmentation map for the source
and the surrounding region is available (lower left panel; gray segment). The segmentation and the flux images are inputs to the
source property determination methods described in Section \ref{sec:source-catalog}, which calculate the source centroid and
Kron aperture (shown as orchid ellipse overlaid on F200W image, upper middle panel). The various circular apertures allow for the \texttt{CIRC} photometric measures about the centroid (upper middle panel, rainbow color circles). An ellipsoidal annulus aperture at a distance of twice
the Kron aperture size is used to determine the local background, and
used to correct some of the photometric measures (i.e., \texttt{KRON}, \texttt{GROWTH}, and \texttt{CIRC\_BSUB}). The curve-of-growth annuli are placed at locations containing 5\% increments of the \texttt{FLUX\_TOT} flux measured within twice the Kron aperture, and as indicated
here for sources with a flat curve-of-growth the maximum aperture
size \texttt{A\_100} where \texttt{FLUX\_TOT} is reached can fall
within even the Kron aperture. The far right panel compares the
background-subtracted flux measurements for the various circular
apertures (rainbow-colored points), the Kron aperture
photometry (orchid point, circularized distance),
and the curve-of-growth (blue line, circularized distances).
\label{fig:photometric-apertures}}
\end{centering}
\end{figure*}

Figure \ref{fig:photometric-apertures} provides an overview of the
various photometric apertures used to generate the catalog entries
for an example source in the JADES GOODS-S mosaic. For each source
detected in the mosaics, the segmentation map (bottom left)
and pixel data (upper left)
are used to define the object centroid (top center) about which
the Kron aperture (top center, magenta ellipse) and circular
apertures (colored circles) are placed for integrating photometry. 
The elliptical annulus used for the background estimate located
at twice the Kron aperture is also indicated (top center),
and is used for background subtraction for the \texttt{KRON},
\texttt{GROWTH}, and \texttt{CIRC\_BSUB} photometric measures
as discussed above.  The bottom middle panel shows the 
curve-of-growth annuli for this source, and as it first reaches
its total flux \texttt{FLUX\_TOT} at a distance smaller than twice
the Kron aperture size the curve-of-growth apertures are placed
at smaller interior distances.  The background-subtracted photometric
measures \texttt{CIRC\_BSUB} (rainbow-colored points), 
Kron photometry (orchid point; plotted at a circularized distance) 
and the curve-of-growth
(blue line; plotted at a circularized distance) are shown in the far right panel; each are consistent at a given distance from the source.
We note that the curve-of-growth provides an empirical two-dimensional
model for the source, and can be used as a proxy for the source 
integrated surface brightness profile. Measures of the light concentration can be estimated from the ratio of the curve-of-growth
aperture sizes, and different estimates of the total flux can
be constructed from the curve-of-growth based on
how the integrated flux changes with distance.
For a detailed analysis of the surface brightness profiles of JADES DR5 sources, we refer the reader to \citet{carreira2026a}. They use Bayesian methods to infer and catalog S\'ersic profile \citep{sersic_1963, sersic_1968} parameters and associated uncertainties for all GOODS-N and GOODS-S DR5 sources in GOODS-N and GOODS-S in each available JWST/NIRCam wide-band filter. 

\section{Photometric Redshifts}
\label{sec:photometric-redshifts}

Photometric redshifts provide a valuable
derived property for objects in the
JADES photometric surveys. To
generate photometric redshift catalogs
for the JADES DR5 sources we
follow closely the methods described in
\citet{hainline2024a} and \citet{rieke2023a},
and use the \texttt{EAZY} code \citep{brammer2008a}
to provide template-based photometric
redshift estimates.
We use the same template set used by
\citet{hainline2024a} that includes the
original \citet{brammer2008a} templates
augmented with line emission, an additional
dusty
template from the \texttt{EAZY}
suite, a high-equivalent width emission line template
from \citet{erb2010a}, and 
templates derived from the JAGUAR
mock galaxy catalogs \citep{williams2018a}.
These templates are shown in Figure 1 of
\citet{hainline2024a}. At high redshift,
the effects of
intergalactic medium absorption from neutral
hydrogen are included in a manner consistent
with the \citet{madau1995a} model.

In using \texttt{EAZY} to compute
photometric redshifts, we first
compute a set of photometric offsets 
derived by fitting templates to the
photometry of a subset of galaxies selected
to have SNR$\sim20-50$ in F115W and 
semimajor axes less than 5 pixels.
We set an error floor on the photometry of 5\%.
In performing these fits to JWST/NIRCam
\texttt{CIRC3} ($r=0.25$")
photometry alone, we found that the template
fits were consistent with no required
rescaling of the photometry for NIRCam filters
to match the SED templates as a
population. This result then improves
on that found in \citet{hainline2024a}, who
found typical photometric offsets of $\sim$2-5\%
required for NIRCam filters, and may originate
from the numerous subsequent
improvements to the image 
quality and cataloging methods. 
Adopting no offsets for NIRCam,
we then determined the photometric offsets for
the HST photometry by re-fitting the HST and NIRCam data jointly
with \texttt{EAZY}. 
The derived photometric offsets
are reported in Table \ref{tab:photometric-offsets},
which were then validated by fitting a subset 
at fixed, spectroscopically-confirmed redshifts
to confirm that the ratio of the template and
observed population SEDs were consistent with
unity in the HST and NIRCam bands. The validated
photometric offsets were then adopted in determining the photometric
redshifts for all
sources by again refitting using \texttt{EAZY}.
We compute separate photometric redshift catalogs for photometry using \texttt{CIRC1} ($r=0.1$") circular
apertures on the filter mosaics (\texttt{PHOTOZ}, see Appendix \ref{sec:photoz-hdu} below) 
and \texttt{KRON\_CONV}
ellipsoidal apertures on the common-PSF mosaics
(\texttt{PHOTOZ\_KRON}, see Appendix 
\ref{sec:photoz-kron-hdu} below). The \texttt{CIRC1}-based
photometry provides high SNR, which allows for photometric
breaks to identified cleanly and improves drop-out selections.
The \texttt{KRON}-based photometry provides better measures
of the total object flux and uses common-PSF images to avoid
chromatic effects for extended objects, with the goal of
fitting SED models with unbiased colors.

For each source, the $\chi^2$ surface computed
from the differences between the best rescaled 
template and the photometric data is recorded.
The minimum of this curve defines the best
photometric redshift $z_a$. Assuming a flat
redshift prior, 
we compute the photometric redshift probability distribution as $p(z) = \exp[-\chi^{2}(z)/2]$
and then renormalize the distribution such that $\int p(z)dz = 1$. 
This computation allows us to record the probability
that sources lie in excess of certain
redshift thresholds or calculate the marginal
percentiles on the cumulative 
redshift probability distributions. 
The shape of the $\chi^2(z)$ also enables the identification of local minima where possible
low-redshift solutions for high-redshift candidate
objects are permitted, and these alternative, 
potentially low-probability solutions are also
recorded.
We elaborate on which
statistics are provided in the JADES DR5 catalogs
in Appendices \ref{sec:photoz-hdu} and \ref{sec:photoz-kron-hdu}.

\begin{deluxetable}{lcc}[!ht]
\tablecaption{\texttt{EAZY}-derived Photometric Offsets \label{tab:photometric-offsets}}
\tablehead{
\colhead{Instrument} & \colhead{Filter} & \colhead{Offset}
}
\startdata
HST/ACS      &            F435W  &   1.16 \\
HST/ACS      &            F606W  &   1.07 \\
HST/ACS      &            F775W  &   1.07  \\
HST/ACS      &            F814W  &   1.07 \\
HST/ACS      &            F850LP &   0.95 \\
JWST/NIRCam  &            All    &   1.0 \\
\enddata
\tablecomments{The photometric offsets are multiplicative factors to the flux in each filter.
The same 
photometric offsets are applied to photometric redshift determinations for
objects in both GOODS-S and GOODS-N.}
\end{deluxetable}

\begin{figure*}[ht!]
\includegraphics[width=1.0\textwidth]{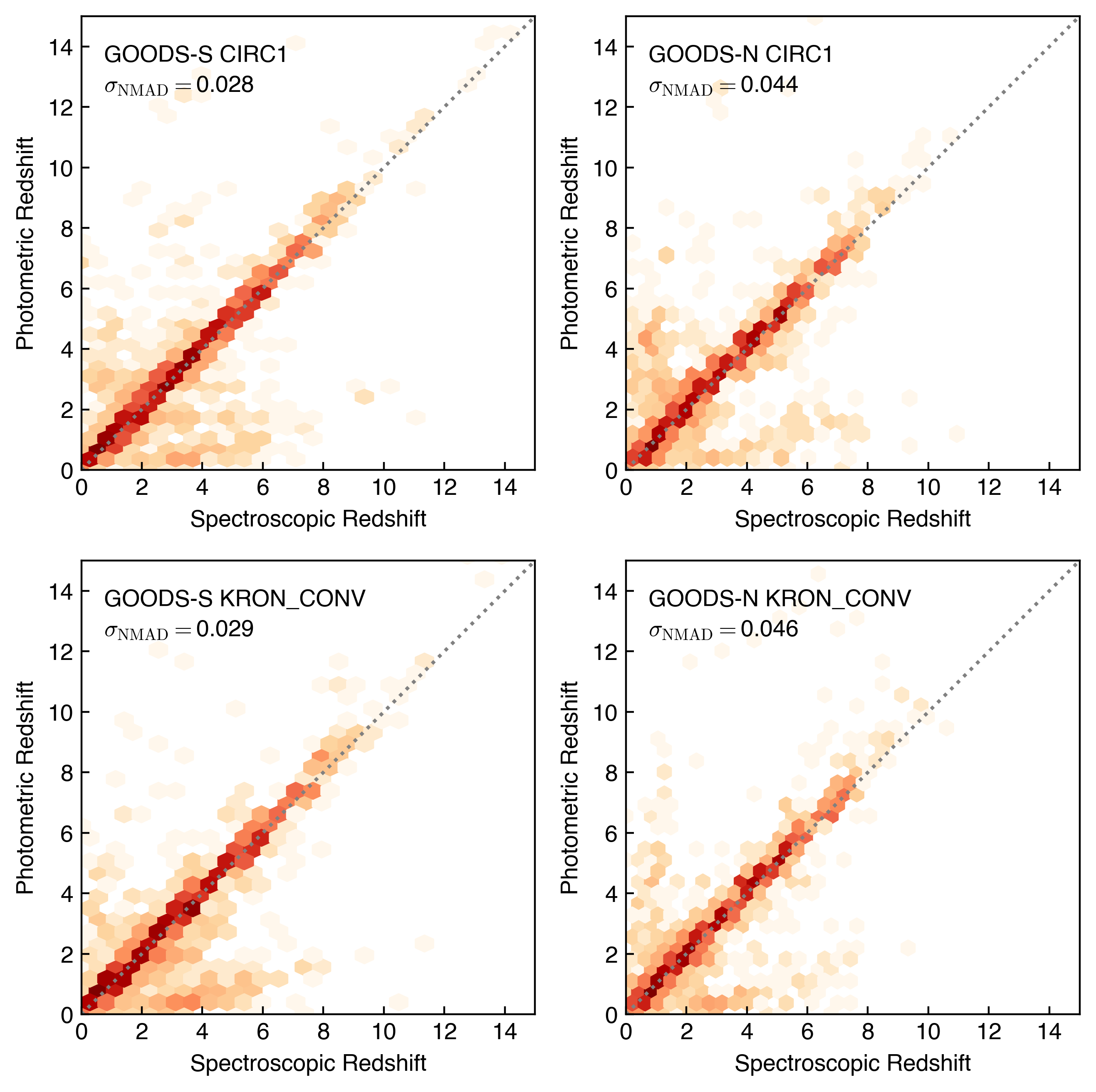}
\caption{Comparison of photometric redshift $z_p$ determined
from the JADES DR5 catalog data using the methods described
in Section \ref{sec:photometric-redshifts} with spectroscopic
redshift $z_s$ for  F444W SNR$>5$ sources. Shown are normalized
two-dimensional histograms of $z_p$ vs. $z_s$ for EAZY \citep{brammer2008a} template-based photometric redshifts using
\texttt{CIRC1} (upper row) or \texttt{KRON\_CONV} photometry (bottom row)
for sources in GOODS-S (left column) and GOODS-N (right column).
The literature references for the spectroscopic redshifts are provided
in Section \ref{sec:photometric-redshifts}. Measures of the fractional normalized mean absolute
deviation (NMAD) $\sigma_{\mathrm{NMAD}}$ in $(1+z)$ between the photometric and spectroscopic redshift
are reported, which are typically 2.8-4.5\%.
\label{fig:photoz}}
\end{figure*}

Figure \ref{fig:photoz} compares the photometric 
redshifts $z_{p}$
and spectroscopic redshifts $z_{s}$ for a subsample of $\approx12,500$ JADES DR5
sources in GOODS-S and GOODS-N. The spectroscopic 
redshifts are taken from a compilation by D. Pusk\'as (priv. comm.).
This compilation includes sources selected from catalogs published
by \citet{danielson2017a},
\citet{urrutia2019a}, \citet{pharo2022a}, \citet{arrabal_haro2023a},
\citet{bacon2023a}, \citet{chisholm2024a}, and \citet{desi2024a},
as well as spectroscopic determinations made from 
the ALMA Spectroscopic Survey in the Hubble Ultra Deep Field
\citep[ASPECS;][]{gonzalez-lopez2019a,boogaard2020a},
the Observing All Phases of Stochastic Star Formation program (OASIS, JWST PID 5997; Looser et al., in prep.),
the NIRSpec Wide GTO Survey \citep{maseda2024a}, 
JWST PID 2198 \citep{barrufet2025a}, JWST PID 4540 \citep{eisenstein2025a}, the JADES Transient Survey \citep[JWST PID 6541;][]{decoursey2025a}, Dark Horse \citep{deugenio2025b}, and FRESCO \citep{oesch2023a} referenced from the literature. These
redshifts are supplemented by the JADES DR4 spectroscopic
catalog \citep{curtis-lake2025a,scholtz2025a} and spectra from prior JADES releases
\citep{bunker2024a,deugenio2025a}. Normalized 
histograms of the locations of F444W SNR$>5$
sources in the $z_{s}$-$z_{p}$ plane are shown for the 
\texttt{CIRC1} (upper row) and \texttt{KRON\_CONV} (bottom row)
photometric redshift determinations for the GOODS-S (left column)
and GOODS-N (right column) subfields within JADES. 
Overall, we find good agreement between the spectroscopic and
photometric redshifts. Following the previous JADES photometric
redshift performance assessment in \citet{rieke2023a}, we 
define the normalized mean absolute deviation (NMAD) scatter
\begin{equation}
\label{eqn:zphot-nmad}
\sigma_{\mathrm{NMAD}} = 1.48 \times \mathrm{median}\left(\left|\frac{\delta z - \mathrm{median}(\delta z)}{1+z_{s}}\right|\right)
\end{equation}
\noindent
where $\delta z\equiv z_s - z_p$. 
We find $\sigma_{\mathrm{NMAD}}=0.028-0.029$ for GOODS-S and
$\sigma_{\mathrm{NMAD}}=0.044-0.045$ for GOODS-N, compared
with the $\sigma_{\mathrm{NMAD}}=0.024$ reported for a smaller
subsample of JADES GOODS-S galaxies in \citet{rieke2023a}.
We note that for our EAZY fits with these templates,
photometric redshift failures occur for both
Balmer/4000\AA{} breaks mimicking Lyman-$\alpha$ breaks and
vice versa, which are apparent as spurs above and below
the $z_p=z_s$ locus. In each panel, the number of catastrophic outliers decreases significantly at $z_s > 6$ where the Lyman-$\alpha$ break is redshifted into the NIRCam wavelength coverage. We provide a more detailed comparison of the spectroscopic and photometric redshifts for high-redshift ($z > 8$) JADES DR5 sources in Hainline et al. (submitted).

\section{Summary and Data Access}
\label{sec:summary}

In this work we have detailed the process
for the generation and cataloging of the
photometric source population in the
JADES Data Release 5 and provided 
detection and deblending images, segmentation
maps, aperture photometry, curve-of-growth, and photometric
redshift catalogs in 35 space-based
filters from James Webb Space Telescope and
Hubble Space Telescope. We briefly summarize
these efforts below.

\begin{itemize}
    \item  Signal-to-noise detection images
    were generated from a stack of long-wavelength
    ($\lambda>2.5\mu$m)
    NIRCam mosaics, supplemented with MIRI or
    HST data where NIRCam coverage was unavailable (Section \ref{sec:detection-image}).
    \item Deblending mosaics were created from
    a stack of red short-wavelength ($\lambda\approx2\mu$m) 
    NIRCam module images, supplemented with detection
    image data where the desired deblending data was unavailable (Section \ref{sec:deblending-image}).
    \item Customized detection and deblending algorithms were
    performed to identify and isolate faint sources in crowded,
    deep extragalactic fields (Section \ref{sec:detection}). The
    source catalogs were manually curated using an interactive 
    web interface based on \texttt{FitsMap} \citep{hausen2022b} (Section \ref{sec:catalog-curation}).
    \item Source properties were determined from the detection
    image data by performing a newly-presented method for
    fast Gaussian regression fits to
    all objects to determine source sizes and apertures for
    Kron photometry (Section \ref{sec:source-catalog}).
    \item Custom model point-spread-functions were generated for the 
    NIRCam images  (Section \ref{sec:nircam-mpsf}) 
    following the method of \citet{ji2024a}, or measured
    empirically for HST mosaics, and then used to create common-PSF
    mosaics (Section \ref{sec:common-psf-mosaics}) and compute aperture corrections.
    \item Sources were annotated with data quality and context
    flags (Section \ref{sec:flagging}) including a new hash encoding method to track the contribution of distinct JWST Programs to data on individual
    sources.
    \item Photometric uncertainty contributions from the sky background
    were computed using a new method that jointly fits the scaling of 
    the RMS uncertainty in apertures of different sizes and uses the 
    source-free inverse variance image as a template for the single-pixel
    uncertainty (Section \ref{sec:uncertainties}).
    \item Forced photometry is performed for 35 space-based filters for $\sim$500,000
    sources in six circular apertures, with and without local background subtraction,
    and on both the intrinsic resolution and common-PSF mosaics (Section \ref{sec:forced-photometry}). Ellipsoidal aperture Kron photometry is also performed for
    every source on both intrinsic resolution and common-PSF mosaics (Section \ref{sec:kron-aperture-photometry}).
    \item For every source, across twenty separate ellipsoidal
    aperture sizes, the photometric
    curve-of-growth is tabulated for every filter (Section \ref{sec:curve-of-growth}).
    These curve-of-growth catalogs allow for users to define their own
    total flux measurements in ellipsoidal apertures for any source in the catalog.
    \item Photometric redshift catalogs are generated using a template-fitting approach
    with the \texttt{EAZY} code \citep{brammer2008a} applied separately to small ($r=0.1$"; \texttt{CIRC1}) circular aperture photometry on intrinsic resolution images and
    Kron aperture photometry on common-PSF images.
\end{itemize}

The JADES DR5 photometric
catalogs will be available along side
the imaging data presented by \benspaper{} at
the Mikulski Archive for Space Telescopes as a High Level Science Product via \dataset[10.17909/8tdj-8n28]{\doi{10.17909/8tdj-8n28}}. The catalogs will also be available through
our interactive FitsMap \citep{hausen2022b}
website at \url{https://jades.idies.jhu.edu}.

\begin{acknowledgments}

BER, BDJ, DJE, CC, JHM, MR, CNAW, and YZ acknowledge support from the NIRCam Science Team contract to the University of Arizona, NAS5-02105. 
Support for JWST programs 3215 and 5015 were provided by NASA through a grant from the Space Telescope Science Institute, which is operated by the Association of Universities for Research in Astronomy, Inc., under NASA contract NAS 5-03127.
ST acknowledges support by the Royal Society Research Grant G125142.
DJE is supported as a Simons Investigator.
S. Alberts acknowledges support from the JWST Mid-Infrared Instrument (MIRI) Science Team Lead, grant 80NSSC18K0555, from NASA Goddard Space Flight Center to the University of Arizona.
S. Arribas acknowledges grant PID2021-127718NB-I00 funded by the Spanish Ministry of Science and Innovation/State Agency of Research (MICIN/AEI/ 10.13039/501100011033).
WMB gratefully acknowledges support from DARK via the DARK fellowship. This work was supported by a research grant (VIL54489) from VILLUM FONDEN.
AJB \& AJC acknowledge funding from the ``FirstGalaxies" Advanced Grant from the European Research Council (ERC) under the European Union’s Horizon 2020 research and innovation programme (Grant agreement No. 789056).
SC acknowledges support by European Union’s HE ERC Starting Grant No. 101040227 - WINGS.
ECL acknowledges support of an STFC Webb Fellowship (ST/W001438/1).
ALD thanks the University of Cambridge Harding Distinguished Postgraduate Scholars Programme and Technology Facilities Council (STFC) Center for Doctoral Training (CDT) in Data intensive science at the University of Cambridge (STFC grant number 2742605) for a PhD studentship.
Funding for RH's research was provided by the Johns Hopkins University, Institute for Data Intensive Engineering and Science (IDIES).
JMH is supported by JWST Program 3215.
RM acknowledges support by the Science and Technology Facilities Council (STFC), by the ERC through Advanced Grant 695671 “QUENCH”, and by the UKRI Frontier Research grant RISEandFALL. RM also acknowledges funding from a research professorship from the Royal Society.
PGP-G acknowledges support from grant PID2022-139567NB-I00 funded by Spanish Ministerio de Ciencia e Innovaci\'on MCIN/AEI/10.13039/501100011033, FEDER, UE.
JAAT acknowledges support from the Simons Foundation.
H\"U acknowledges funding by the European Union (ERC APEX, 101164796). Views and opinions expressed are however those of the authors only and do not necessarily reflect those of the European Union or the European Research Council Executive Agency. Neither the European Union nor the granting authority can be held responsible for them.
LW acknowledges support from the Gavin Boyle Fellowship at the Kavli Institute for Cosmology, Cambridge and from the Kavli Foundation.
The research of CCW is supported by NOIRLab, which is managed by the Association of Universities for Research in Astronomy (AURA) under a cooperative agreement with the National Science Foundation.

This work is based on observations made with the NASA/ESA/CSA James Webb Space Telescope. The data were obtained from the Mikulski Archive for Space Telescopes at the Space Telescope Science Institute, which is operated by the Association of Universities for Research in Astronomy, Inc., under NASA contract NAS 5-03127 for JWST.
JADES DR5 includes NIRCam data from JWST programs 1176, 1180, 1181, 1210, 1264, 1283, 1286, 1287, 1895, 1963, 2079, 2198, 2514, 2516, 2674, 3215, 3577, 3990, 4540, 4762, 5398, 5997, 6434, 6511, and 6541.
The authors acknowledge the teams of programs  1895, 1963, 2079, 2514, 3215, 3577, 3990, 6434, and 6541 for developing their observing program with a zero-exclusive-access period.
The authors acknowledge use of the {\it lux} supercomputer at UC Santa Cruz, funded by NSF MRI grant AST 1828315.
This research made use of \texttt{photutils}, an \texttt{astropy} package for
detection and photometry of astronomical sources (Bradley et al.
2025).
\end{acknowledgments}

\facilities{JWST(NIRCam and MIRI), HST(ACS and WFC3)}

\software{\texttt{astropy} \citep{astropy2013a,astropy2018a,astropy2022a},
          \texttt{EAZY} \citep{brammer2008a},
          \texttt{FitsMap} \citep{hausen2022b},
          \texttt{photutils} \citep{bradley2025a},
          \texttt{Source Extractor} \citep{bertin1996a},
          }

\appendix

\section{JADES Data Release 5 Photometric Catalog Format}
\label{sec:catalog-format}

This appendix details the format of the High Level Science Product versions of the JADES DR5 photometric catalogs. These catalogs are provided as collections of tables in multi-extension Flexible Image Transport System \citep[FITS;][]{wells1981a,pence2010a} files.
Section \ref{sec:filters-hdu} presents the
\texttt{FILTERS} Header Data Unit (HDU) 
that describes the DR5 filter properties
and the corresponding
circular aperture corrections.
The \texttt{FLAG} HDU presented in Section
\ref{sec:flag-hdu} provides information on the
local data quality and context around each
source, proximate sources, and the JWST programs
contributing data to the photometric measurements
for each object.
Section \ref{sec:size-hdu} describes
the \texttt{SIZE} HDU that contains information on
the source centroids, segmentation bounding 
boxes, sizes, position angles, and
structural properties.
In Sections \ref{sec:circ-hdu}-\ref{sec:circ-bsub-conv-hdu} the HDUs containing circular aperture
photometry measures are presented.
The HDUs containing
Kron photometry measured on the unconvolved and
common PSF images are detailed in
Sections \ref{sec:kron-hdu} and \ref{sec:kron-conv-hdu}, respectively.
Section \ref{sec:miri-hdu} presents the
HDU that contains all the MIRI photometry,
measured in both circular and Kron apertures.
The HUDs containing the photometric redshift 
catalogs are detailed in Sections \ref{sec:photoz-hdu} and \ref{sec:photoz-kron-hdu}.
Owing to their data volume,
the curve-of-growth catalogs (e.g., \texttt{GROWTH} and \texttt{GROWTH\_CONV} are provided
as separate multi-HDU FITS files. The
structural contents of these files are presented
in Sections \ref{sec:growth-catalog} and
\ref{sec:growth-conv-catalog}.
 
\subsection{\texttt{FILTERS} Header Data Unit}
\label{sec:filters-hdu}

The \texttt{FILTERS} HDU provide information on the
various band passes used in the JADES DR5 photometric
catalogs. Table \ref{tab:filters-hdu} lists
each field in the HDU, their units and datatype, and 
brief descriptions.
The filters include
HST ACS F435W, F606W, F775W, F814W, and
F850LP filters, HST WFC3 F105W, F125W, F140W, 
and F160W filters, JWST NIRCam F070W, F090W,
F115W, F150W, F162M, F182M, F200W, F210M,
F250M, F277W, F300M, F335M, F356W, F410M,
F430M, F444W, F460M, and F480M, and
JWST MIRI F560W, F770W, F1000W, F1280W, F1500W,
F1800W, F2100W, and F2550W.
Each filter is specified in the table by its
unique \texttt{BAND} name, and its
pivot wavelength (\texttt{WAVE\_PIVOT}), and
the 
minimum and maximum effective wavelengths
(\texttt{WAVE\_MIN} and \texttt{WAVE\_MAX}).
The circular aperture radius (\texttt{R\_CIRC0}) encircling 80\% of the total PSF flux and the aperture
corrections for all the circular aperture
sizes for each band are also provided.

\begin{deluxetable}{lccl}[!ht]
\tablewidth{0pt}
\tablecaption{\texttt{FILTERS} Header Data Unit \label{tab:filters-hdu}}
\tablehead{
\colhead{Field} & \colhead{Units} & \colhead{Datatype} & \colhead{Description}
}
\startdata
\texttt{BAND}  & --- & \texttt{str}   & Telescope bandpass name (e.g., F090W)\\
\texttt{WAVE\_PIVOT}  & $\mu$m & \texttt{float32} & Pivot wavelength of bandpass \\
\texttt{BANDWIDTH}  & $\mu$m & \texttt{float32} & Bandwidth of bandpass \\
\texttt{WAVE\_MIN}  & $\mu$m & \texttt{float32} & Minimum effective wavelength of bandpass \\
\texttt{WAVE\_MAX}  & $\mu$m & \texttt{float32} & Maximum effective wavelength of bandpass \\
\texttt{R\_CIRC0}    & arcsec & \texttt{float32} & Radius encircling 80\% of bandpass PSF energy \\
\texttt{AC\_CIRC1}   & --- & \texttt{float32} & Aperture correction applied to bandpass \texttt{CIRC1} photometry\\
\texttt{AC\_CIRC2}   & --- & \texttt{float32} & Aperture correction applied to bandpass \texttt{CIRC2} photometry\\
\texttt{AC\_CIRC3}   & --- & \texttt{float32} & Aperture correction applied to bandpass \texttt{CIRC3} photometry\\
\texttt{AC\_CIRC4}   & --- & \texttt{float32} & Aperture correction applied to bandpass \texttt{CIRC4} photometry\\
\texttt{AC\_CIRC5}   & --- & \texttt{float32} & Aperture correction applied to bandpass \texttt{CIRC5} photometry\\
\texttt{AC\_CIRC6}   & --- & \texttt{float32} & Aperture correction applied to bandpass \texttt{CIRC6} photometry\\
\enddata
\tablecomments{The aperture corrections are applied as multiplicative factors to circular aperture flux measurements.}
\end{deluxetable}

\newpage

\subsection{\texttt{FLAG} Header Data Unit}
\label{sec:flag-hdu}

Table \ref{sec:flag-hdu} provides the details of the
\texttt{FLAG} HDU, which provides data quality
and context information for each source.
The source \texttt{ID}, right ascension \texttt{RA}
and declination \texttt{DEC} are provided for
each object, and this essential
information is replicated in each subsequent HDU for convenience.
For each \texttt{BAND} the total exposure time (\texttt{\{BAND\}\_TEXP}) and
inverse variance of the sky background (\texttt{\{BAND\}\_WHT}) at the source location is tabulated. The number of bad or
missing pixels within a region defined by the
source bounding box plus a surrounding 10-pixel frame
in each filter is included as \texttt{\{BAND\}\_FLAG}.
The bright neighbor flag (\texttt{FLAG\_BN}; see Section \ref{sec:bright-neighbors}), parent segment ID (\texttt{PARENT\_ID}; see Section \ref{sec:parent-segmentation-identification}), and program ID bithash (\texttt{PID\_HASH}; see Section \ref{sec:bithash} and \benspaper{})
are also provided for each source. 

\begin{deluxetable}{lccl}[!ht]
\tablewidth{0pt}
\tablecaption{\texttt{FLAG} Header Data Unit \label{tab:flag-hdu}}
\tablehead{
\colhead{Field} & \colhead{Units} & \colhead{Datatype} & \colhead{Description}
}
\startdata
\texttt{ID}  & --- & \texttt{int32}   & Unique source identifier \\
\texttt{RA}  & deg & \texttt{float32} & Right ascension \\
\texttt{DEC} & deg & \texttt{float32} & Declination \\
\texttt{\{BAND\}\_FLAG}   & pix & \texttt{int32} & Number of bad or empty pixels within the region of the object's bounding box plus a 10-pixel frame\\
\texttt{\{BAND\}\_TEXP}   & sec & \texttt{float32} & Exposure time in \texttt{\{BAND\}} at source location\\
\texttt{\{BAND\}\_WHT}   & (MJy/sr)$^{-2}$ & \texttt{float32} & Estimated inverse variance of sky background at source location\\
\texttt{FLAG\_BN}    & --- & \texttt{int32} & Value 1 indicates neighbor $\geq2\times$ brighter, value 2 indicates neighbor $\geq10\times$ brighter\\
\texttt{PARENT\_ID}  & --- & \texttt{int32} & Unique source identifier in blended segmentation map\\
\texttt{PID\_HASH}   & --- & \texttt{int32} & Bithash for JWST Program IDs contributing to this object's data\\
\enddata
\end{deluxetable}

\vspace{-1cm}
\subsection{\texttt{SIZE} Header Data Unit}
\label{sec:size-hdu}

The \texttt{SIZE} HDU contains information on the object
size, sky coordinates, image centroid, bounding box, Kron 
aperture properties, and detection image pixel flux distribution.
Table \ref{tab:size-hdu} lists the contents of the \texttt{SIZE} HDU.
Sky positions (right ascension \texttt{RA} and declination \texttt{DEC})
are provided, corresponding to the centroid pixel positions \texttt{X} and \texttt{Y} determined as described in Section \ref{sec:centroiding}. The pixel barycenter of the light distribution is
also provided in the \texttt{XC} and \texttt{YC}. The pixel
bounding box of each source segmentation is recorded in the
\texttt{BBOX\_XMIN}, \texttt{BBOX\_XMAX}, \texttt{BBOX\_YMIN},
and \texttt{BBOX\_YMAX} fields. The semimajor (\texttt{A}) and semiminor  
(\texttt{B}) sizes determined from the Gaussian regression fits,
their position angle \texttt{THETA}, and 
the unscaled Kron radius \texttt{R\_KRON\_U} (see Section \ref{sec:kron-aperture-photometry}) are provided. The full-width at half-maximum \texttt{FWHM} converted from the sizes \texttt{A} and \texttt{B} is
reported. The Gini coefficient of the light distribution within the
segmentation of each source \citep[i.e.,][]{lotz2004a} is noted
in the \texttt{GINI} field.

\begin{deluxetable}{lccl}[!ht]
\tablewidth{0pt}
\tablecaption{\texttt{SIZE} Header Data Unit \label{tab:size-hdu}}
\tablehead{
\colhead{Field} & \colhead{Units} & \colhead{Datatype} & \colhead{Description}
}
\startdata
\texttt{ID}  & --- & \texttt{int32}   & Unique source identifier \\
\texttt{RA}  & deg & \texttt{float32} & Right ascension \\
\texttt{DEC} & deg & \texttt{float32} & Declination \\
\texttt{X}   & pix & \texttt{float32} & Centroid x-pixel position \\
\texttt{Y}   & pix & \texttt{float32} & Centroid y-pixel position \\
\texttt{XC}  & pix & \texttt{float32} & Barycenter x-pixel position \\
\texttt{YC}  & pix & \texttt{float32} & Barycenter y-pixel position \\
\texttt{BBOX\_XMIN}  & pix & \texttt{int32} & Segmentation bounding box minimum x pixel \\
\texttt{BBOX\_XMAX}  & pix & \texttt{int32} & Segmentation bounding box maximum x pixel \\
\texttt{BBOX\_YMIN}  & pix & \texttt{int32} & Segmentation bounding box minimum y pixel \\
\texttt{BBOX\_YMAX}  & pix & \texttt{int32} & Segmentation bounding box maximum y pixel \\
\texttt{A}          & arcsec & \texttt{float32} & 2D Gaussian model standard deviation semimajor axis \\
\texttt{B}          & arcsec & \texttt{float32} & 2D Gaussian model standard deviation semiminor axis \\
\texttt{THETA}      & deg & \texttt{float32} & 2D Gaussian model position angle of semimajor axis relative to image x-axis\\
\texttt{R\_KRON\_U} & --- & \texttt{float32} & Unscaled Kron radius (minimum 1.4)\\
\texttt{FWHM}       & arcsec & \texttt{float32} & Circularized full width at half maximum determined from \texttt{A} and \texttt{B}\\
\texttt{GINI}       & --- & \texttt{float32} & Gini coefficient computed from detection signal image within the source segmentation\\
\enddata
\end{deluxetable}

\newpage
\subsection{\texttt{CIRC} Header Data Unit}
\label{sec:circ-hdu}

Table \ref{tab:circ-hdu} presents the contents of the
\texttt{CIRC} HDU of the photometric catalog. This HDU
records the aperture-corrected
circular aperture fluxes for each source as measured
from the unconvolved image mosaics
about the source centroids and without any additional
background subtraction applied to the image pixel
values. Aperture fluxes in seven apertures
are reported, with \texttt{CIRC1}-\texttt{CIRC6}
corresponding to aperture radii of
$r=[0.1",0.15",0.25",0.3",0.35",0.5"]$ and
\texttt{CIRC0} corresponding to the aperture
radius for each filter that encircles 80\% of
the PSF flux. The fields with the \texttt{\_e} suffixes
indicate uncertainties
measured from the uncertainty model regression
(see Section \ref{sec:uncertainties}) and those with
\texttt{\_ei} suffixes indicate uncertainties
measured from the mosaic \texttt{ERR} image where available.

\begin{deluxetable}{lccp{10cm}}[!ht]
\tablewidth{0pt}
\tablecaption{\texttt{CIRC} Header Data Unit \label{tab:circ-hdu}}
\tablehead{
\colhead{Field} & \colhead{Units} & \colhead{Datatype} & \colhead{Description}
}
\startdata
\texttt{ID}  & --- & \texttt{int32}   & Unique source identifier \\
\texttt{RA}  & deg & \texttt{float32} & Right ascension \\
\texttt{DEC} & deg & \texttt{float32} & Declination \\
\texttt{\{BAND\}\_CIRC0}    & nJy & \texttt{float32} & Aperture-corrected flux for filter \texttt{\{BAND\}} within a circular aperture enclosing 80\% of the PSF energy\\
\texttt{\{BAND\}\_CIRC0\_e} & nJy & \texttt{float32} & Aperture-corrected flux uncertainty for filter \texttt{\{BAND\}} determined from uncertainty model regression\\
\texttt{\{BAND\}\_CIRC0\_ei} & nJy & \texttt{float32} & Aperture-corrected flux uncertainty determined from mosaic \texttt{ERR} image\\
\texttt{\{BAND\}\_CIRC1}    & nJy & \texttt{float32} & Aperture-corrected flux for filter \texttt{\{BAND\}} within $r=0.1$" circular aperture\\
\texttt{\{BAND\}\_CIRC1\_e} & nJy & \texttt{float32} & Aperture-corrected flux uncertainty for filter \texttt{\{BAND\}} determined from uncertainty model regression\\
\texttt{\{BAND\}\_CIRC1\_ei} & nJy & \texttt{float32} & Aperture-corrected flux uncertainty determined from mosaic \texttt{ERR} image\\
\texttt{\{BAND\}\_CIRC2}    & nJy & \texttt{float32} & Aperture-corrected flux for filter \texttt{\{BAND\}} within $r=0.15$" circular aperture\\
\texttt{\{BAND\}\_CIRC2\_e} & nJy & \texttt{float32} & Aperture-corrected flux uncertainty for filter \texttt{\{BAND\}} determined from uncertainty model regression. \\
\texttt{\{BAND\}\_CIRC2\_ei} & nJy & \texttt{float32} & Aperture-corrected flux uncertainty determined from mosaic \texttt{ERR} image\\ 
\texttt{\{BAND\}\_CIRC3}    & nJy & \texttt{float32} & Aperture-corrected flux for filter \texttt{\{BAND\}} within $r=0.25$" circular aperture\\
\texttt{\{BAND\}\_CIRC3\_e} & nJy & \texttt{float32} & Aperture-corrected flux uncertainty for filter \texttt{\{BAND\}} determined from uncertainty model regression. \\
\texttt{\{BAND\}\_CIRC3\_ei} & nJy & \texttt{float32} & Aperture-corrected flux uncertainty determined from mosaic \texttt{ERR} image\\ 
\texttt{\{BAND\}\_CIRC4}    & nJy & \texttt{float32} & Aperture-corrected flux for filter \texttt{\{BAND\}} within $r=0.3$" circular aperture\\
\texttt{\{BAND\}\_CIRC4\_e} & nJy & \texttt{float32} & Aperture-corrected flux uncertainty for filter \texttt{\{BAND\}} determined from uncertainty model regression\\
\texttt{\{BAND\}\_CIRC4\_ei} & nJy & \texttt{float32} & Aperture-corrected flux uncertainty determined from mosaic \texttt{ERR} image\\ 
\texttt{\{BAND\}\_CIRC5}    & nJy & \texttt{float32} & Aperture-corrected flux for filter \texttt{\{BAND\}} within $r=0.35$" circular aperture\\
\texttt{\{BAND\}\_CIRC5\_e} & nJy & \texttt{float32} & Aperture-corrected flux uncertainty for filter \texttt{\{BAND\}} determined from uncertainty model regression\\
\texttt{\{BAND\}\_CIRC5\_ei} & nJy & \texttt{float32} & Aperture-corrected flux uncertainty determined from mosaic \texttt{ERR} image. \\ 
\texttt{\{BAND\}\_CIRC6}    & nJy & \texttt{float32} & Aperture-corrected flux for filter \texttt{\{BAND\}} within $r=0.5$" circular aperture.\\
\texttt{\{BAND\}\_CIRC6\_e} & nJy & \texttt{float32} & Aperture-corrected flux uncertainty for filter \texttt{\{BAND\}} determined from uncertainty model regression\\
\texttt{\{BAND\}\_CIRC6\_ei} & nJy & \texttt{float32} & Aperture-corrected flux uncertainty determined from mosaic \texttt{ERR} image\\ 
\enddata
\tablecomments{The \texttt{\{BAND\}} prefix in the field names indicate for which filter the measurement was performed (e.g., \texttt{F090W\_CIRC1} corresponds to the $r=0.1"$ circular aperture flux for the object in the \texttt{F090W} filter). The fluxes reported in the \texttt{CIRC} HDU reflect direct measurements on the image mosaics about the source centroids with no additional background subtraction applied.}
\end{deluxetable}

\newpage
\subsection{\texttt{CIRC\_BSUB} Header Data Unit}
\label{sec:circ-bsub-hdu}

The \texttt{CIRC\_BSUB} HDU provides the
background-subtracted, aperture-corrected circular aperture
flux for each object measured from the unconvolved image mosaics.
Table \ref{tab:circ-bsub-hdu} lists the fields provided in the HDU.
The definitions of these fields largely follow
those in the \texttt{CIRC} HDU, except the local 
background is measured in an annulus at radii $1.5"<r<1.55"$
and subtracted from the circular aperture flux. The
background removed from each aperture in each
filter is provided in the fields with the \texttt{\_bkg}
suffix.

\begin{deluxetable}{lccp{10cm}}[!ht]
\tablewidth{0pt}
\tablecaption{\texttt{CIRC\_BSUB} Header Data Unit \label{tab:circ-bsub-hdu}}
\tablehead{
\colhead{Field} & \colhead{Units} & \colhead{Datatype} & \colhead{Description}
}
\startdata
\texttt{ID}  & --- & \texttt{int32}   & Unique source identifier \\
\texttt{RA}  & deg & \texttt{float32} & Right ascension \\
\texttt{DEC} & deg & \texttt{float32} & Declination \\
\texttt{\{BAND\}\_CIRC0}    & nJy & \texttt{float32} & Aperture-corrected, background-subtracted flux for filter \texttt{\{BAND\}} within a circular aperture enclosing 80\% of the PSF energy\\
\texttt{\{BAND\}\_CIRC0\_bkg}    & nJy & \texttt{float32} & Background subtraction value for \texttt{\{BAND\}\_CIRC0}\\
\texttt{\{BAND\}\_CIRC0\_e} & nJy & \texttt{float32} & Aperture-corrected flux uncertainty for filter \texttt{\{BAND\}} determined from uncertainty model regression\\
\texttt{\{BAND\}\_CIRC0\_ei} & nJy & \texttt{float32} & Aperture-corrected flux uncertainty determined from mosaic \texttt{ERR} image\\
\texttt{\{BAND\}\_CIRC1}    & nJy & \texttt{float32} & Aperture-corrected, background-subtracted flux for filter \texttt{\{BAND\}} within $r=0.1$" circular aperture\\
\texttt{\{BAND\}\_CIRC1\_bkg}    & nJy & \texttt{float32} & Background subtraction value for \texttt{\{BAND\}\_CIRC1}\\
\texttt{\{BAND\}\_CIRC1\_e} & nJy & \texttt{float32} & Aperture-corrected flux uncertainty for filter \texttt{\{BAND\}} determined from uncertainty model regression\\
\texttt{\{BAND\}\_CIRC1\_ei} & nJy & \texttt{float32} & Aperture-corrected flux uncertainty determined from mosaic \texttt{ERR} image\\
\texttt{\{BAND\}\_CIRC2}    & nJy & \texttt{float32} & Aperture-corrected, background-subtracted flux for filter \texttt{\{BAND\}} within $r=0.15$" circular aperture\\
\texttt{\{BAND\}\_CIRC2\_bkg}    & nJy & \texttt{float32} & Background subtraction value for \texttt{\{BAND\}\_CIRC2}\\
\texttt{\{BAND\}\_CIRC2\_e} & nJy & \texttt{float32} & Aperture-corrected flux uncertainty for filter \texttt{\{BAND\}} determined from uncertainty model regression\\
\texttt{\{BAND\}\_CIRC2\_ei} & nJy & \texttt{float32} & Aperture-corrected flux uncertainty determined from mosaic \texttt{ERR} image\\ 
\texttt{\{BAND\}\_CIRC3}    & nJy & \texttt{float32} & Aperture-corrected, background-subtracted flux for filter \texttt{\{BAND\}} within $r=0.25$" circular aperture\\
\texttt{\{BAND\}\_CIRC3\_bkg}    & nJy & \texttt{float32} & Background subtraction value for \texttt{\{BAND\}\_CIRC3}\\
\texttt{\{BAND\}\_CIRC3\_e} & nJy & \texttt{float32} & Aperture-corrected flux uncertainty for filter \texttt{\{BAND\}} determined from uncertainty model regression\\
\texttt{\{BAND\}\_CIRC3\_ei} & nJy & \texttt{float32} & Aperture-corrected flux uncertainty determined from mosaic \texttt{ERR} image\\ 
\texttt{\{BAND\}\_CIRC4}    & nJy & \texttt{float32} & Aperture-corrected, background-subtracted flux for filter \texttt{\{BAND\}} within $r=0.3$" circular aperture\\
\texttt{\{BAND\}\_CIRC4\_bkg}    & nJy & \texttt{float32} & Background subtraction value for \texttt{\{BAND\}\_CIRC4}\\
\texttt{\{BAND\}\_CIRC4\_e} & nJy & \texttt{float32} & Aperture-corrected flux uncertainty for filter \texttt{\{BAND\}} determined from uncertainty model regression\\
\texttt{\{BAND\}\_CIRC4\_ei} & nJy & \texttt{float32} & Aperture-corrected flux uncertainty determined from mosaic \texttt{ERR} image\\ 
\texttt{\{BAND\}\_CIRC5}    & nJy & \texttt{float32} & Aperture-corrected, background-subtracted flux for filter \texttt{\{BAND\}} within $r=0.35$" circular aperture\\
\texttt{\{BAND\}\_CIRC5\_bkg}    & nJy & \texttt{float32} & Background subtraction value for \texttt{\{BAND\}\_CIRC5}\\
\texttt{\{BAND\}\_CIRC5\_e} & nJy & \texttt{float32} & Aperture-corrected flux uncertainty for filter \texttt{\{BAND\}} determined from uncertainty model regression\\
\texttt{\{BAND\}\_CIRC5\_ei} & nJy & \texttt{float32} & Aperture-corrected flux uncertainty determined from mosaic \texttt{ERR} image\\ 
\texttt{\{BAND\}\_CIRC6}    & nJy & \texttt{float32} & Aperture-corrected, background-subtracted flux for filter \texttt{\{BAND\}} within $r=0.5$" circular aperture\\
\texttt{\{BAND\}\_CIRC6\_bkg}    & nJy & \texttt{float32} & Background subtraction value for \texttt{\{BAND\}\_CIRC6}\\
\texttt{\{BAND\}\_CIRC6\_e} & nJy & \texttt{float32} & Aperture-corrected flux uncertainty for filter \texttt{\{BAND\}} determined from uncertainty model regression\\
\texttt{\{BAND\}\_CIRC6\_ei} & nJy & \texttt{float32} & Aperture-corrected flux uncertainty determined from mosaic \texttt{ERR} image\\ 
\enddata
\tablecomments{The \texttt{\{BAND\}} prefix in the field names indicate for which filter the measurement was performed (e.g., \texttt{F090W\_CIRC1} corresponds to the $r=0.1"$ circular aperture flux for the object in the \texttt{F090W} filter). The fluxes reported in the \texttt{CIRC\_BSUB} HDU reflect measurements on the image mosaics about the source centroids, after a local background estimated from an annulus at radii $1.5"<r<1.55"$ is subtracted from the aperture fluxes.
The subtracted background values are provided in the fields labeled
with the \texttt{\_bkg} suffix.}
\end{deluxetable}

\newpage
\subsection{\texttt{CIRC\_CONV} Header Data Unit}
\label{sec:circ-conv-hdu}

The aperture-corrected circular aperture flux and uncertainties
measured from the common PSF images in each filter 
are provided in the \texttt{CIRC\_CONV} HDU. Table \ref{tab:circ-conv-hdu}
lists the fields in this HDU, which are defined the
same manner as for the \texttt{CIRC} HDU. The
aperture corrections for the common PSF images blueward of F444W
are determined from the target F444W mPSF.

\begin{deluxetable}{lccp{10cm}}[!ht]
\tablewidth{0pt}
\tablecaption{\texttt{CIRC\_CONV} Header Data Unit \label{tab:circ-conv-hdu}}
\tablehead{
\colhead{Field} & \colhead{Units} & \colhead{Datatype} & \colhead{Description}
}
\startdata
\texttt{ID}  & --- & \texttt{int32}   & Unique source identifier \\
\texttt{RA}  & deg & \texttt{float32} & Right ascension \\
\texttt{DEC} & deg & \texttt{float32} & Declination \\
\texttt{\{BAND\}\_CIRC0}    & nJy & \texttt{float32} & Aperture-corrected flux for filter \texttt{\{BAND\}} common-PSF image within a circular aperture enclosing 80\% of the PSF energy\\
\texttt{\{BAND\}\_CIRC0\_e} & nJy & \texttt{float32} & Aperture-corrected flux uncertainty for filter \texttt{\{BAND\}} determined from uncertainty model regression\\
\texttt{\{BAND\}\_CIRC0\_ei} & nJy & \texttt{float32} & Aperture-corrected flux uncertainty determined from common-PSF \texttt{ERR} image\\
\texttt{\{BAND\}\_CIRC1}    & nJy & \texttt{float32} & Aperture-corrected flux for filter \texttt{\{BAND\}} common-PSF image  within $r=0.1$" circular aperture\\
\texttt{\{BAND\}\_CIRC1\_e} & nJy & \texttt{float32} & Aperture-corrected flux uncertainty for filter \texttt{\{BAND\}} determined from uncertainty model regression\\
\texttt{\{BAND\}\_CIRC1\_ei} & nJy & \texttt{float32} & Aperture-corrected flux uncertainty determined from common-PSF \texttt{ERR} image\\
\texttt{\{BAND\}\_CIRC2}    & nJy & \texttt{float32} & Aperture-corrected flux for filter \texttt{\{BAND\}} common-PSF image  within $r=0.15$" circular aperture\\
\texttt{\{BAND\}\_CIRC2\_e} & nJy & \texttt{float32} & Aperture-corrected flux uncertainty for filter \texttt{\{BAND\}} determined from uncertainty model regression. \\
\texttt{\{BAND\}\_CIRC2\_ei} & nJy & \texttt{float32} & Aperture-corrected flux uncertainty determined from mosaic \texttt{ERR} image\\ 
\texttt{\{BAND\}\_CIRC3}    & nJy & \texttt{float32} & Aperture-corrected flux for filter \texttt{\{BAND\}} common-PSF image  within $r=0.25$" circular aperture\\
\texttt{\{BAND\}\_CIRC3\_e} & nJy & \texttt{float32} & Aperture-corrected flux uncertainty for filter \texttt{\{BAND\}} determined from uncertainty model regression. \\
\texttt{\{BAND\}\_CIRC3\_ei} & nJy & \texttt{float32} & Aperture-corrected flux uncertainty determined from common-PSF \texttt{ERR} image\\ 
\texttt{\{BAND\}\_CIRC4}    & nJy & \texttt{float32} & Aperture-corrected flux for filter \texttt{\{BAND\}} common-PSF image  within $r=0.3$" circular aperture\\
\texttt{\{BAND\}\_CIRC4\_e} & nJy & \texttt{float32} & Aperture-corrected flux uncertainty for filter \texttt{\{BAND\}} determined from uncertainty model regression\\
\texttt{\{BAND\}\_CIRC4\_ei} & nJy & \texttt{float32} & Aperture-corrected flux uncertainty determined from mosaic \texttt{ERR} image\\ 
\texttt{\{BAND\}\_CIRC5}    & nJy & \texttt{float32} & Aperture-corrected flux for filter \texttt{\{BAND\}} common-PSF image  within $r=0.35$" circular aperture\\
\texttt{\{BAND\}\_CIRC5\_e} & nJy & \texttt{float32} & Aperture-corrected flux uncertainty for filter \texttt{\{BAND\}} determined from uncertainty model regression\\
\texttt{\{BAND\}\_CIRC5\_ei} & nJy & \texttt{float32} & Aperture-corrected flux uncertainty determined from common-PSF \texttt{ERR} image. \\ 
\texttt{\{BAND\}\_CIRC6}    & nJy & \texttt{float32} & Aperture-corrected flux for filter \texttt{\{BAND\}} common-PSF image  within $r=0.5$" circular aperture.\\
\texttt{\{BAND\}\_CIRC6\_e} & nJy & \texttt{float32} & Aperture-corrected flux uncertainty for filter \texttt{\{BAND\}} determined from uncertainty model regression\\
\texttt{\{BAND\}\_CIRC6\_ei} & nJy & \texttt{float32} & Aperture-corrected flux uncertainty determined from common-PSF \texttt{ERR} image\\ 
\enddata
\tablecomments{The fields in the \texttt{CIRC\_CONV} HDU are
defined in the same manner as for the \texttt{CIRC} HDU (see Table \ref{tab:circ-hdu}, but are measured from common PSF mosaics for each filter.}
\end{deluxetable}

\newpage
\subsection{\texttt{CIRC\_BSUB\_CONV} Header Data Unit}
\label{sec:circ-bsub-conv-hdu}

Table \ref{tab:circ-bsub-conv-hdu} lists the contents of the
\texttt{CIRC\_BSUB\_CONV} HDU, which provides the
aperture-corrected circular aperture fluxes and uncertainties
measured on the common PSF images after a local background
subtraction is applied. As with the values in the \texttt{CIRC\_CONV}
HDU, the background subtraction applied to the flux fields in this HDU 
are measured in an annulus extending over radii $1.5"<r<1.55"$ and
the actual background corrections applied to each object in each
filter are recorded in the fields with the \texttt{\_bkg} suffix.
The aperture corrections are computed from the target F444W mPSF.

\begin{deluxetable}{lccp{10cm}}[!ht]
\tablewidth{0pt}
\tablecaption{\texttt{CIRC\_BSUB\_CONV} Header Data Unit \label{tab:circ-bsub-conv-hdu}}
\tablehead{
\colhead{Field} & \colhead{Units} & \colhead{Datatype} & \colhead{Description}
}
\startdata
\texttt{ID}  & --- & \texttt{int32}   & Unique source identifier \\
\texttt{RA}  & deg & \texttt{float32} & Right ascension \\
\texttt{DEC} & deg & \texttt{float32} & Declination \\
\texttt{\{BAND\}\_CIRC0}    & nJy & \texttt{float32} & Aperture-corrected, background-subtracted flux for filter \texttt{\{BAND\}} common-PSF image within a circular aperture enclosing 80\% of the PSF energy\\
\texttt{\{BAND\}\_CIRC0\_bkg}    & nJy & \texttt{float32} & Common PSF image background subtraction value for \texttt{\{BAND\}\_CIRC0}\\
\texttt{\{BAND\}\_CIRC0\_e} & nJy & \texttt{float32} & Aperture-corrected flux uncertainty for filter \texttt{\{BAND\}} determined from uncertainty model regression applied to common-PSF image \\
\texttt{\{BAND\}\_CIRC0\_ei} & nJy & \texttt{float32} & Aperture-corrected flux uncertainty determined from common-PSF \texttt{ERR} image\\
\texttt{\{BAND\}\_CIRC1}    & nJy & \texttt{float32} & Aperture-corrected, background-subtracted flux for filter \texttt{\{BAND\}} common-PSF image within $r=0.1$" circular aperture\\
\texttt{\{BAND\}\_CIRC1\_bkg}    & nJy & \texttt{float32} & Common PSF image background subtraction value for \texttt{\{BAND\}\_CIRC1}\\
\texttt{\{BAND\}\_CIRC1\_e} & nJy & \texttt{float32} & Aperture-corrected flux uncertainty for filter \texttt{\{BAND\}} determined from uncertainty model regression applied to common-PSF image \\
\texttt{\{BAND\}\_CIRC1\_ei} & nJy & \texttt{float32} & Aperture-corrected flux uncertainty determined from common-PSF \texttt{ERR} image\\
\texttt{\{BAND\}\_CIRC2}    & nJy & \texttt{float32} & Aperture-corrected, background-subtracted flux for filter \texttt{\{BAND\}} common-PSF image within $r=0.15$" circular aperture\\
\texttt{\{BAND\}\_CIRC2\_bkg}    & nJy & \texttt{float32} & Common PSF image background subtraction value for \texttt{\{BAND\}\_CIRC2}\\
\texttt{\{BAND\}\_CIRC2\_e} & nJy & \texttt{float32} & Aperture-corrected flux uncertainty for filter \texttt{\{BAND\}} determined from uncertainty model regression applied to common-PSF image \\
\texttt{\{BAND\}\_CIRC2\_ei} & nJy & \texttt{float32} & Aperture-corrected flux uncertainty determined from common-PSF \texttt{ERR} image\\ 
\texttt{\{BAND\}\_CIRC3}    & nJy & \texttt{float32} & Aperture-corrected, background-subtracted flux for filter \texttt{\{BAND\}} common-PSF image within $r=0.25$" circular aperture\\
\texttt{\{BAND\}\_CIRC3\_bkg}    & nJy & \texttt{float32} & Common PSF image background subtraction value for \texttt{\{BAND\}\_CIRC3}\\
\texttt{\{BAND\}\_CIRC3\_e} & nJy & \texttt{float32} & Aperture-corrected flux uncertainty for filter \texttt{\{BAND\}} determined from uncertainty model regression applied to common-PSF image \\
\texttt{\{BAND\}\_CIRC3\_ei} & nJy & \texttt{float32} & Aperture-corrected flux uncertainty determined from common-PSF \texttt{ERR} image\\ 
\texttt{\{BAND\}\_CIRC4}    & nJy & \texttt{float32} & Aperture-corrected, background-subtracted flux for filter \texttt{\{BAND\}} common-PSF image within $r=0.3$" circular aperture\\
\texttt{\{BAND\}\_CIRC4\_bkg}    & nJy & \texttt{float32} & Common PSF image background subtraction value for \texttt{\{BAND\}\_CIRC4}\\
\texttt{\{BAND\}\_CIRC4\_e} & nJy & \texttt{float32} & Aperture-corrected flux uncertainty for filter \texttt{\{BAND\}} determined from uncertainty model regression applied to common-PSF image \\
\texttt{\{BAND\}\_CIRC4\_ei} & nJy & \texttt{float32} & Aperture-corrected flux uncertainty determined from common-PSF \texttt{ERR} image\\ 
\texttt{\{BAND\}\_CIRC5}    & nJy & \texttt{float32} & Aperture-corrected, background-subtracted flux for filter \texttt{\{BAND\}} common-PSF image within $r=0.35$" circular aperture\\
\texttt{\{BAND\}\_CIRC5\_bkg}    & nJy & \texttt{float32} & Common PSF image background subtraction value for \texttt{\{BAND\}\_CIRC5}\\
\texttt{\{BAND\}\_CIRC5\_e} & nJy & \texttt{float32} & Aperture-corrected flux uncertainty for filter \texttt{\{BAND\}} determined from uncertainty model regression applied to common-PSF image \\
\texttt{\{BAND\}\_CIRC5\_ei} & nJy & \texttt{float32} & Aperture-corrected flux uncertainty determined from common-PSF \texttt{ERR} image\\ 
\texttt{\{BAND\}\_CIRC6}    & nJy & \texttt{float32} & Aperture-corrected, background-subtracted flux for filter \texttt{\{BAND\}} common-PSF image within $r=0.5$" circular aperture\\
\texttt{\{BAND\}\_CIRC6\_bkg}    & nJy & \texttt{float32} & Common PSF image background subtraction value for \texttt{\{BAND\}\_CIRC6}\\
\texttt{\{BAND\}\_CIRC6\_e} & nJy & \texttt{float32} & Aperture-corrected flux uncertainty for filter \texttt{\{BAND\}} determined from uncertainty model regression applied to common-PSF image \\
\texttt{\{BAND\}\_CIRC6\_ei} & nJy & \texttt{float32} & Aperture-corrected flux uncertainty determined from common-PSF \texttt{ERR} image\\ 
\enddata
\tablecomments{The fields in this table follow the same definitions as the properties listed in Table \ref{tab:circ-conv-hdu} and are measured on the common PSF images, but the flux values reported here include a local background subtraction determined from an annulus extending over radii $1.5"<r<1.55"$. The aperture corrections in this HDU are determined from the target F444W mPSF.}
\end{deluxetable}

\newpage
\subsection{\texttt{KRON} Header Data Unit}
\label{sec:kron-hdu}

The \texttt{KRON} HDU contains information about the ellipsoidal
aperture \citet{kron1980a} photometry for each source, measured
as described in Section \ref{sec:kron-aperture-photometry}.
Table \ref{tab:kron-hdu} lists the quantities recorded in the HDU.
Object properties are provided for a Kron parameter $k=2.5$ (\texttt{\_KRON} suffix)
and $k=1.4$ (\texttt{\_KRON\_S} suffix).
The semimajor and semiminor axes of the ellipsoidal
Kron aperture are provided 
along with the position angle measured counter-clockwise relative to the image $x$-axis. The
Kron flux in each filter have a local background subtraction applied,
and the background (quantities with \texttt{\_bkg} suffix) 
is measured in an 4-pixel-wide ellipsoidal aperture located
at twice the $k=2.5$ Kron aperture with the same axis ratio.
Uncertainties measured from the regression model (\texttt{\_e} suffix) 
and from the \texttt{ERR} mosaic  (\texttt{\_ei} suffix) are also provided.

\begin{deluxetable}{lccp{10cm}}[!ht]
\tablewidth{0pt}
\tablecaption{\texttt{KRON} Header Data Unit \label{tab:kron-hdu}}
\tablehead{
\colhead{Field} & \colhead{Units} & \colhead{Datatype} & \colhead{Description}
}
\startdata
\texttt{ID}  & --- & \texttt{int32}   & Unique source identifier \\
\texttt{RA}  & deg & \texttt{float32} & Right ascension \\
\texttt{DEC} & deg & \texttt{float32} & Declination \\
\texttt{A\_KRON} & arcsec & \texttt{float32} & Semimajor axis size of aperture defined with Kron parameter $k=2.5$\\
\texttt{B\_KRON} & arcsec & \texttt{float32} & Semiminor axis size of aperture defined with Kron parameter $k=2.5$\\
\texttt{THETA\_KRON} & deg & \texttt{float32} & Position angle of Kron aperture semimajor axis relative to image $x$-axis, for Kron parameter $k=2.5$\\
\texttt{A\_KRON\_S} & arcsec & \texttt{float32} & Semimajor axis size of aperture defined with Kron parameter $k=1.4$\\
\texttt{B\_KRON\_S} & arcsec & \texttt{float32} & Semiminor axis size of aperture defined with Kron parameter $k=1.4$\\
\texttt{THETA\_KRON\_S} & deg & \texttt{float32} & Position angle of Kron aperture semimajor axis relative to image x axis, for Kron parameter $k=1.4$\\
\texttt{\{BAND\}\_KRON}    & nJy & \texttt{float32} & Aperture-corrected, background-subtracted flux for filter \texttt{\{BAND\}} within the aperture with Kron parameter $k=2.5$\\
\texttt{\{BAND\}\_KRON\_bkg}    & nJy & \texttt{float32} & Background subtraction value for \texttt{\{BAND\}\_KRON} with Kron parameter $k=2.5$\\
\texttt{\{BAND\}\_KRON\_e} & nJy & \texttt{float32} & Aperture-corrected flux uncertainty for filter \texttt{\{BAND\}} determined from uncertainty model regression using Kron parameter $k=2.5$\\
\texttt{\{BAND\}\_KRON\_ei} & nJy & \texttt{float32} & Aperture-corrected flux uncertainty determined from mosaic \texttt{ERR} image for aperture size with Kron parameter $k=2.5$\\
\texttt{\{BAND\}\_KRON\_S}    & nJy & \texttt{float32} & Aperture-corrected, background-subtracted flux for filter \texttt{\{BAND\}} within the aperture with Kron parameter $k=1.4$\\
\texttt{\{BAND\}\_KRON\_S\_bkg}    & nJy & \texttt{float32} & Background subtraction value for \texttt{\{BAND\}\_KRON} with Kron parameter $k=1.4$\\
\texttt{\{BAND\}\_KRON\_S\_e} & nJy & \texttt{float32} & Aperture-corrected flux uncertainty for filter \texttt{\{BAND\}} determined from uncertainty model regression using Kron parameter $k=1.4$\\
\texttt{\{BAND\}\_KRON\_S\_ei} & nJy & \texttt{float32} & Aperture-corrected flux uncertainty determined from mosaic \texttt{ERR} image using Kron parameter $k=1.4$\\
\enddata
\tablecomments{The fields with a \texttt{\_S} suffix use a Kron parameter
$k=1.4$, while those without the suffix use $k=2.5$. See Section \ref{sec:kron-aperture-photometry} and Appendix \ref{sec:kron-hdu} for more details.}
\end{deluxetable}

\newpage
\subsection{\texttt{KRON\_CONV} Header Data Unit}
\label{sec:kron-conv-hdu}

Table \ref{tab:kron-conv-hdu} provides the contents of the
\texttt{KRON\_CONV} HDU that reports the ellipsoidal 
\citet{kron1980a} photometry (see Section \ref{sec:kron-aperture-photometry})
for each object performed on the
common PSF images (see Section \ref{sec:common-psf-mosaics}).
The field definitions follow those in Section \ref{sec:kron-hdu}
and Table \ref{tab:kron-hdu}, including the Kron parameter value designations (e.g., \texttt{\_S} suffix for quantities using $k=1.4$ and $k=2.5$ otherwise) and background subtraction method, but the reported values have been measured
on the common PSF mosaics.

\begin{deluxetable}{lccp{10cm}}[!ht]
\tablewidth{0pt}
\tablecaption{\texttt{KRON\_CONV} Header Data Unit \label{tab:kron-conv-hdu}}
\tablehead{
\colhead{Field} & \colhead{Units} & \colhead{Datatype} & \colhead{Description}
}
\startdata
\texttt{ID}  & --- & \texttt{int32}   & Unique source identifier \\
\texttt{RA}  & deg & \texttt{float32} & Right ascension \\
\texttt{DEC} & deg & \texttt{float32} & Declination \\
\texttt{A\_KRON} & arcsec & \texttt{float32} & Semimajor axis size of aperture defined with Kron parameter $k=2.5$\\
\texttt{B\_KRON} & arcsec & \texttt{float32} & Semiminor axis size of aperture defined with Kron parameter $k=2.5$\\
\texttt{THETA\_KRON} & deg & \texttt{float32} & Position angle of Kron aperture semimajor axis relative to image x axis, for Kron parameter $k=2.5$\\
\texttt{A\_KRON\_S} & arcsec & \texttt{float32} & Semimajor axis size of aperture defined with Kron parameter $k=1.4$\\
\texttt{B\_KRON\_S} & arcsec & \texttt{float32} & Semiminor axis size of aperture defined with Kron parameter $k=1.4$\\
\texttt{THETA\_KRON\_S} & deg & \texttt{float32} & Position angle of Kron aperture semimajor axis relative to image x axis, for Kron parameter $k=1.4$\\
\texttt{\{BAND\}\_KRON}    & nJy & \texttt{float32} & Aperture-corrected, background-subtracted flux for filter \texttt{\{BAND\}} common-PSF image within the aperture with Kron parameter $k=2.5$\\
\texttt{\{BAND\}\_KRON\_bkg}    & nJy & \texttt{float32} & Common PSF image background subtraction value for \texttt{\{BAND\}\_KRON} with Kron parameter $k=2.5$\\
\texttt{\{BAND\}\_KRON\_e} & nJy & \texttt{float32} & Aperture-corrected flux uncertainty for filter \texttt{\{BAND\}} determined from uncertainty model regression applied to common-PSF image using Kron parameter $k=2.5$\\
\texttt{\{BAND\}\_KRON\_ei} & nJy & \texttt{float32} & Aperture-corrected flux uncertainty determined from  common-PSF  \texttt{ERR} image using Kron parameter $k=2.5$\\
\texttt{\{BAND\}\_KRON\_S}    & nJy & \texttt{float32} & Aperture-corrected, background-subtracted flux for filter \texttt{\{BAND\}} common-PSF image within the aperture with Kron parameter $k=1.4$\\
\texttt{\{BAND\}\_KRON\_S\_bkg}    & nJy & \texttt{float32} & Common PSF image background subtraction value for \texttt{\{BAND\}\_KRON} with Kron parameter $k=1.4$\\
\texttt{\{BAND\}\_KRON\_S\_e} & nJy & \texttt{float32} & Aperture-corrected flux uncertainty for filter \texttt{\{BAND\}} determined from uncertainty model regression applied to common-PSF image using Kron parameter $k=1.4$\\
\texttt{\{BAND\}\_KRON\_S\_ei} & nJy & \texttt{float32} & Aperture-corrected flux uncertainty determined from  common-PSF  \texttt{ERR} image using Kron parameter $k=1.4$\\
\enddata
\tablecomments{The fields with a \texttt{\_S} suffix use a Kron parameter
$k=1.4$, while those without the suffix use $k=2.5$. See Section \ref{sec:kron-aperture-photometry} and Appendix \ref{sec:kron-conv-hdu} for more details.}
\end{deluxetable}

\newpage
\subsection{\texttt{MIRI} Header Data Unit}
\label{sec:miri-hdu}

The \texttt{MIRI} HDU contains the photometric measurements performed on the 
JWST/MIRI mosaic images. All these measurements are performed on 
unconvolved mosaics. The HDU contains fields that record circular aperture photometry
with (\texttt{\_BSUB} suffix) and without (no additional suffix) background
subtraction applied. The ellipsoidal \citet{kron1980a} photometry for the
MIRI filters is included, using Kron parameters of $k=1.4$ (\texttt{\_S} suffix)
and $k=2.5$ (no additional suffix). The Kron-related parameters are defined as
in the \texttt{KRON} HDU detailed above in Section \ref{sec:kron-hdu}.

\begin{deluxetable*}{lccp{10cm}}[!ht]
\centerwidetable
\tablewidth{0pt}
\tablecaption{\texttt{MIRI} Header Data Unit \label{tab:miri-hdu}}
\tablehead{
\colhead{Field} & \colhead{Units} & \colhead{Datatype} & \colhead{Description}
}
\startdata
\texttt{ID}  & --- & \texttt{int32}   & Unique source identifier \\
\texttt{RA}  & deg & \texttt{float32} & Right ascension \\
\texttt{DEC} & deg & \texttt{float32} & Declination \\
\texttt{\{BAND\}\_CIRC\{X\}}    & nJy & \texttt{float32} & Aperture-corrected flux for filter \texttt{\{BAND\}} within a circular aperture \texttt{CIRC\{X\}}, with \texttt{\{X\}}$\in[0-8]$\\
\texttt{\{BAND\}\_CIRC\{X\}\_bkg}    & nJy & \texttt{float32} & Background subtraction value for \texttt{\{BAND\}\_CIRC\{X\}}\\
\texttt{\{BAND\}\_CIRC\{X\}\_e} & nJy & \texttt{float32} & Aperture-corrected flux uncertainty for filter \texttt{\{BAND\}} determined from uncertainty model regression\\
\texttt{\{BAND\}\_CIRC\{X\}\_ei} & nJy & \texttt{float32} & Aperture-corrected flux uncertainty determined from mosaic \texttt{ERR} image\\
\texttt{\{BAND\}\_CIRC\{X\}\_BSUB}    & nJy & \texttt{float32} & Aperture-corr., background-sub. flux for filter \texttt{\{BAND\}} within a circular aperture \texttt{CIRC\{X\}}, with \texttt{\{X\}}$\in[0-8]$\\
\texttt{\{BAND\}\_CIRC\{X\}\_bkg\_BSUB}    & nJy & \texttt{float32} & Background subtraction value for \texttt{\{BAND\}\_CIRCX\_BSUB}\\
\texttt{\{BAND\}\_CIRC\{X\}\_e\_BSUB} & nJy & \texttt{float32} & Aperture-corrected flux uncertainty for filter \texttt{\{BAND\}} determined from uncertainty model regression\\
\texttt{\{BAND\}\_CIRC\{X\}\_ei\_BSUB} & nJy & \texttt{float32} & Aperture-corrected flux uncertainty determined from mosaic \texttt{ERR} image\\
\texttt{A\_KRON} & arcsec & \texttt{float32} & Semimajor axis size of aperture defined with Kron parameter $k=2.5$\\
\texttt{B\_KRON} & arcsec & \texttt{float32} & Semiminor axis size of aperture defined with Kron parameter $k=2.5$\\
\texttt{THETA\_KRON} & deg & \texttt{float32} & Position angle of Kron aperture semimajor axis relative to image $x$-axis, for Kron parameter $k=2.5$\\
\texttt{A\_KRON\_S} & arcsec & \texttt{float32} & Semimajor axis size of aperture defined with Kron parameter $k=1.4$\\
\texttt{B\_KRON\_S} & arcsec & \texttt{float32} & Semiminor axis size of aperture defined with Kron parameter $k=1.4$\\
\texttt{THETA\_KRON\_S} & deg & \texttt{float32} & Position angle of Kron aperture semimajor axis relative to image x axis, for Kron parameter $k=1.4$\\
\texttt{\{BAND\}\_KRON}    & nJy & \texttt{float32} & Aperture-corrected, background-subtracted flux for filter \texttt{\{BAND\}} within the aperture with Kron parameter $k=2.5$\\
\texttt{\{BAND\}\_KRON\_bkg}    & nJy & \texttt{float32} & Background subtraction value for \texttt{\{BAND\}\_KRON} with Kron parameter $k=2.5$\\
\texttt{\{BAND\}\_KRON\_e} & nJy & \texttt{float32} & Aperture-corrected flux uncertainty for filter \texttt{\{BAND\}} determined from uncertainty model regression using Kron parameter $k=2.5$\\
\texttt{\{BAND\}\_KRON\_ei} & nJy & \texttt{float32} & Aperture-corrected flux uncertainty determined from mosaic \texttt{ERR} image for aperture size with Kron parameter $k=2.5$\\
\texttt{\{BAND\}\_KRON\_S}    & nJy & \texttt{float32} & Aperture-corrected, background-subtracted flux for filter \texttt{\{BAND\}} within the aperture with Kron parameter $k=1.4$\\
\texttt{\{BAND\}\_KRON\_S\_bkg}    & nJy & \texttt{float32} & Background subtraction value for \texttt{\{BAND\}\_KRON} with Kron parameter $k=1.4$\\
\texttt{\{BAND\}\_KRON\_S\_e} & nJy & \texttt{float32} & Aperture-corrected flux uncertainty for filter \texttt{\{BAND\}} determined from uncertainty model regression using Kron parameter $k=1.4$\\
\texttt{\{BAND\}\_KRON\_S\_ei} & nJy & \texttt{float32} & Aperture-corrected flux uncertainty determined from mosaic \texttt{ERR} image using Kron parameter $k=1.4$\\
\enddata
\tablecomments{The circular aperture and Kron photometry parameters for this \texttt{MIRI} HDU follow those for the \texttt{CIRC}, \texttt{CIRC\_BSUB}, and \texttt{KRON} HDUs. Quantities with a \texttt{\_BSUB} suffix have background subtraction applied. The Kron photometry fields with an \texttt{\_S} suffix
use a Kron parameter $k=1.4$, while fields without the suffix use $k=2.5$.}
\end{deluxetable*}

\newpage
\subsection{\texttt{PHOTOZ} Header Data Unit}
\label{sec:photoz-hdu}

Table \ref{tab:photoz-hdu} details the contents of the \texttt{PHOTOZ}
HDU that provides photometric redshifts determined from the $r=0.1"$ radius circular
aperture photometry (\texttt{CIRC1}) performed on the unconvolved image mosaics.
In addition to the source identifier \texttt{ID}, photometric redshift
information inferred from the EAZY \citep{brammer2008a} analysis
discussed in Section \ref{sec:photometric-redshifts} is presented.
This information includes the EAZY best-fit redshift \texttt{z\_a}, the 
maximum likelihood redshift texttt{z\_ml}, the redshift \texttt{z\_peak} at
the peak of the $\exp(-\chi^2)$ surface, confidence
intervals on the photometric redshift, the probability of the source
lying at selected redshifts, and information on potential low-redshift solutions.
The binned photometric redshift distribution for each source is also provided
in the \texttt{Prob\_z\_bins} array for each \texttt{z\_bins} redshift bin.

\begin{deluxetable*}{lccp{10cm}}[!ht]
\centerwidetable
\tablewidth{0pt}
\tablecaption{\texttt{PHOTOZ} Header Data Unit \label{tab:photoz-hdu}}
\tablehead{
\colhead{Field} & \colhead{Units} & \colhead{Datatype} & \colhead{Description}
}
\startdata
\texttt{ID}       & --- & \texttt{int32}   & Unique source identifier \\
\texttt{z\_a}     & --- & \texttt{float32} & Photometric redshift\\
\texttt{z\_ml}    & --- & \texttt{float32} & Maximum-likelihood photometric redshift\\
\texttt{chi\_a}   & --- & \texttt{float32} & $\chi^2$ value associated with photometric redshift\\
\texttt{l68}      & --- & \texttt{float32} & Lower 68\% confidence interval on photometric redshift\\
\texttt{u68}      & --- & \texttt{float32} & Upper 68\% confidence interval on photometric redshift\\
\texttt{l95}      & --- & \texttt{float32} & Lower 95\% confidence interval on photometric redshift\\
\texttt{u95}      & --- & \texttt{float32} & Upper 95\% confidence interval on photometric redshift\\
\texttt{l99}      & --- & \texttt{float32} & Lower 99\% confidence interval on photometric redshift\\
\texttt{u99}      & --- & \texttt{float32} & Upper 99\% confidence interval on photometric redshift\\
\texttt{nfilt}    & --- & \texttt{float32} & Number of photometric filters used to constrain SED for photometric redshift\\
\texttt{z\_peak}  & --- & \texttt{float32} & Value of photometric redshift at peak of $\exp -\chi^2$ \\
\texttt{chi\_peak}  & --- & \texttt{float32} & Value of $\chi^2$ at peak of $\exp -\chi^2$ \\
\texttt{z025}  & --- & \texttt{float32} & Photometric redshift at marginal 2.5\% cumulative probability \\
\texttt{z160}  & --- & \texttt{float32} & Photometric redshift at marginal 16.0\% cumulative probability \\
\texttt{z500}  & --- & \texttt{float32} & Median of marginal photometric redshift probability distribution\\
\texttt{z840}  & --- & \texttt{float32} & Photometric redshift at marginal 84.0\% cumulative probability \\
\texttt{z975}  & --- & \texttt{float32} & Photometric redshift at marginal 97.5\% cumulative probability \\
\texttt{Prob\_gt\_5}  & --- & \texttt{float32} & Integrated probability of photometric redshift $z>5$\\
\texttt{Prob\_gt\_6}  & --- & \texttt{float32} & Integrated probability of photometric redshift $z>6$\\
\texttt{Prob\_gt\_7}  & --- & \texttt{float32} & Integrated probability of photometric redshift $z>7$\\
\texttt{Prob\_gt\_8}  & --- & \texttt{float32} & Integrated probability of photometric redshift $z>8$\\
\texttt{Prob\_gt\_9}  & --- & \texttt{float32} & Integrated probability of photometric redshift $z>9$\\
\texttt{chisq\_z\_lt\_7}  & --- & \texttt{float32} & Minimum $\chi^2$ at photometric redshift $z<7$\\
\texttt{z\_chisq\_z\_lt\_7}  & --- & \texttt{float32} & Photometric redshift at minimum $\chi^2$ at photometric redshift $z<7$\\
\texttt{chisq\_z\_lt\_6}  & --- & \texttt{float32} & Minimum $\chi^2$ at photometric redshift $z<6$\\
\texttt{z\_chisq\_z\_lt\_6}  & --- & \texttt{float32} & Photometric redshift at minimum $\chi^2$ at photometric redshift $z<6$\\
\texttt{chisq\_z\_lt\_5}  & --- & \texttt{float32} & Minimum $\chi^2$ at photometric redshift $z<5$\\
\texttt{z\_chisq\_z\_lt\_5}  & --- & \texttt{float32} & Photometric redshift at minimum $\chi^2$ at photometric redshift $z<5$\\
\texttt{chisq\_z\_lt\_4}  & --- & \texttt{float32} & Minimum $\chi^2$ at photometric redshift $z<4$\\
\texttt{z\_chisq\_z\_lt\_4}  & --- & \texttt{float32} & Photometric redshift at minimum $\chi^2$ at photometric redshift $z<4$\\
\texttt{z\_bins}  & --- & 22$\times$\texttt{float32} & Redshift bins for photometric redshift probability distribution\\
\texttt{Prob\_z\_bins}  & --- & 22$\times$\texttt{float32} & Binned photometric redshift probability distribution\\
\enddata
\tablecomments{This HDU reports photometric redshift information inferred from $r=0.1"$ radius circular aperture (\texttt{CIRC1} photometry performed on the unconvolved image mosaics, as described in Section \ref{sec:photometric-redshifts} and Appendix \ref{sec:photoz-hdu}.}
\end{deluxetable*}

\newpage
\subsection{\texttt{PHOTOZ\_KRON} Header Data Unit}
\label{sec:photoz-kron-hdu}

The \texttt{PHOTOZ\_KRON} HDU reports photometric redshift information
inferred from applying EAZY \citep{brammer2008a} to the 
\citet{kron1980a} ellipsoidal aperture photometry performed on
common PSF images. The details of the photometric redshift measurements
are discussed in Section \ref{sec:photometric-redshifts}.
The properties in this HDU follow the same format and units of
the \texttt{PHOTOZ} HDU described in Appendix \ref{sec:photoz-hdu}.

\begin{deluxetable*}{lccp{10cm}}[!ht]
\centerwidetable
\tablewidth{0pt}
\tablecaption{\texttt{PHOTOZ\_KRON} Header Data Unit \label{tab:photoz-kron-hdu}}
\tablehead{
\colhead{Field} & \colhead{Units} & \colhead{Datatype} & \colhead{Description}
}
\startdata
\texttt{ID}       & --- & \texttt{int32}   & Unique source identifier \\
\texttt{z\_a}     & --- & \texttt{float32} & Photometric redshift\\
\texttt{z\_ml}    & --- & \texttt{float32} & Maximum-likelihood photometric redshift\\
\texttt{chi\_a}   & --- & \texttt{float32} & $\chi^2$ value associated with photometric redshift\\
\texttt{l68}      & --- & \texttt{float32} & Lower 68\% confidence interval on photometric redshift\\
\texttt{u68}      & --- & \texttt{float32} & Upper 68\% confidence interval on photometric redshift\\
\texttt{l95}      & --- & \texttt{float32} & Lower 95\% confidence interval on photometric redshift\\
\texttt{u95}      & --- & \texttt{float32} & Upper 95\% confidence interval on photometric redshift\\
\texttt{l99}      & --- & \texttt{float32} & Lower 99\% confidence interval on photometric redshift\\
\texttt{u99}      & --- & \texttt{float32} & Upper 99\% confidence interval on photometric redshift\\
\texttt{nfilt}    & --- & \texttt{float32} & Number of photometric filters used to constrain SED for photometric redshift\\
\texttt{z\_peak}  & --- & \texttt{float32} & Value of photometric redshift at peak of $\exp -\chi^2$ \\
\texttt{chi\_peak}  & --- & \texttt{float32} & Value of $\chi^2$ at peak of $\exp -\chi^2$ \\
\texttt{z025}  & --- & \texttt{float32} & Photometric redshift at marginal 2.5\% cumulative probability \\
\texttt{z160}  & --- & \texttt{float32} & Photometric redshift at marginal 16.0\% cumulative probability \\
\texttt{z500}  & --- & \texttt{float32} & Median of marginal photometric redshift probability distribution\\
\texttt{z840}  & --- & \texttt{float32} & Photometric redshift at marginal 84.0\% cumulative probability \\
\texttt{z975}  & --- & \texttt{float32} & Photometric redshift at marginal 97.5\% cumulative probability \\
\texttt{Prob\_gt\_5}  & --- & \texttt{float32} & Integrated probability of photometric redshift $z>5$\\
\texttt{Prob\_gt\_6}  & --- & \texttt{float32} & Integrated probability of photometric redshift $z>6$\\
\texttt{Prob\_gt\_7}  & --- & \texttt{float32} & Integrated probability of photometric redshift $z>7$\\
\texttt{Prob\_gt\_8}  & --- & \texttt{float32} & Integrated probability of photometric redshift $z>8$\\
\texttt{Prob\_gt\_9}  & --- & \texttt{float32} & Integrated probability of photometric redshift $z>9$\\
\texttt{chisq\_z\_lt\_7}  & --- & \texttt{float32} & Minimum $\chi^2$ at photometric redshift $z<7$\\
\texttt{z\_chisq\_z\_lt\_7}  & --- & \texttt{float32} & Photometric redshift at minimum $\chi^2$ at photometric redshift $z<7$\\
\texttt{chisq\_z\_lt\_6}  & --- & \texttt{float32} & Minimum $\chi^2$ at photometric redshift $z<6$\\
\texttt{z\_chisq\_z\_lt\_6}  & --- & \texttt{float32} & Photometric redshift at minimum $\chi^2$ at photometric redshift $z<6$\\
\texttt{chisq\_z\_lt\_5}  & --- & \texttt{float32} & Minimum $\chi^2$ at photometric redshift $z<5$\\
\texttt{z\_chisq\_z\_lt\_5}  & --- & \texttt{float32} & Photometric redshift at minimum $\chi^2$ at photometric redshift $z<5$\\
\texttt{chisq\_z\_lt\_4}  & --- & \texttt{float32} & Minimum $\chi^2$ at photometric redshift $z<4$\\
\texttt{z\_chisq\_z\_lt\_4}  & --- & \texttt{float32} & Photometric redshift at minimum $\chi^2$ at photometric redshift $z<4$\\
\texttt{z\_bins}  & --- & 22$\times$\texttt{float32} & Redshift bins for photometric redshift probability distribution\\
\texttt{Prob\_z\_bins}  & --- & 22$\times$\texttt{float32} & Binned photometric redshift probability distribution\\
\enddata
\tablecomments{This HDU reports photometric redshift information inferred from ellipsoidal \citet{kron1980a} photometry performed on the common PSF image mosaics, as described in Section \ref{sec:photometric-redshifts} and Appendix \ref{sec:photoz-kron-hdu}. The common PSF mosaics are described in Section \ref{sec:common-psf-mosaics} and the Kron
photometry method is detailed in Section \ref{sec:kron-aperture-photometry}.}
\end{deluxetable*}

\newpage
\subsection{\texttt{GROWTH} Catalog}
\label{sec:growth-catalog}

The curve-of-growth measurements described in Section \ref{sec:curve-of-growth}
result in a large number of fields for each object for each of the 35 mosaic
images. As a result, the curve-of-growth measurements are separated into a 
distinct High Level Science Product catalog with separate HDUs for each
\texttt{BAND} labeled by its filter name (e.g., \texttt{F070W}).
Table \ref{tab:growth-band-hdu} details the contents of these 35 HDUs in the
\texttt{GROWTH} catalog, measured from the unconvolved mosaic images.
Each source identifier and sky position is provided, along with the
total flux within the outer aperture of the curve of growth defined
by twice the Kron aperture (see Sections \ref{sec:kron-aperture-photometry}
and \ref{sec:curve-of-growth}). The Kron aperture semimajor and semiminor
axes are reported (\texttt{A\_KRON} and \texttt{B\_KRON}), as well as the
position angle \texttt{THETA} measured counter clockwise from the image $x$-axis.
The flux uncertainty determined from
model regression (\texttt{\_e} suffix) and the \texttt{ERR} image (\texttt{\_ei} suffix)
are provided, including a field \texttt{F\_SKY\_e}) recording the
contribution of the sky noise to the total uncertainty. The curve-of-growth
measurements have a local background subtraction applied, which is 
reported in \texttt{F\_TOT\_bkg} and measured from the annuli defined by
inner and outer semimajor axes (\texttt{A\_IN\_BKG} and \texttt{A\_OUT\_BKG},
respectively) and the Kron aperture axis ratio \texttt{Q}. The apertures
containing 5-100\% of the total flux are reported as \texttt{A\_5} through
\texttt{A\_100} in 5\% increments.

\begin{deluxetable}{lccp{10cm}}[!ht]
\tablewidth{0pt}
\tablecaption{\texttt{GROWTH} \texttt{\{BAND\}} Header Data Unit \label{tab:growth-band-hdu}}
\tablehead{
\colhead{Field} & \colhead{Units} & \colhead{Datatype} & \colhead{Description}
}
\startdata
\texttt{ID}  & --- & \texttt{int32}   & Unique source identifier \\
\texttt{RA}  & deg & \texttt{float32} & Right ascension \\
\texttt{DEC} & deg & \texttt{float32} & Declination \\
\texttt{F\_TOT} & nJy & \texttt{float32} & Total flux within twice the Kron aperture determined from \texttt{\{BAND\}} mosaic\\
\texttt{F\_SKY\_e} & nJy & \texttt{float32} & Contribution of the sky background to the \texttt{F\_TOT} flux uncertainty, determined from the uncertainty regression model.\\
\texttt{F\_TOT\_ei} & nJy & \texttt{float32} & Uncertainty estimate for \texttt{F\_TOT} measured from the mosaic \texttt{ERR} HDU.\\
\texttt{F\_TOT\_bkg} & nJy & \texttt{float32} & Background subtracted from the \texttt{F\_TOT} flux.\\
\texttt{A\_KRON} & arcsec & \texttt{float32} & Source aperture semimajor axis determined with Kron parameter $k=2.5$\\
\texttt{B\_KRON} & arcsec & \texttt{float32} & Source aperture semiminor axis determined with Kron parameter $k=2.5$\\
\texttt{A\_IN\_BKG} & arcsec & \texttt{float32} & Inner ellipsoidal annulus aperture semimajor axis used to determine background.\\
\texttt{A\_OUT\_BKG} & arcsec & \texttt{float32} & Outer ellipsoidal annulus aperture semimajor axis used to determine background.\\
\texttt{Q} & --- & \texttt{float32} & Axis ratio of ellipsoidal annulus used to determine background.\\
\texttt{THETA} & deg & \texttt{float32} & Position angle of apertures relative to image x axis\\
\texttt{A\_5} & arcsec & \texttt{float32} & Semimajor axis of ellipsoidal aperture containing 5\% of \texttt{F\_TOT}\\
\texttt{A\_10} & arcsec & \texttt{float32} & Semimajor axis of ellipsoidal aperture  containing 10\% of \texttt{F\_TOT}\\
\texttt{A\_15} & arcsec & \texttt{float32} & Semimajor axis of ellipsoidal aperture  containing 15\% of \texttt{F\_TOT}\\
\texttt{A\_20} & arcsec & \texttt{float32} & Semimajor axis of ellipsoidal aperture  containing 20\% of \texttt{F\_TOT}\\
\texttt{A\_25} & arcsec & \texttt{float32} & Semimajor axis of ellipsoidal aperture  containing 25\% of \texttt{F\_TOT}\\
\texttt{A\_30} & arcsec & \texttt{float32} & Semimajor axis of ellipsoidal aperture  containing 30\% of \texttt{F\_TOT}\\
\texttt{A\_35} & arcsec & \texttt{float32} & Semimajor axis of ellipsoidal aperture  containing 35\% of \texttt{F\_TOT}\\
\texttt{A\_40} & arcsec & \texttt{float32} & Semimajor axis of ellipsoidal aperture  containing 40\% of \texttt{F\_TOT}\\
\texttt{A\_45} & arcsec & \texttt{float32} & Semimajor axis of ellipsoidal aperture  containing 45\% of \texttt{F\_TOT}\\
\texttt{A\_50} & arcsec & \texttt{float32} & Semimajor axis of ellipsoidal aperture  containing 50\% of \texttt{F\_TOT}\\
\texttt{A\_55} & arcsec & \texttt{float32} & Semimajor axis of ellipsoidal aperture  containing 55\% of \texttt{F\_TOT}\\
\texttt{A\_60} & arcsec & \texttt{float32} & Semimajor axis of ellipsoidal aperture  containing 60\% of \texttt{F\_TOT}\\
\texttt{A\_65} & arcsec & \texttt{float32} & Semimajor axis of ellipsoidal aperture  containing 65\% of \texttt{F\_TOT}\\
\texttt{A\_70} & arcsec & \texttt{float32} & Semimajor axis of ellipsoidal aperture  containing 70\% of \texttt{F\_TOT}\\
\texttt{A\_75} & arcsec & \texttt{float32} & Semimajor axis of ellipsoidal aperture  containing 75\% of \texttt{F\_TOT}\\
\texttt{A\_80} & arcsec & \texttt{float32} & Semimajor axis of ellipsoidal aperture  containing 80\% of \texttt{F\_TOT}\\
\texttt{A\_85} & arcsec & \texttt{float32} & Semimajor axis of ellipsoidal aperture  containing 85\% of \texttt{F\_TOT}\\
\texttt{A\_90} & arcsec & \texttt{float32} & Semimajor axis of ellipsoidal aperture  containing 90\% of \texttt{F\_TOT}\\
\texttt{A\_95} & arcsec & \texttt{float32} & Semimajor axis of ellipsoidal aperture  containing 95\% of \texttt{F\_TOT}\\
\texttt{A\_100} & arcsec & \texttt{float32} & Semimajor axis of ellipsoidal aperture  containing 100\% of \texttt{F\_TOT}\\
\enddata
\tablecomments{The \texttt{GROWTH} catalog consists of 35 HDUs, one for each filter. The above
Table lists the properties of these 35 HDUs and are measured from the unconvolved image mosaics as described in Section \ref{sec:curve-of-growth} and Appendix \ref{sec:growth-catalog}.}
\end{deluxetable}

\newpage

\subsection{\texttt{GROWTH\_CONV} Catalog}
\label{sec:growth-conv-catalog}

Owing to the large number of measurements per object and the
large number of filters in the dataset,
the curve-of-growth measurements on the common PSF images are 
provided in a standalone \texttt{GROWTH\_CONV} catalog.
In this catalog, each of the 35 filters has a corresponding
HDU with contents detailed in Table \ref{tab:growth-conv-band-hdu}.
The properties of these \texttt{GROWTH\_CONV} catalog HDUs
are defined in the same manner as for the \texttt{GROWTH} catalog,
as described in Section \ref{sec:curve-of-growth} and Appendix \ref{sec:growth-catalog} above. These properties include the
semimajor axes of apertures containing 5-100\% of the total
flux \texttt{F\_TOT}
of the object in increments of 5\% and the axis ratio of the
apertures. Flux uncertainty information is also provided for each source.
We note that only filters blueward of F444W are brought to a common PSF
approximating the F444W PSF, while longer wavelength filters are
left unconvolved.

\begin{deluxetable}{lccp{10cm}}[!ht]
\tablewidth{0pt}
\tablecaption{\texttt{GROWTH\_CONV} \texttt{\{BAND\}} Header Data Unit \label{tab:growth-conv-band-hdu}}
\tablehead{
\colhead{Field} & \colhead{Units} & \colhead{Datatype} & \colhead{Description}
}
\startdata
\texttt{ID}  & --- & \texttt{int32}   & Unique source identifier \\
\texttt{RA}  & deg & \texttt{float32} & Right ascension \\
\texttt{DEC} & deg & \texttt{float32} & Declination \\
\texttt{F\_TOT} & nJy & \texttt{float32} & Total flux within twice the Kron aperture determined from common-PSF mosaic in \texttt{\{BAND\}}\\
\texttt{F\_SKY\_e} & nJy & \texttt{float32} & Contribution of the sky background to the \texttt{F\_TOT} flux uncertainty, determined from the uncertainty regression model.\\
\texttt{F\_TOT\_ei} & nJy & \texttt{float32} & Uncertainty estimate for \texttt{F\_TOT} measured from the mosaic \texttt{ERR} HDU.\\
\texttt{F\_TOT\_bkg} & nJy & \texttt{float32} & Background subtracted from the \texttt{F\_TOT} flux.\\
\texttt{A\_KRON} & arcsec & \texttt{float32} & Source aperture semimajor axis determined with Kron parameter $k=2.5$\\
\texttt{B\_KRON} & arcsec & \texttt{float32} & Source aperture semiminor axis determined with Kron parameter $k=2.5$\\
\texttt{A\_IN\_BKG} & arcsec & \texttt{float32} & Inner ellipsoidal annulus aperture semimajor axis used to determine background.\\
\texttt{A\_OUT\_BKG} & arcsec & \texttt{float32} & Outer ellipsoidal annulus aperture semimajor axis used to determine background.\\
\texttt{Q} & --- & \texttt{float32} & Axis ratio of ellipsoidal annulus used to determine background.\\
\texttt{THETA} & deg & \texttt{float32} & Position angle of apertures relative to image x axis\\
\texttt{A\_5} & arcsec & \texttt{float32} & Semimajor axis of ellipsoidal aperture containing 5\% of \texttt{F\_TOT}\\
\texttt{A\_10} & arcsec & \texttt{float32} & Semimajor axis of ellipsoidal aperture  containing 10\% of \texttt{F\_TOT}\\
\texttt{A\_15} & arcsec & \texttt{float32} & Semimajor axis of ellipsoidal aperture  containing 15\% of \texttt{F\_TOT}\\
\texttt{A\_20} & arcsec & \texttt{float32} & Semimajor axis of ellipsoidal aperture  containing 20\% of \texttt{F\_TOT}\\
\texttt{A\_25} & arcsec & \texttt{float32} & Semimajor axis of ellipsoidal aperture  containing 25\% of \texttt{F\_TOT}\\
\texttt{A\_30} & arcsec & \texttt{float32} & Semimajor axis of ellipsoidal aperture  containing 30\% of \texttt{F\_TOT}\\
\texttt{A\_35} & arcsec & \texttt{float32} & Semimajor axis of ellipsoidal aperture  containing 35\% of \texttt{F\_TOT}\\
\texttt{A\_40} & arcsec & \texttt{float32} & Semimajor axis of ellipsoidal aperture  containing 40\% of \texttt{F\_TOT}\\
\texttt{A\_45} & arcsec & \texttt{float32} & Semimajor axis of ellipsoidal aperture  containing 45\% of \texttt{F\_TOT}\\
\texttt{A\_50} & arcsec & \texttt{float32} & Semimajor axis of ellipsoidal aperture  containing 50\% of \texttt{F\_TOT}\\
\texttt{A\_55} & arcsec & \texttt{float32} & Semimajor axis of ellipsoidal aperture  containing 55\% of \texttt{F\_TOT}\\
\texttt{A\_60} & arcsec & \texttt{float32} & Semimajor axis of ellipsoidal aperture  containing 60\% of \texttt{F\_TOT}\\
\texttt{A\_65} & arcsec & \texttt{float32} & Semimajor axis of ellipsoidal aperture  containing 65\% of \texttt{F\_TOT}\\
\texttt{A\_70} & arcsec & \texttt{float32} & Semimajor axis of ellipsoidal aperture  containing 70\% of \texttt{F\_TOT}\\
\texttt{A\_75} & arcsec & \texttt{float32} & Semimajor axis of ellipsoidal aperture  containing 75\% of \texttt{F\_TOT}\\
\texttt{A\_80} & arcsec & \texttt{float32} & Semimajor axis of ellipsoidal aperture  containing 80\% of \texttt{F\_TOT}\\
\texttt{A\_85} & arcsec & \texttt{float32} & Semimajor axis of ellipsoidal aperture  containing 85\% of \texttt{F\_TOT}\\
\texttt{A\_90} & arcsec & \texttt{float32} & Semimajor axis of ellipsoidal aperture  containing 90\% of \texttt{F\_TOT}\\
\texttt{A\_95} & arcsec & \texttt{float32} & Semimajor axis of ellipsoidal aperture  containing 95\% of \texttt{F\_TOT}\\
\texttt{A\_100} & arcsec & \texttt{float32} & Semimajor axis of ellipsoidal aperture  containing 100\% of \texttt{F\_TOT}\\
\enddata
\tablecomments{The \texttt{GROWTH\_CONV} catalog consists of 35 HDUs, one for each filter. The above
Table lists the properties of these 35 HDUs and are measured from the common PSF image mosaics as described in Sections \ref{sec:common-psf-mosaics} and \ref{sec:curve-of-growth} and Appendix \ref{sec:growth-conv-catalog}. Note that only filters blueward of F444W are brought to a common PSF approximating F444W, while longer wavelength filters are left unconvolved.}
\end{deluxetable}


\bibliography{main}{}
\bibliographystyle{aasjournalv7}


\end{document}